\ifdefined\PRformat
\documentclass[usenatbib, twocolumn, nofootinbib, superscriptaddress, longbibliography, reprint, prd]{revtex4-1}
\else
    \documentclass[a4paper,11pt]{article}
    \pdfoutput=1
    \usepackage{jcappub}
\fi

\usepackage{adjustbox}
\usepackage{orcidlink}
\usepackage{makecell}
\usepackage{multirow}
\usepackage{bigints}
\usepackage{natbib}
\usepackage{color}
\usepackage{wrapfig}
\usepackage{amsmath,amsfonts,bm}
\usepackage{hyperref}
\usepackage[T1]{fontenc}
\usepackage{graphicx}   
\usepackage[mathlines]{lineno}
\usepackage{wrapfig}
\usepackage{fontawesome}
\usepackage{enumitem}
\usepackage[normalem]{ulem}  

\newcommand{\abstracttext}
{
We present the first reconstruction of the cosmic microwave background (CMB) lensing potential from the SPT-3G Summer survey using two years of temperature data. The Summer survey has a total area of approximately 2640 deg$^2$, split into three fields covering 1210, 570, and 860 deg$^2$, respectively. 
A joint analysis of the three Summer fields yields a lensing amplitude of $A^{\rm comb} = { 1.014} \pm 0.053$ relative to a fiducial \textit{Planck} 2018 $\Lambda$CDM cosmology for the multipole range $50 < L < 2000$. 
These early results from the SPT-3G Summer survey highlight the potential for increasing the  signal-to-noise ratio when combining the Summer fields with the SPT-3G Main and Wide fields for a total SPT-3G survey area of $\sim$ 10 000 deg$^2$.
}

\newcommand{\tittext}{CMB Lensing Reconstruction Using Two Years of Temperature Data from the SPT-3G Summer Survey}

\ifdefined\PRformat
\else

\fi
\defcitealias{smith17}{SF17}
\defcitealias{ferraro18}{FS18}
\defcitealias{raghunathan23}{RO23}

\begin{document}

\title{\tittext}
\ifdefined\PRformat
    \author{K.~Levy}
    \affiliation{School of Physics, University of Melbourne, Parkville, VIC 3010, Australia}
    \author{S.~Raghunathan\,\orcidlink{0000-0003-1405-378X}}
    \affiliation{Department of Physics \& Astronomy, University of California, One Shields Avenue, Davis, CA 95616, USA}
    \affiliation{Center for AstroPhysical Surveys, National Center for Supercomputing Applications, Urbana, IL, 61801, USA}
    \author{F.~Guidi\,\orcidlink{0000-0001-7593-3962}}
    \affiliation{Department of Physics \& Astronomy, University of California, One Shields Avenue, Davis, CA 95616, USA}
    \affiliation{Sorbonne Universit\'e, CNRS, UMR 7095, Institut d'Astrophysique de Paris, 98 bis bd Arago, 75014 Paris, France}
    \author{Y.~Li\,\orcidlink{0000-0002-4820-1122}}
    \affiliation{Kavli Institute for Cosmological Physics, University of Chicago, 5640 South Ellis Avenue, Chicago, IL, 60637, USA}
    \author{E.~Anderes\,\orcidlink{0009-0003-3245-3979}}
    \affiliation{Department of Statistics, University of California, One Shields Avenue, Davis, CA 95616, USA}
    \author{A.~J.~Anderson\,\orcidlink{0000-0002-4435-4623}}
    \affiliation{Fermi National Accelerator Laboratory, MS209, P.O. Box 500, Batavia, IL, 60510, USA}
    \affiliation{Kavli Institute for Cosmological Physics, University of Chicago, 5640 South Ellis Avenue, Chicago, IL, 60637, USA}
    \affiliation{Department of Astronomy and Astrophysics, University of Chicago, 5640 South Ellis Avenue, Chicago, IL, 60637, USA}
    \author{B.~Ansarinejad}
    \affiliation{School of Physics, University of Melbourne, Parkville, VIC 3010, Australia}
    \author{M.~Archipley\,\orcidlink{0000-0002-0517-9842}}
    \affiliation{Department of Astronomy and Astrophysics, University of Chicago, 5640 South Ellis Avenue, Chicago, IL, 60637, USA}
    \affiliation{Kavli Institute for Cosmological Physics, University of Chicago, 5640 South Ellis Avenue, Chicago, IL, 60637, USA}
    \author{L.~Balkenhol\,\orcidlink{0000-0001-6899-1873}}
    \affiliation{Sorbonne Universit\'e, CNRS, UMR 7095, Institut d'Astrophysique de Paris, 98 bis bd Arago, 75014 Paris, France}
    \author{D.~R.~Barron\,\orcidlink{0000-0002-1623-5651}}
    \affiliation{Department of Physics and Astronomy, University of New Mexico, Albuquerque, NM, 87131, USA}
    \author{K.~Benabed}
    \affiliation{Sorbonne Universit\'e, CNRS, UMR 7095, Institut d'Astrophysique de Paris, 98 bis bd Arago, 75014 Paris, France}
    \author{A.~N.~Bender\,\orcidlink{0000-0001-5868-0748}}
    \affiliation{High-Energy Physics Division, Argonne National Laboratory, 9700 South Cass Avenue, Lemont, IL, 60439, USA}
    \affiliation{Kavli Institute for Cosmological Physics, University of Chicago, 5640 South Ellis Avenue, Chicago, IL, 60637, USA}
    \affiliation{Department of Astronomy and Astrophysics, University of Chicago, 5640 South Ellis Avenue, Chicago, IL, 60637, USA}
    \author{B.~A.~Benson\,\orcidlink{0000-0002-5108-6823}}
    \affiliation{Fermi National Accelerator Laboratory, MS209, P.O. Box 500, Batavia, IL, 60510, USA}
    \affiliation{Kavli Institute for Cosmological Physics, University of Chicago, 5640 South Ellis Avenue, Chicago, IL, 60637, USA}
    \affiliation{Department of Astronomy and Astrophysics, University of Chicago, 5640 South Ellis Avenue, Chicago, IL, 60637, USA}
    \author{F.~Bianchini\,\orcidlink{0000-0003-4847-3483}}
    \affiliation{Kavli Institute for Particle Astrophysics and Cosmology, Stanford University, 452 Lomita Mall, Stanford, CA, 94305, USA}
    \affiliation{Department of Physics, Stanford University, 382 Via Pueblo Mall, Stanford, CA, 94305, USA}
    \affiliation{SLAC National Accelerator Laboratory, 2575 Sand Hill Road, Menlo Park, CA, 94025, USA}
    \author{L.~E.~Bleem\,\orcidlink{0000-0001-7665-5079}}
    \affiliation{High-Energy Physics Division, Argonne National Laboratory, 9700 South Cass Avenue, Lemont, IL, 60439, USA}
    \affiliation{Kavli Institute for Cosmological Physics, University of Chicago, 5640 South Ellis Avenue, Chicago, IL, 60637, USA}
    \affiliation{Department of Astronomy and Astrophysics, University of Chicago, 5640 South Ellis Avenue, Chicago, IL, 60637, USA}
    \author{S.~Bocquet\,\orcidlink{0000-0002-4900-805X}}
    \affiliation{University Observatory, Faculty of Physics, LMU Munich, Scheinerstr.~1, 81679 Munich, Germany}
    \author{F.~R.~Bouchet\,\orcidlink{0000-0002-8051-2924}}
    \affiliation{Sorbonne Universit\'e, CNRS, UMR 7095, Institut d'Astrophysique de Paris, 98 bis bd Arago, 75014 Paris, France}
    \author{E.~Camphuis\,\orcidlink{0000-0003-3483-8461}}
    \affiliation{Sorbonne Universit\'e, CNRS, UMR 7095, Institut d'Astrophysique de Paris, 98 bis bd Arago, 75014 Paris, France}
    \author{M.~G.~Campitiello}
    \affiliation{High-Energy Physics Division, Argonne National Laboratory, 9700 South Cass Avenue, Lemont, IL, 60439, USA}
    \author{J.~E.~Carlstrom\,\orcidlink{0000-0002-2044-7665}}
    \affiliation{Kavli Institute for Cosmological Physics, University of Chicago, 5640 South Ellis Avenue, Chicago, IL, 60637, USA}
    \affiliation{Enrico Fermi Institute, University of Chicago, 5640 South Ellis Avenue, Chicago, IL, 60637, USA}
    \affiliation{Department of Physics, University of Chicago, 5640 South Ellis Avenue, Chicago, IL, 60637, USA}
    \affiliation{High-Energy Physics Division, Argonne National Laboratory, 9700 South Cass Avenue, Lemont, IL, 60439, USA}
    \affiliation{Department of Astronomy and Astrophysics, University of Chicago, 5640 South Ellis Avenue, Chicago, IL, 60637, USA}
    \author{J.~Carron\,\orcidlink{0000-0002-5751-1392}}
    \affiliation{Istituto ricerche solari Aldo e Cele Dacc\`o (IRSOL), Faculty of Informatics, Universit\`a della Svizzera italiana, 6605 Locarno, Switzerland}
    \affiliation{Universit\'e de Gen\`eve, D\'epartement de Physique Th\'eorique, 24 Quai Ansermet, CH-1211 Gen\`eve 4, Switzerland}
    \author{C.~L.~Chang}
    \affiliation{High-Energy Physics Division, Argonne National Laboratory, 9700 South Cass Avenue, Lemont, IL, 60439, USA}
    \affiliation{Kavli Institute for Cosmological Physics, University of Chicago, 5640 South Ellis Avenue, Chicago, IL, 60637, USA}
    \affiliation{Department of Astronomy and Astrophysics, University of Chicago, 5640 South Ellis Avenue, Chicago, IL, 60637, USA}
    \author{P.~M.~Chichura\,\orcidlink{0000-0002-5397-9035}}
    \affiliation{Department of Physics, University of Chicago, 5640 South Ellis Avenue, Chicago, IL, 60637, USA}
    \affiliation{Kavli Institute for Cosmological Physics, University of Chicago, 5640 South Ellis Avenue, Chicago, IL, 60637, USA}
    \author{A.~Chokshi}
    \affiliation{Department of Astronomy and Astrophysics, University of Chicago, 5640 South Ellis Avenue, Chicago, IL, 60637, USA}
    \author{T.-L.~Chou\,\orcidlink{0000-0002-3091-8790}}
    \affiliation{Department of Astronomy and Astrophysics, University of Chicago, 5640 South Ellis Avenue, Chicago, IL, 60637, USA}
    \affiliation{Kavli Institute for Cosmological Physics, University of Chicago, 5640 South Ellis Avenue, Chicago, IL, 60637, USA}
    \affiliation{National Taiwan University, No. 1, Sec. 4, Roosevelt Road, Taipei 106319, Taiwan}
    \author{A.~Coerver\,\orcidlink{0000-0002-2707-1672}}
    \affiliation{Department of Physics, University of California, Berkeley, CA, 94720, USA}
    \author{T.~M.~Crawford\,\orcidlink{0000-0001-9000-5013}}
    \affiliation{Department of Astronomy and Astrophysics, University of Chicago, 5640 South Ellis Avenue, Chicago, IL, 60637, USA}
    \affiliation{Kavli Institute for Cosmological Physics, University of Chicago, 5640 South Ellis Avenue, Chicago, IL, 60637, USA}
    \author{C.~Daley\,\orcidlink{0000-0002-3760-2086}}
    \affiliation{Universit\'e Paris-Saclay, Universit\'e Paris Cit\'e, CEA, CNRS, AIM, 91191, Gif-sur-Yvette, France}
    \affiliation{Department of Astronomy, University of Illinois Urbana-Champaign, 1002 West Green Street, Urbana, IL, 61801, USA}
    \author{T.~de~Haan\,\orcidlink{0000-0001-5105-9473}}
    \affiliation{High Energy Accelerator Research Organization (KEK), Tsukuba, Ibaraki 305-0801, Japan}
    \author{K.~R.~Dibert}
    \affiliation{Department of Astronomy and Astrophysics, University of Chicago, 5640 South Ellis Avenue, Chicago, IL, 60637, USA}
    \affiliation{Kavli Institute for Cosmological Physics, University of Chicago, 5640 South Ellis Avenue, Chicago, IL, 60637, USA}
    \author{M.~A.~Dobbs}
    \affiliation{Department of Physics and McGill Space Institute, McGill University, 3600 Rue University, Montreal, Quebec H3A 2T8, Canada}
    \affiliation{Canadian Institute for Advanced Research, CIFAR Program in Gravity and the Extreme Universe, Toronto, ON, M5G 1Z8, Canada}
    \author{M.~Doohan}
    \affiliation{School of Physics, University of Melbourne, Parkville, VIC 3010, Australia}
    \author{D.~Dutcher\,\orcidlink{0000-0002-9962-2058}}
    \affiliation{Joseph Henry Laboratories of Physics, Jadwin Hall, Princeton University, Princeton, NJ 08544, USA}
    \author{C.~Feng}
    \affiliation{Department of Astronomy, University of Science and Technology of China, Hefei 230026, China}
    \affiliation{School of Astronomy and Space Science, University of Science and Technology of China, Hefei 230026}
    \affiliation{Department of Physics, University of Illinois Urbana-Champaign, 1110 West Green Street, Urbana, IL, 61801, USA}
    \author{K.~R.~Ferguson\,\orcidlink{0000-0002-4928-8813}}
    \affiliation{Department of Physics and Astronomy, University of California, Los Angeles, CA, 90095, USA}
    \affiliation{Department of Physics and Astronomy, Michigan State University, East Lansing, MI 48824, USA}
    \author{N.~C.~Ferree\,\orcidlink{0000-0002-7130-7099}}
    \affiliation{California Institute of Technology, 1200 East California Boulevard., Pasadena, CA, 91125, USA}
    \affiliation{Kavli Institute for Particle Astrophysics and Cosmology, Stanford University, 452 Lomita Mall, Stanford, CA, 94305, USA}
    \affiliation{Department of Physics, Stanford University, 382 Via Pueblo Mall, Stanford, CA, 94305, USA}
    \author{K.~Fichman}
    \affiliation{Department of Physics, University of Chicago, 5640 South Ellis Avenue, Chicago, IL, 60637, USA}
    \affiliation{Kavli Institute for Cosmological Physics, University of Chicago, 5640 South Ellis Avenue, Chicago, IL, 60637, USA}
    \author{A.~Foster\,\orcidlink{0000-0002-7145-1824}}
    \affiliation{Joseph Henry Laboratories of Physics, Jadwin Hall, Princeton University, Princeton, NJ 08544, USA}
    \author{S.~Galli}
    \affiliation{Sorbonne Universit\'e, CNRS, UMR 7095, Institut d'Astrophysique de Paris, 98 bis bd Arago, 75014 Paris, France}
    \author{A.~E.~Gambrel}
    \affiliation{Kavli Institute for Cosmological Physics, University of Chicago, 5640 South Ellis Avenue, Chicago, IL, 60637, USA}
    \author{A.~K.~Gao}
    \affiliation{Department of Physics, University of Illinois Urbana-Champaign, 1110 West Green Street, Urbana, IL, 61801, USA}
    \author{F.~Ge}
    \affiliation{California Institute of Technology, 1200 East California Boulevard., Pasadena, CA, 91125, USA}
    \affiliation{Kavli Institute for Particle Astrophysics and Cosmology, Stanford University, 452 Lomita Mall, Stanford, CA, 94305, USA}
    \affiliation{Department of Physics, Stanford University, 382 Via Pueblo Mall, Stanford, CA, 94305, USA}
    \affiliation{Department of Physics \& Astronomy, University of California, One Shields Avenue, Davis, CA 95616, USA}
    \author{S.~Guns}
    \affiliation{Department of Physics, University of California, Berkeley, CA, 94720, USA}
    \author{N.~W.~Halverson}
    \affiliation{CASA, Department of Astrophysical and Planetary Sciences, University of Colorado, Boulder, CO, 80309, USA }
    \affiliation{Department of Physics, University of Colorado, Boulder, CO, 80309, USA}
    \author{E.~Hivon\,\orcidlink{0000-0003-1880-2733}}
    \affiliation{Sorbonne Universit\'e, CNRS, UMR 7095, Institut d'Astrophysique de Paris, 98 bis bd Arago, 75014 Paris, France}
    \author{G.~P.~Holder\,\orcidlink{0000-0002-0463-6394}}
    \affiliation{Department of Physics, University of Illinois Urbana-Champaign, 1110 West Green Street, Urbana, IL, 61801, USA}
    \author{W.~L.~Holzapfel}
    \affiliation{Department of Physics, University of California, Berkeley, CA, 94720, USA}
    \author{J.~C.~Hood}
    \affiliation{Kavli Institute for Cosmological Physics, University of Chicago, 5640 South Ellis Avenue, Chicago, IL, 60637, USA}
    \author{A.~Hryciuk}
    \affiliation{Department of Physics, University of Chicago, 5640 South Ellis Avenue, Chicago, IL, 60637, USA}
    \affiliation{Kavli Institute for Cosmological Physics, University of Chicago, 5640 South Ellis Avenue, Chicago, IL, 60637, USA}
    \author{N.~Huang\,\orcidlink{0000-0003-3595-0359}}
    \affiliation{Department of Physics, University of California, Berkeley, CA, 94720, USA}
    \author{T.~Jhaveri}
    \affiliation{Department of Astronomy and Astrophysics, University of Chicago, 5640 South Ellis Avenue, Chicago, IL, 60637, USA}
    \affiliation{Kavli Institute for Cosmological Physics, University of Chicago, 5640 South Ellis Avenue, Chicago, IL, 60637, USA}
    \author{F.~K\'eruzor\'e}
    \affiliation{High-Energy Physics Division, Argonne National Laboratory, 9700 South Cass Avenue, Lemont, IL, 60439, USA}
    \author{A.~R.~Khalife\,\orcidlink{0000-0002-8388-4950}}
    \affiliation{Sorbonne Universit\'e, CNRS, UMR 7095, Institut d'Astrophysique de Paris, 98 bis bd Arago, 75014 Paris, France}
    \author{L.~Knox}
    \affiliation{Department of Physics \& Astronomy, University of California, One Shields Avenue, Davis, CA 95616, USA}
    \author{K.~Kornoelje}
    \affiliation{Department of Astronomy and Astrophysics, University of Chicago, 5640 South Ellis Avenue, Chicago, IL, 60637, USA}
    \affiliation{Kavli Institute for Cosmological Physics, University of Chicago, 5640 South Ellis Avenue, Chicago, IL, 60637, USA}
    \affiliation{High-Energy Physics Division, Argonne National Laboratory, 9700 South Cass Avenue, Lemont, IL, 60439, USA}
    \author{C.-L.~Kuo}
    \affiliation{Kavli Institute for Particle Astrophysics and Cosmology, Stanford University, 452 Lomita Mall, Stanford, CA, 94305, USA}
    \affiliation{Department of Physics, Stanford University, 382 Via Pueblo Mall, Stanford, CA, 94305, USA}
    \affiliation{SLAC National Accelerator Laboratory, 2575 Sand Hill Road, Menlo Park, CA, 94025, USA}
    \author{A.~E.~Lowitz\,\orcidlink{0000-0002-4747-4276}}
    \affiliation{Kavli Institute for Cosmological Physics, University of Chicago, 5640 South Ellis Avenue, Chicago, IL, 60637, USA}
    \author{C.~Lu}
    \affiliation{Department of Physics, University of Illinois Urbana-Champaign, 1110 West Green Street, Urbana, IL, 61801, USA}
    \author{G.~P.~Lynch\,\orcidlink{0009-0004-3143-1708}}
    \affiliation{Department of Physics \& Astronomy, University of California, One Shields Avenue, Davis, CA 95616, USA}
    \author{T.~J.~Maccarone\,\orcidlink{0000-0003-0976-4755}}
    \affiliation{Department of Physics \& Astronomy, Box 41051, Texas Tech University, Lubbock TX 79409-1051, USA}
    \author{A.~S.~Maniyar\,\orcidlink{0000-0002-4617-9320}}
    \affiliation{Kavli Institute for Particle Astrophysics and Cosmology, Stanford University, 452 Lomita Mall, Stanford, CA, 94305, USA}
    \affiliation{Department of Physics, Stanford University, 382 Via Pueblo Mall, Stanford, CA, 94305, USA}
    \affiliation{SLAC National Accelerator Laboratory, 2575 Sand Hill Road, Menlo Park, CA, 94025, USA}
    \author{E.~S.~Martsen}
    \affiliation{Department of Astronomy and Astrophysics, University of Chicago, 5640 South Ellis Avenue, Chicago, IL, 60637, USA}
    \affiliation{Kavli Institute for Cosmological Physics, University of Chicago, 5640 South Ellis Avenue, Chicago, IL, 60637, USA}
    \author{F.~Menanteau}
    \affiliation{Department of Astronomy, University of Illinois Urbana-Champaign, 1002 West Green Street, Urbana, IL, 61801, USA}
    \affiliation{Center for AstroPhysical Surveys, National Center for Supercomputing Applications, Urbana, IL, 61801, USA}
    \author{M.~Millea\,\orcidlink{0000-0001-7317-0551}}
    \affiliation{Department of Physics, University of California, Berkeley, CA, 94720, USA}
    \author{J.~Montgomery}
    \affiliation{Department of Physics and McGill Space Institute, McGill University, 3600 Rue University, Montreal, Quebec H3A 2T8, Canada}
    \author{Y.~Nakato}
    \affiliation{Department of Physics, Stanford University, 382 Via Pueblo Mall, Stanford, CA, 94305, USA}
    \author{T.~Natoli}
    \affiliation{Kavli Institute for Cosmological Physics, University of Chicago, 5640 South Ellis Avenue, Chicago, IL, 60637, USA}
    \author{Y.~Omori\,\orcidlink{0000-0002-0963-7310}}
    \affiliation{Department of Astronomy and Astrophysics, University of Chicago, 5640 South Ellis Avenue, Chicago, IL, 60637, USA}
    \affiliation{Kavli Institute for Cosmological Physics, University of Chicago, 5640 South Ellis Avenue, Chicago, IL, 60637, USA}
    \affiliation{NSF-Simons AI Institute for the Sky (SkAI), 172 E. Chestnut St., Chicago, IL 60611, USA}
    \author{A.~Ouellette\,\orcidlink{0000-0003-0170-5638}}
    \affiliation{Department of Physics, University of Illinois Urbana-Champaign, 1110 West Green Street, Urbana, IL, 61801, USA}
    \author{Z.~Pan\,\orcidlink{0000-0002-6164-9861}}
    \affiliation{High-Energy Physics Division, Argonne National Laboratory, 9700 South Cass Avenue, Lemont, IL, 60439, USA}
    \affiliation{Kavli Institute for Cosmological Physics, University of Chicago, 5640 South Ellis Avenue, Chicago, IL, 60637, USA}
    \affiliation{Department of Physics, University of Chicago, 5640 South Ellis Avenue, Chicago, IL, 60637, USA}
    \author{P.~Paschos}
    \affiliation{Enrico Fermi Institute, University of Chicago, 5640 South Ellis Avenue, Chicago, IL, 60637, USA}
    \author{K.~A.~Phadke\,\orcidlink{0000-0001-7946-557X}}
    \affiliation{Department of Astronomy, University of Illinois Urbana-Champaign, 1002 West Green Street, Urbana, IL, 61801, USA}
    \affiliation{Center for AstroPhysical Surveys, National Center for Supercomputing Applications, Urbana, IL, 61801, USA}
    \affiliation{NSF-Simons AI Institute for the Sky (SkAI), 172 E. Chestnut St., Chicago, IL 60611, USA}
    \author{A.~W.~Pollak}
    \affiliation{Department of Astronomy and Astrophysics, University of Chicago, 5640 South Ellis Avenue, Chicago, IL, 60637, USA}
    \author{K.~Prabhu}
    \affiliation{Department of Physics \& Astronomy, University of California, One Shields Avenue, Davis, CA 95616, USA}
    \author{W.~Quan\,\orcidlink{0009-0002-2589-5501}}
    \affiliation{High-Energy Physics Division, Argonne National Laboratory, 9700 South Cass Avenue, Lemont, IL, 60439, USA}
    \affiliation{Department of Physics, University of Chicago, 5640 South Ellis Avenue, Chicago, IL, 60637, USA}
    \affiliation{Kavli Institute for Cosmological Physics, University of Chicago, 5640 South Ellis Avenue, Chicago, IL, 60637, USA}
    \author{M.~Rahimi}
    \affiliation{School of Physics, University of Melbourne, Parkville, VIC 3010, Australia}
    \author{A.~Rahlin\,\orcidlink{0000-0003-3953-1776}}
    \affiliation{Department of Astronomy and Astrophysics, University of Chicago, 5640 South Ellis Avenue, Chicago, IL, 60637, USA}
    \affiliation{Kavli Institute for Cosmological Physics, University of Chicago, 5640 South Ellis Avenue, Chicago, IL, 60637, USA}
    \author{C.~L.~Reichardt\,\orcidlink{0000-0003-2226-9169}}
    \affiliation{School of Physics, University of Melbourne, Parkville, VIC 3010, Australia}
    \author{M.~Rouble}
    \affiliation{Department of Physics and McGill Space Institute, McGill University, 3600 Rue University, Montreal, Quebec H3A 2T8, Canada}
    \author{J.~E.~Ruhl}
    \affiliation{Department of Physics, Case Western Reserve University, Cleveland, OH, 44106, USA}
    \author{A.~C.~Silva~Oliveira\,\orcidlink{0000-0001-5755-5865}}
    \affiliation{California Institute of Technology, 1200 East California Boulevard., Pasadena, CA, 91125, USA}
    \affiliation{Kavli Institute for Particle Astrophysics and Cosmology, Stanford University, 452 Lomita Mall, Stanford, CA, 94305, USA}
    \affiliation{Department of Physics, Stanford University, 382 Via Pueblo Mall, Stanford, CA, 94305, USA}
    \author{A.~Simpson}
    \affiliation{Department of Astronomy and Astrophysics, University of Chicago, 5640 South Ellis Avenue, Chicago, IL, 60637, USA}
    \affiliation{Kavli Institute for Cosmological Physics, University of Chicago, 5640 South Ellis Avenue, Chicago, IL, 60637, USA}
    \author{J.~A.~Sobrin\,\orcidlink{0000-0001-6155-5315}}
    \affiliation{Department of Physics, Villanova University, 800 E Lancaster Ave, Villanova, PA 19085, USA}
    \affiliation{Fermi National Accelerator Laboratory, MS209, P.O. Box 500, Batavia, IL, 60510, USA}
    \author{A.~A.~Stark}
    \affiliation{Center for Astrophysics \textbar{} Harvard \& Smithsonian, 60 Garden Street, Cambridge, MA, 02138, USA}
    \author{J.~Stephen}
    \affiliation{Enrico Fermi Institute, University of Chicago, 5640 South Ellis Avenue, Chicago, IL, 60637, USA}
    \author{C.~Tandoi\,\orcidlink{0000-0002-2077-6004}}
    \affiliation{Department of Astronomy, University of Illinois Urbana-Champaign, 1002 West Green Street, Urbana, IL, 61801, USA}
    \author{C.~Trendafilova}
    \affiliation{Center for AstroPhysical Surveys, National Center for Supercomputing Applications, Urbana, IL, 61801, USA}
    \author{J.~D.~Vieira\,\orcidlink{0000-0001-7192-3871}}
    \affiliation{Department of Astronomy, University of Illinois Urbana-Champaign, 1002 West Green Street, Urbana, IL, 61801, USA}
    \affiliation{Department of Physics, University of Illinois Urbana-Champaign, 1110 West Green Street, Urbana, IL, 61801, USA}
    \affiliation{Center for AstroPhysical Surveys, National Center for Supercomputing Applications, Urbana, IL, 61801, USA}
    \author{A.~G.~Vieregg\,\orcidlink{0000-0002-4528-9886}}
    \affiliation{Kavli Institute for Cosmological Physics, University of Chicago, 5640 South Ellis Avenue, Chicago, IL, 60637, USA}
    \affiliation{Department of Astronomy and Astrophysics, University of Chicago, 5640 South Ellis Avenue, Chicago, IL, 60637, USA}
    \affiliation{Enrico Fermi Institute, University of Chicago, 5640 South Ellis Avenue, Chicago, IL, 60637, USA}
    \affiliation{Department of Physics, University of Chicago, 5640 South Ellis Avenue, Chicago, IL, 60637, USA}
    \author{A.~Vitrier\,\orcidlink{0009-0009-3168-092X}}
    \affiliation{Sorbonne Universit\'e, CNRS, UMR 7095, Institut d'Astrophysique de Paris, 98 bis bd Arago, 75014 Paris, France}
    \author{Y.~Wan}
    \affiliation{Department of Astronomy, University of Illinois Urbana-Champaign, 1002 West Green Street, Urbana, IL, 61801, USA}
    \affiliation{Center for AstroPhysical Surveys, National Center for Supercomputing Applications, Urbana, IL, 61801, USA}
    \author{N.~Whitehorn\,\orcidlink{0000-0002-3157-0407}}
    \affiliation{Department of Physics and Astronomy, Michigan State University, East Lansing, MI 48824, USA}
    \author{W.~L.~K.~Wu\,\orcidlink{0000-0001-5411-6920}}
    \affiliation{California Institute of Technology, 1200 East California Boulevard., Pasadena, CA, 91125, USA}
    \affiliation{Kavli Institute for Particle Astrophysics and Cosmology, Stanford University, 452 Lomita Mall, Stanford, CA, 94305, USA}
    \affiliation{SLAC National Accelerator Laboratory, 2575 Sand Hill Road, Menlo Park, CA, 94025, USA}
    \author{M.~R.~Young}
    \affiliation{Fermi National Accelerator Laboratory, MS209, P.O. Box 500, Batavia, IL, 60510, USA}
    \affiliation{Kavli Institute for Cosmological Physics, University of Chicago, 5640 South Ellis Avenue, Chicago, IL, 60637, USA}
    \author{J.~A.~Zebrowski}
    \affiliation{Kavli Institute for Cosmological Physics, University of Chicago, 5640 South Ellis Avenue, Chicago, IL, 60637, USA}
    \affiliation{Department of Astronomy and Astrophysics, University of Chicago, 5640 South Ellis Avenue, Chicago, IL, 60637, USA}
    \affiliation{Fermi National Accelerator Laboratory, MS209, P.O. Box 500, Batavia, IL, 60510, USA}
    \collaboration{SPT-3G Collaboration}
    \noaffiliation
\else
    \author[a]{K.~Levy,}
    \author[b,c]{S.~Raghunathan,\orcidlink{0000-0003-1405-378X}}
    \author[b,d]{F.~Guidi,\orcidlink{0000-0001-7593-3962}}
    \author[e]{Y.~Li,\orcidlink{0000-0002-4820-1122}}
    \author[f]{E.~Anderes,\orcidlink{0009-0003-3245-3979}}
    \author[g,e,h]{A.~J.~Anderson,\orcidlink{0000-0002-4435-4623}}
    \author[a]{B.~Ansarinejad,}
    \author[h,e]{M.~Archipley,\orcidlink{0000-0002-0517-9842}}
    \author[d]{L.~Balkenhol,\orcidlink{0000-0001-6899-1873}}
    \author[i]{D.~R.~Barron,\orcidlink{0000-0002-1623-5651}}
    \author[d]{K.~Benabed,}
    \author[j,e,h]{A.~N.~Bender,\orcidlink{0000-0001-5868-0748}}
    \author[g,e,h]{B.~A.~Benson,\orcidlink{0000-0002-5108-6823}}
    \author[k,l,m]{F.~Bianchini,\orcidlink{0000-0003-4847-3483}}
    \author[j,e,h]{L.~E.~Bleem,\orcidlink{0000-0001-7665-5079}}
    \author[n]{S.~Bocquet,\orcidlink{0000-0002-4900-805X}}
    \author[d]{F.~R.~Bouchet,\orcidlink{0000-0002-8051-2924}}
    \author[d]{E.~Camphuis,\orcidlink{0000-0003-3483-8461}}
    \author[j]{M.~G.~Campitiello,}
    \author[e,o,p,j,h]{J.~E.~Carlstrom,\orcidlink{0000-0002-2044-7665}}
    \author[q,r]{J.~Carron,\orcidlink{0000-0002-5751-1392}}
    \author[j,e,h]{C.~L.~Chang,}
    \author[p,e]{P.~M.~Chichura,\orcidlink{0000-0002-5397-9035}}
    \author[h]{A.~Chokshi,}
    \author[h,e,s]{T.-L.~Chou,\orcidlink{0000-0002-3091-8790}}
    \author[t]{A.~Coerver,\orcidlink{0000-0002-2707-1672}}
    \author[h,e]{T.~M.~Crawford,\orcidlink{0000-0001-9000-5013}}
    \author[u,v]{C.~Daley,\orcidlink{0000-0002-3760-2086}}
    \author[w]{T.~de~Haan,\orcidlink{0000-0001-5105-9473}}
    \author[h,e]{K.~R.~Dibert,}
    \author[x,y]{M.~A.~Dobbs,}
    \author[a]{M.~Doohan,}
    \author[z]{D.~Dutcher,\orcidlink{0000-0002-9962-2058}}
    \author[aa,bb,cc]{C.~Feng,}
    \author[dd,ee]{K.~R.~Ferguson,\orcidlink{0000-0002-4928-8813}}
    \author[ff,k,l]{N.~C.~Ferree,\orcidlink{0000-0002-7130-7099}}
    \author[p,e]{K.~Fichman,}
    \author[z]{A.~Foster,\orcidlink{0000-0002-7145-1824}}
    \author[d]{S.~Galli,}
    \author[e]{A.~E.~Gambrel,}
    \author[cc]{A.~K.~Gao,}
    \author[ff,k,l,b]{F.~Ge,}
    \author[t]{S.~Guns,}
    \author[gg,hh]{N.~W.~Halverson,}
    \author[d]{E.~Hivon,\orcidlink{0000-0003-1880-2733}}
    \author[cc]{G.~P.~Holder,\orcidlink{0000-0002-0463-6394}}
    \author[t]{W.~L.~Holzapfel,}
    \author[e]{J.~C.~Hood,}
    \author[p,e]{A.~Hryciuk,}
    \author[t]{N.~Huang,\orcidlink{0000-0003-3595-0359}}
    \author[h,e]{T.~Jhaveri,}
    \author[j]{F.~K\'eruzor\'e,}
    \author[d]{A.~R.~Khalife,\orcidlink{0000-0002-8388-4950}}
    \author[b]{L.~Knox,}
    \author[h,e,j]{K.~Kornoelje,}
    \author[k,l,m]{C.-L.~Kuo,}
    \author[e]{A.~E.~Lowitz,\orcidlink{0000-0002-4747-4276}}
    \author[cc]{C.~Lu,}
    \author[b]{G.~P.~Lynch,\orcidlink{0009-0004-3143-1708}}
    \author[ii]{T.~J.~Maccarone,\orcidlink{0000-0003-0976-4755}}
    \author[k,l,m]{A.~S.~Maniyar,\orcidlink{0000-0002-4617-9320}}
    \author[h,e]{E.~S.~Martsen,}
    \author[v,c]{F.~Menanteau,}
    \author[t]{M.~Millea,\orcidlink{0000-0001-7317-0551}}
    \author[x]{J.~Montgomery,}
    \author[l]{Y.~Nakato,}
    \author[e]{T.~Natoli,}
    \author[cc]{A.~Ouellette,\orcidlink{0000-0003-0170-5638}}
    \author[j,e,p]{Z.~Pan,\orcidlink{0000-0002-6164-9861}}
    \author[o]{P.~Paschos,}
    \author[v,c,jj]{K.~A.~Phadke,\orcidlink{0000-0001-7946-557X}}
    \author[h]{A.~W.~Pollak,}
    \author[b]{K.~Prabhu,}
    \author[j,p,e]{W.~Quan,\orcidlink{0009-0002-2589-5501}}
    \author[a]{M.~Rahimi,}
    \author[h,e]{A.~Rahlin,\orcidlink{0000-0003-3953-1776}}
    \author[a]{C.~L.~Reichardt,\orcidlink{0000-0003-2226-9169}}
    \author[x]{M.~Rouble,}
    \author[kk]{J.~E.~Ruhl,}
    \author[ff,k,l]{A.~C.~Silva~Oliveira,\orcidlink{0000-0001-5755-5865}}
    \author[h,e]{A.~Simpson,}
    \author[ll,g]{J.~A.~Sobrin,\orcidlink{0000-0001-6155-5315}}
    \author[mm]{A.~A.~Stark,}
    \author[o]{J.~Stephen,}
    \author[v]{C.~Tandoi,\orcidlink{0000-0002-2077-6004}}
    \author[c]{C.~Trendafilova,}
    \author[v,cc,c]{J.~D.~Vieira,\orcidlink{0000-0001-7192-3871}}
    \author[e,h,o,p]{A.~G.~Vieregg,\orcidlink{0000-0002-4528-9886}}
    \author[d]{A.~Vitrier,\orcidlink{0009-0009-3168-092X}}
    \author[v,c]{Y.~Wan,}
    \author[ee]{N.~Whitehorn,\orcidlink{0000-0002-3157-0407}}
    \author[ff,k,m]{W.~L.~K.~Wu,\orcidlink{0000-0001-5411-6920}}
    \author[g,e]{M.~R.~Young}
    \author[e,h,g]{and J.~A.~Zebrowski}
    
    \affiliation[a]{School of Physics, University of Melbourne, Parkville, VIC 3010, Australia}
    \affiliation[b]{Department of Physics \& Astronomy, University of California, One Shields Avenue, Davis, CA 95616, USA}
    \affiliation[c]{Center for AstroPhysical Surveys, National Center for Supercomputing Applications, Urbana, IL, 61801, USA}
    \affiliation[d]{Sorbonne Universit\'e, CNRS, UMR 7095, Institut d'Astrophysique de Paris, 98 bis bd Arago, 75014 Paris, France}
    \affiliation[e]{Kavli Institute for Cosmological Physics, University of Chicago, 5640 South Ellis Avenue, Chicago, IL, 60637, USA}
    \affiliation[f]{Department of Statistics, University of California, One Shields Avenue, Davis, CA 95616, USA}
    \affiliation[g]{Fermi National Accelerator Laboratory, MS209, P.O. Box 500, Batavia, IL, 60510, USA}
    \affiliation[h]{Department of Astronomy and Astrophysics, University of Chicago, 5640 South Ellis Avenue, Chicago, IL, 60637, USA}
    \affiliation[i]{Department of Physics and Astronomy, University of New Mexico, Albuquerque, NM, 87131, USA}
    \affiliation[j]{High-Energy Physics Division, Argonne National Laboratory, 9700 South Cass Avenue, Lemont, IL, 60439, USA}
    \affiliation[k]{Kavli Institute for Particle Astrophysics and Cosmology, Stanford University, 452 Lomita Mall, Stanford, CA, 94305, USA}
    \affiliation[l]{Department of Physics, Stanford University, 382 Via Pueblo Mall, Stanford, CA, 94305, USA}
    \affiliation[m]{SLAC National Accelerator Laboratory, 2575 Sand Hill Road, Menlo Park, CA, 94025, USA}
    \affiliation[n]{University Observatory, Faculty of Physics, LMU Munich, Scheinerstr.~1, 81679 Munich, Germany}
    \affiliation[o]{Enrico Fermi Institute, University of Chicago, 5640 South Ellis Avenue, Chicago, IL, 60637, USA}
    \affiliation[p]{Department of Physics, University of Chicago, 5640 South Ellis Avenue, Chicago, IL, 60637, USA}
    \affiliation[q]{Istituto ricerche solari Aldo e Cele Dacc\`o (IRSOL), Faculty of Informatics, Universit\`a della Svizzera italiana, 6605 Locarno, Switzerland}
    \affiliation[r]{Universit\'e de Gen\`eve, D\'epartement de Physique Th\'eorique, 24 Quai Ansermet, CH-1211 Gen\`eve 4, Switzerland}
    \affiliation[s]{National Taiwan University, No. 1, Sec. 4, Roosevelt Road, Taipei 106319, Taiwan}
    \affiliation[t]{Department of Physics, University of California, Berkeley, CA, 94720, USA}
    \affiliation[u]{Universit\'e Paris-Saclay, Universit\'e Paris Cit\'e, CEA, CNRS, AIM, 91191, Gif-sur-Yvette, France}
    \affiliation[v]{Department of Astronomy, University of Illinois Urbana-Champaign, 1002 West Green Street, Urbana, IL, 61801, USA}
    \affiliation[w]{High Energy Accelerator Research Organization (KEK), Tsukuba, Ibaraki 305-0801, Japan}
    \affiliation[x]{Department of Physics and McGill Space Institute, McGill University, 3600 Rue University, Montreal, Quebec H3A 2T8, Canada}
    \affiliation[y]{Canadian Institute for Advanced Research, CIFAR Program in Gravity and the Extreme Universe, Toronto, ON, M5G 1Z8, Canada}
    \affiliation[z]{Joseph Henry Laboratories of Physics, Jadwin Hall, Princeton University, Princeton, NJ 08544, USA}
    \affiliation[aa]{Department of Astronomy, University of Science and Technology of China, Hefei 230026, China}
    \affiliation[bb]{School of Astronomy and Space Science, University of Science and Technology of China, Hefei 230026}
    \affiliation[cc]{Department of Physics, University of Illinois Urbana-Champaign, 1110 West Green Street, Urbana, IL, 61801, USA}
    \affiliation[dd]{Department of Physics and Astronomy, University of California, Los Angeles, CA, 90095, USA}
    \affiliation[ee]{Department of Physics and Astronomy, Michigan State University, East Lansing, MI 48824, USA}
    \affiliation[ff]{California Institute of Technology, 1200 East California Boulevard., Pasadena, CA, 91125, USA}
    \affiliation[gg]{CASA, Department of Astrophysical and Planetary Sciences, University of Colorado, Boulder, CO, 80309, USA }
    \affiliation[hh]{Department of Physics, University of Colorado, Boulder, CO, 80309, USA}
    \affiliation[ii]{Department of Physics \& Astronomy, Box 41051, Texas Tech University, Lubbock TX 79409-1051, USA}
    \affiliation[jj]{NSF-Simons AI Institute for the Sky (SkAI), 172 E. Chestnut St., Chicago, IL 60611, USA}
    \affiliation[kk]{Department of Physics, Case Western Reserve University, Cleveland, OH, 44106, USA}
    \affiliation[ll]{Department of Physics, Villanova University, 800 E Lancaster Ave, Villanova, PA 19085, USA}
    \affiliation[mm]{Center for Astrophysics \textbar{} Harvard \& Smithsonian, 60 Garden Street, Cambridge, MA, 02138, USA}
\fi
\ifdefined\PRformat
    \begin{abstract}
    \abstracttext{}
    \end{abstract}
\else
    \abstract{\abstracttext}
\fi

\maketitle

\section{Introduction}

\label{sec:introduction}
Gravitational lensing of the cosmic microwave background (CMB) by the large-scale matter distribution provides a powerful probe of cosmology (e.g., \citealp{lewis06}). It encodes information about both the expansion history of the Universe and the growth of large-scale structure.

While measurements of the primary CMB anisotropy power spectra constrain mostly the amplitude of the fluctuations at the redshift of last scattering ($z \approx 1100$; e.g., \citealp{planck18-5, louis25, camphuis25}), measurements of the gravitational deflection of CMB photons, known as CMB lensing (e.g. \citealp{blanchard87, lewis06}), provides information about all the structure between the last scattering surface and today. Gravitational lensing of the CMB causes arcminute-scale deflections which imprint a distinctive non-Gaussian signal in the CMB maps. 
The lensing induced correlations between spherical harmonic modes of the CMB anisotropies can be leveraged to produce a  map of the projected matter distribution of the Universe. The CMB lensing signal is most efficiently lensed by matter at intermediate redshifts, typically for redshifts $0.5 < z < 5$, with the mean redshift of the lensing kernel being around $z \sim 2$ and the peak of the distribution being around $z \sim 1$ \citep{lewis06}. This covers redshifts when structure growth is suppressed by massive neutrinos and the cosmic expansion is driven by dark energy. 
Therefore, CMB lensing measurements provide constraints on key cosmological parameters, including the amplitude of matter density fluctuations and the matter density parameter, the sum of neutrino masses, and parameters governing dark energy (e.g., \citealp{carron22, qu24, madhavacheril24, ge25, qu25}). Additionally, characterization of the lensing potential is crucial for disentangling the lensing induced \textit{B}-mode power from those arising from primordial gravitational waves (e.g., \citealp{kamionkowski16, bicep21}).

Since gravitational lensing directly probes baryonic and dark matter, it has proven to be a powerful tool to measure the large-scale matter distribution. Lensing measurements from galaxy surveys extract the lensing potential through a statistical analysis of the image distortions induced on background galaxies by the intervening large-scale structure, known as cosmic shear (e.g., \citealp{kaiser92, bernardeau97, kilbinger15}). One disadvantage of such measurement is that the signal-to-noise ratio (SNR) of the background galaxies that can be used for cosmic shear measurements decreases with redshift. It is therefore challenging to probe the matter distribution at high redshifts ($z \gtrsim 1$). Additionally, uncertainties in the photometric redshift estimation of the background galaxies, which increase with redshift, lead to additional errors in the reconstruction of the lensing power spectrum. The CMB, on the other hand, is a diffuse light source that covers the entire sky, has well-understood statistical properties, as well as a high, precisely known redshift, making it a powerful tool to reconstruct the projected matter density at high redshifts. 
{Moreover, cross-checks between CMB and optical lensing surveys, which are subject to different instrumental and astrophysical systematics, provide a valuable test of the consistency of structure growth measurements (e.g., \citealp{kim24}).}

The first detection of CMB lensing (3.4$\sigma$ detection) was made by cross-correlating CMB maps from the Wilkinson Microwave Anisotropy Probe (WMAP) satellite with radio galaxy counts from the NRAO VLA SkySurvey \citep[NVSS;][]{smith07}. The first detection from CMB data alone was obtained by the Atacama Cosmology Telescope (ACT) collaboration (4$\sigma$ detection; \citealp{das11}). Besides the lensing measurements from ACT \citep{das11, das14, sherwin17, darwish21, madhavacheril24, qu24}, several other experiments have detected the lensing signal, such as the South Pole Telescope \citep[SPT;][]{engelen12, story15, omori17, wu19, millea21, pan23, ge25}, \textit{Planck} \citep{planck14-17, planck16-15, planck20-8, carron22}, POLARBEAR \citep{ade14, faundez20}, and BICEP \& \textit{Keck Array} \citep{bicep2keck16, ade23}. To this point, the highest-SNR measurements of the CMB lensing power spectrum comes from a combination of the \textit{Planck} PR4 \texttt{NPIPE} maps \citep{carron22}, the ACT DR6 dataset \citep{qu24}, and 2-year polarization-only data from the SPT-3G Main field \citep{ge25}. Each survey independently delivers a lensing measurement with comparable statistical significance, reaching a SNR of roughly $\sim$ 40. Combining the lensing bandpowers yields a lensing SNR of 61 \citep{qu25}.

The SPT-3G observing program consists of the deep Main field and the wider but shallower Summer and Wide fields, providing complementary depth and sky coverage \citep{prabhu24}. 
In this work, we use a curved-sky quadratic estimator (QE; \citealt{okamoto03}) technique to reconstruct the lensing potential using the first 2 years of CMB temperature data from the SPT-3G Summer survey. 
While the white noise levels in the Summer fields are $\sim 3\text{--}4$ times higher than those of the SPT-3G Main survey, the total sky area of the Summer fields is roughly double that of the 1500 deg$^2$ Main field. 
Between these factors and analysis choices, the overall lensing measurement uncertainty is approximately double that achieved in the upcoming lensing analysis of the 2-year Main survey dataset (Omori et al., in prep.). 
Given this, we leave cosmological interpretation of lensing measurements to that work. 
Future work will combine lensing measurements from the 4-year Summer survey with those from the Main and Wide surveys. 
With the full set of Summer field data soon to be available, this work provides a first quality assessment of the SPT-3G Summer fields.

Throughout this paper, we assume a spatially flat $\Lambda$CDM \textit{Planck} 2018 cosmology \citep{planck18-6} with a Hubble constant $H_0 = 67.4 \text{ km s}^{-1}\text{Mpc}^{-1}$, baryon density $\Omega_b h^2 = 0.0224$, dark matter density $\Omega_c h^2 = 0.120$, matter density parameter $\Omega_m = 0.315$, primordial power spectrum with an amplitude $A_s = 2.101\times 10^{-9}$ and scalar spectral index $n_s = 0.965$, matter fluctuation amplitude $\sigma_8 = 0.811$, optical depth $\tau = 0.054$, and a single massive neutrino with $m_\nu = 0.06 \text{ eV}$.

This paper is structured as follows: the SPT-3G Summer survey and the simulations used in this work are described in Section \ref{sec:data} and Section \ref{sec:simulations}, respectively; the method used to reconstruct the lensing potential is explained in Section \ref{sec:method}; the main results of our analysis are provided in Section \ref{sec:results}, including the lensing potential maps, systematic checks concerning analysis choices, null tests, and the final lensing power spectra; in Section \ref{sec:conclusion}, we summarize and discuss our results.

\section{SPT-3G Summer Survey}
\label{sec:data}
In this section, we briefly describe the SPT-3G experiment, the SPT-3G Summer survey and the data reduction pipeline used to go from the raw time-ordered data (TOD) to the maps used for the lensing analysis.

\subsection{Instrument} 
The South Pole Telescope\footnote{\url{https://pole.uchicago.edu/public/Home.html}} \citep{carlstrom11} is a 10-meter diameter  telescope located at the Amundsen–Scott South Pole Station in Antarctica. 
SPT-3G \citep{benson14, bender18, sobrin18, sobrin22} is the third generation SPT CMB receiver and was installed on the telescope in 2017.
SPT-3G contains $\sim$ 16 000 polarization-sensitive detectors at frequency bands centered around 95, 150, and 220 GHz.
At these frequencies, the 10-meter primary mirror provides arcminute-scale resolution.

\subsection{Observations} 
In February 2018, SPT-3G began mapping the 1500 $\mathrm{deg}^2$ Main field. Besides the Main field, three additional fields, referred to as Summer fields, were observed from 2019 to 2023. Observations of these fields were conducted during the austral summer (December to March), a period when the Sun is above the horizon and close enough to the Main field to be detected through the telescope’s sidelobes. The Summer survey spans approximately 2640 $\mathrm{deg}^2$, split into:

\begin{itemize}[align=parleft, left=0pt]
    \item Summer-A: A 1210 deg$^2$ field extending from 50$^\circ$ to 100$^\circ$ RA and $-28^\circ$ to $-63^\circ$ Dec, divided for the 2019/20 observations into 6 sub-fields centered at $-29.75^\circ$, $-33.25^\circ$, $-38.5^\circ$, $-45.5^\circ$, $-52.5^\circ$, and $-59.5^\circ$ Dec,
    \item Summer-B: A 570 deg$^2$ field extending from 0$^\circ$ to 50$^\circ$ RA and $-28^\circ$ to $-42^\circ$ Dec, divided into 4 sub-fields centered at $-29.75^\circ$, $-33.25^\circ$, $-36.75^\circ$, and $-40.25^\circ$ Dec,
    \item Summer-C: A 860 deg$^2$ field extending from 150$^\circ$ to 225$^\circ$ RA and $-28^\circ$ to $-42^\circ$ Dec, divided into 4 sub-fields centered at $-29.75^\circ$, $-33.25^\circ$, $-36.75^\circ$, and $-40.25^\circ$ Dec.
\end{itemize} 

{The division into sub-fields is done to limit variations in the detector responsivity throughout a single observation.} Each sub-field is observed using a raster scanning strategy: starting from the lowest declination, the telescope completes a left and right scan across the full RA range at constant declination and then takes a $12.5'$ step in declination until the maximum elevation is reached. In this work, we use temperature data taken during the 2019–2020 and 2020–2021 austral summers. We refer to these observations as the SPT-3G 2-year Summer dataset. Although CMB polarization data can yield robust lensing reconstructions since extragalactic foregrounds are largely unpolarized \citep{datta19, gupta19}, we restrict our lensing analysis to temperature data, which dominates the overall lensing SNR in these maps. 

\subsection{Data Reduction} 
\label{sec_data_reduction}
We follow a similar approach as previous SPT analyses to transfer the raw TOD into the final coadded maps (e.g., \citealp{dutcher21, quan26}). 
{We designed a comprehensive set of null tests to identify systematic effects specific to the summer-field data, including contamination from the Sun, Moon, and ground pickup, as well as scan-direction, temporal, and detector-wafer-dependent effects. The data pass all tests, with no evidence for significant residual systematic contamination. A detailed description of these tests and their results will be presented in a forthcoming publication on the summer-field maps and power spectra.} 
The 2-year Summer maps will be released in early 2027. 

\textit{Time-ordered Data Processing}: First, the TOD from each detector is converted to CMB temperature units. To reduce the impact of the atmospheric noise and thermal drifts in the detector wafers, a 29th-order Legendre polynomial fit is subtracted from the TOD and modes with $\ell_x < 300$ are removed, where $\ell_x$ denotes the angular wavenumber along the scan direction. Additionally, a low-pass filter removing modes with $\ell_x > 6144$ is applied to prevent aliasing when binning the TOD into map pixels later on. 
Before applying the above filters, we mask sources above 20 mJy to avoid filtering artifacts. 

\textit{Coadded Maps and Temperature Calibration}: After filtering, the TOD for each sub-field are coadded into maps with $N_{\rm side} = 2048$ \citep{gorski05}. The weights used for coadding are given by the inverse-variance of the detector noise to reduce the noise in the final map. The maps are calibrated by computing the cross-spectra between the SPT-3G maps with \textit{Planck} 2018 maps for the multipole range $\ell \in [500,1000]$. We mask all the point sources that were masked during map-making and correct for the beams, pixel window functions, and transfer functions. The resulting recalibration factors are applied to each subfield before stitching them together to get the full Summer-field maps. 
{The telescope beam response function is derived from observations of Saturn and bright point sources in the CMB fields, following a procedure similar to that described in \citet{camphuis25}.}
The observed white noise levels for 95, 150, and 220 GHz in the full coadded maps are 12, 11, 40 $\mu$K-arcmin for Summer-A; 13, 14, 49 $\mu$K-arcmin for Summer-B; and 13, 12, 44 $\mu$K-arcmin for Summer-C. 

\textit{Point Source and Cluster Mitigation}: Since the lensing pipeline used in our analysis (see Section \ref{sec:method}) picks up non-Gaussian signals, contributions from point sources and galaxy clusters have to be mitigated. Unlike previous SPT lensing analyses, which relied solely on inpainting and masking, we introduce a point source template subtraction step.
The point source templates are obtained from the SPT-3G point source lists by converting the measured fluxes of sources with $S > 6$ mJy at 150 GHz into thermodynamic units for each frequency band. 
{We only template-subtract sources with $S > 6\text{ mJy}$ because the sky simulations used in this work (see Section \ref{sec:simulations}) already include residual Gaussian power for unmasked sources with $S \le 6\text{ mJy}$. We convolve the template maps with the telescope beam and pass it through a mock-observation pipeline to apply the same timestream filtering ($\ell_x$ cuts, low-pass filtering, and polynomial subtraction) as applied to the actual data. This ensures the subtracted template accurately matches the filtering artifacts present in the observed maps.}
Point source template subtraction is advantageous as it removes the bulk of point source contamination while largely preserving the CMB lensing signal and avoiding an increase in the mean field (see Eq.~\ref{eq:mf_bias}). While this approach works well for sources with $S \leq 20$ mJy at 150\,GHz, residual mismatches between the true source emission and the templates for sources with higher fluxes introduce spurious non-Gaussian features that bias the lensing reconstruction. Consequently, these sources, as well as galaxy clusters detected with a SNR $>9$, are addressed with inpainting and masking, consistent with previous analyses.

We apply an inpainting method similar to those used by \citet{benoit13, raghunathan19}. 
{We define an inner region, $R \leq R_1$ (where $R_1$ corresponds to the ``Inpainting Radius'' in Table~\ref{tab:masking}), and an outer region, $R_1 < R \leq R_2$ with $R_2 = 25'$, around each point source and galaxy cluster.} The inpainted pixel values within the inner region $\hat{T}_1$ are determined based on the values within the outer region $T_2$ using constrained Gaussian realizations:

\begin{align}
    \hat{T}_1 = \tilde{T}_1 + \hat{\boldsymbol{C}}_{12}\hat{\boldsymbol{C}}_{22}^{-1}(T_2-\tilde{T}_2) \;,
\end{align}
where $\tilde{T}_1$ and $\tilde{T}_2$ refer to pixel values from a random Gaussian realization in the inner and outer regions, respectively. $\hat{\boldsymbol{C}}_{XY}$ refers to the covariance matrix between two regions, $X$ and $Y$, of the CMB fields. We use 20 000 simulations to calculate the covariance matrix. The Gaussian realizations have a size of $180' \times 180'$ and include the lensed CMB, as well as the same foreground power, noise power, and transfer function as the data. 

We inpaint point sources with flux densities in the range $20 < S \leq 100$ mJy and galaxy clusters with $5 < \mathrm{SNR} \leq 9$. For brighter point sources and clusters, the affected regions are too extended for inpainting to provide a reliable reconstruction. These objects are subsequently removed through masking.

\textit{Final Maps for Lensing Reconstruction}:
We combine the template-subtracted, inpainted maps from each frequency band into a minimum-variance map using a harmonic-space internal linear combination (ILC) technique (e.g.,\citealp{tegmark96, tegmark03, delabrouille03}). The weights are optimized to minimize the total variance from extragalactic signals, calculated using \textsc{Agora} simulations \citep{omori24}, and the experimental noise. \textsc{Agora} simulations are N-body simulations including tSZ, kSZ, CIB, radio sources and weak lensing components capturing the non-Gaussianity of the extragalactic foregrounds and the lensing potential. 

Finally, we apply a mask for point sources with $S > 100$ mJy and for clusters with SNR $> 9$. The specific radii used for masking are summarized in Table~\ref{tab:masking}. We apply a cosine taper with a radius of $45'$ to the border of the mask, and a $15'$ cosine taper to the point source and cluster holes in the mask. In total, we lose 11, 17, and 12\% of the sky area for Summer-A, Summer-B, and Summer-C due to boundary and point source masking as well as inpainting.

\begin{table}
  \centering
  \begin{tabular}{l|l|l|l} 
    \hline
    \ifdefined\PRformat
    \else
          \makecell{Source Type} & \makecell{Flux $S$ [mJy] / \\ Cluster SNR} & \makecell{Inpainting \\ Radius} & \makecell{Masking \\ Radius} \\
        \hline
        \makecell{Point \\ Sources} &  \makecell{ $ S \leq 20$ \\ $20 < S \leq 100$ \\ $100 < S \leq 1000$ \\ $S > 1000$} &  \makecell{$-$ \\ $5'$ \\ $-$ \\ $-$} &  \makecell{$-$ \\ $-$ \\ $7'$ \\ $10'$} \\
        \hline
        \makecell{Galaxy \\ Clusters} & \makecell{$5 <$ SNR $\leq 9$ \\ SNR $> 9$} & \makecell{ $5'$ \\ $-$} & \makecell{$-$ \\ $7'$} \\
    \fi
    \hline
  \end{tabular}
  \caption{Inpainting and masking radii for point sources and galaxy clusters as a function of flux or SNR measured at 150 GHz. The contamination by point sources with flux values $6<S\leq20$ mJy is mitigated by subtracting a point source template. The simulations used for the lensing pipeline include point source power up to 6 mJy.}
  \label{tab:masking}
\end{table}


\section{Simulations}
\label{sec:simulations}
Simulations are used to estimate the filter transfer functions and to compute bias corrections and uncertainties for the reconstruction (see Section \ref{sec:method}). 
The simulations include the lensed or unlensed primary CMB and foreground signals. 
The sky simulations are convolved with the beam of the corresponding SPT-3G frequency band before being passed through the mock-observing pipeline to create $N_\text{side} = 2048$ simulations capturing the filtering of the data. 
{The mock-observation pipeline applies the identical filtering and processing steps used on the data to the simulations. These include the TOD processing and binning that produce the final sky maps, as described in \S\ref{sec_data_reduction}. Simulated noise realizations are then added to the resulting maps.}

The unlensed CMB temperature and lensing potential power spectra are generated using CAMB \citep{lewis05} and the cosmological parameters highlighted in Section \ref{sec:introduction}. We use HEALPix \citep{gorski05} to generate Gaussian realizations with $N_\text{side} = 8192$ of the unlensed CMB and lensing potential, and LensPix \citep{lewis00} to obtain the lensed CMB realizations.
Besides the lensed CMB signal, we include Gaussian realizations of extragalactic foregrounds. These are generated using auto- and cross-power spectra derived from 95, 150, and 220 GHz \textsc{Agora} simulations. The simulations are calibrated to the measured SPT-3G Main field power spectra{,}\footnote{While the simulations are calibrated to the Main field, the mismatch between the Summer field data and the simulations is within the statistical uncertainties of the Summer field power spectra. Moreover, our lensing pipeline uses a realization-dependent bias subtraction ($N^{0, \rm RD}_{L}$), ensuring the results are robust to such residual data-simulation discrepancies (see Section~\ref{sec:method}).} where galaxy clusters with an SZ detection significance $\xi > 10$ and point sources with a 150 GHz flux density exceeding 6 mJy have been masked. We do not include any Galactic foregrounds, as their impact on the reconstructed lensing potential is negligible given the location of the fields and the angular scales used in the reconstruction.


The transfer function for each Summer field and frequency band is estimated as the ratio of the power spectrum of the mock-observed simulations to that of the corresponding input simulations:

\begin{align}
    \gamma_\ell = \frac{\langle T^{\rm out}_{\ell m}T^{\rm out*}_{\ell m} \rangle}{\langle T^{\rm in}_{\ell m}T^{\rm in*}_{\ell m} \rangle} \;,
    \label{eq:transfer_function}
\end{align}
where the average is taken over all $\ell m$ pairs at a given $\ell$ over 100 simulations. These transfer functions are deconvolved from both the data and the mock observations prior to constructing the ILC maps.

Owing to the limited number of observations of the Summer fields in the first two years ($\sim$100), it is not possible to construct a sufficiently large set of independent noise realizations directly from the data, as is done for the Main field, where thousands of observations are available. Instead, we generate noise simulations using a procedure that largely follows the “tiled method” described in \citet{atkins23}. For each Summer field, we split the data into four bundles such that the weights of each subfield are uniformly distributed at 150 GHz, and construct null maps by differencing the left-going (L) and right-going (R) scan-direction splits.\footnote{Although the time-constant effect has not been corrected in the maps, we find the signal leakage in the L-R difference map to be negligible for the purpose of noise modeling.} From these null maps, we jointly model across all the frequency bands and Stokes parameters. To capture the spatial variation of the noise properties across the field, we divide the maps into $5^\circ \times 3^\circ$ tiles in RA and Dec. Within each tile, we perform a singular value decomposition (SVD) to produce a set of SVD maps, which represent the uncorrelated eigenmodes of the joint noise covariance. This transformation decorrelates the noise between the different frequency and polarization channels, allowing each resulting component to be modeled independently. For each tile, we compute the 2D power spectrum of each SVD map directly from the HEALPix grid using non-uniform fast Fourier transform methods. We then generate noise realizations for each bundle separately prior to coaddition. Although we only use the coadded map in this lensing analysis, modeling and sampling the noise at the bundle level is crucial for reducing the bias in the 4-point estimator caused by the sample variance in the noise model estimation. We find that noise simulations produced by this procedure reproduce the statistical properties of the coadded maps sufficiently well (see Section \ref{sec:lr}), eliminating the need for the "cross-estimator" approach to explicitly avoid the instrumental-noise bias in the lensing reconstruction \citep{madhavacheril20, madhavacheril24, qu24}.

Finally, we construct inpainted ILC maps and apply the same boundary and point source masks used for the data, ensuring that the simulations and data are treated consistently.

\section{CMB Lensing Analysis}
\label{sec:method}
As the CMB photons propagate towards us from the last-scattering surface, they are deflected by the intervening gravitational potential. The lensed CMB temperature field $T(\hat{\boldsymbol{n}})$ along the line of sight $\hat{\boldsymbol{n}}$ is given by a surface-brightness-conserving remapping of the underlying unlensed temperature field $\tilde{T}(\hat{\boldsymbol{n}})$:

\begin{align}
    T(\hat{\boldsymbol{n}}) = \tilde{T}[\hat{\boldsymbol{n}} + \boldsymbol{\alpha}(\hat{\boldsymbol{n}})] \;,
    \label{eq:cmb_lensing}
\end{align}
where $\boldsymbol{\alpha}(\hat{\boldsymbol{n}})$ denotes the deflection angle due to the gravitational potential between the last-scattering surface and the observer. 

The deflection angle can be expressed as the gradient of the lensing potential $\phi(\hat{\boldsymbol{n}})$, which is related to the 3D gravitational potential $\psi(\chi\hat{\boldsymbol{n}}, \eta_0-\chi)$ through: 

\begin{align}
\phi(\hat{\boldsymbol{n}}) = - \frac{2}{c^2}\int_{0}^{\chi_{\rm{CMB}}}\text{d}\chi \frac{f(\chi_{\rm{CMB}}-\chi)}{f(\chi_{\rm{CMB}})f(\chi)}\psi(\chi\hat{\boldsymbol{n}}, \eta_0-\chi)\;,
\label{eq:lensing_potential}
\end{align}
where $\chi$ is the comoving distance along the line of sight, $\chi_{\rm{CMB}}$ is the comoving distance to the surface of last scattering, $f(\chi)$ is the comoving angular diameter distance, and $\eta_0-\chi$ is the conformal time at which a CMB photon would have been at the position $\chi\hat{\boldsymbol{n}}$. 

Taking the Laplacian of the lensing potential leads to the convergence field $\kappa(\hat{\boldsymbol{n}})$:

\begin{align}
\kappa(\hat{\boldsymbol{n}}) = -\frac{1}2\nabla^2 \phi(\hat{\boldsymbol{n}}) \; ,
\label{eq:kappa_phi}
\end{align}
which describes an over-density ($\kappa > 0$) or under-density ($\kappa < 0$) that stretches or contracts the observed CMB pattern.

Gravitational lensing of the CMB introduces correlations between previously uncorrelated CMB modes, which can be used to reconstruct the lensing potential. Considering an ensemble of lensed CMB temperature fields, each field lensed by the same lensing potential, 
lensing will add off-diagonal terms to the covariance matrix, 
with the change in the covariance given to first order by \citep{okamoto03}:


\ifdefined\PRformat
    \begin{widetext}
    \begin{align}
     \Delta\langle T_{\ell_1 m_1} T_{\ell_2m_2}\rangle_{\rm {CMB}} = &\sum_{LM}(-1)^{M}\begin{pmatrix}
    \ell_1 & \ell_2 & L\\
    m_1 & m_2 & -M
    \end{pmatrix}
     W_{\ell_1 \ell_2 L} \phi_{LM}
    \; ,
    \end{align}
    \label{eq:lensing_correlation}
    \end{widetext}
\else
    \begin{align}
     \Delta\langle T_{\ell_1 m_1} T_{\ell_2m_2}\rangle_{\rm {CMB}} = &\sum_{LM}(-1)^{M}\begin{pmatrix}
    \ell_1 & \ell_2 & L\\
    m_1 & m_2 & -M
    \end{pmatrix}
     W_{\ell_1 \ell_2 L} \phi_{LM}
    \; ,
    \label{eq:lensing_correlation}
    \end{align}
\fi
where $T_{\ell_1 m_1}$ and $T_{\ell_2 m_2}$ are the spherical harmonic coefficients of the lensed temperature field, and the term in parentheses refers to the Wigner 3$j$ symbol. 
 The spherical harmonic coefficients of the lensing potential field are labeled $\phi_{LM}$. The weight function $W_{\ell_1\ell_2L}$ is given by:

\ifdefined\PRformat
    \begin{widetext}
        \begin{align}\nonumber
            W_{\ell_1\ell_2L} = -\tilde{C}^{TT}_{\ell_1}\sqrt{\frac{(2\ell_1+1)(2\ell_2+1)(2L+1)}{4\pi}} \sqrt{L(L+1)\ell_1(\ell_1+1)}  \left( \frac{1+(-1)^{\ell_1+\ell_2+L}}{2} \right) \begin{pmatrix}
            \ell_1 & \ell_2 & L\\
            1 & 0 & -1
            \end{pmatrix}+ (\ell_1 \leftrightarrow \ell_2)\;,
        \end{align}  
        \label{eq:lensing_weights}
    \end{widetext}
\else
    \begin{align} \nonumber
        W_{\ell_1\ell_2L} = -\tilde{C}^{TT}_{\ell_1}\sqrt{\frac{(2\ell_1+1)(2\ell_2+1)(2L+1)}{4\pi}} \sqrt{L(L+1)\ell_1(\ell_1+1)} \\
        \times \left( \frac{1+(-1)^{\ell_1+\ell_2+L}}{2} \right) \begin{pmatrix}
        \ell_1 & \ell_2 & L\\
        1 & 0 & -1
        \end{pmatrix}+ (\ell_1 \leftrightarrow \ell_2)\;,
    \label{eq:lensing_weights}
    \end{align}
\fi
with $\tilde{C}^{TT}_{\ell}$ being the power spectrum of the unlensed CMB temperature field and $\leftrightarrow$ denoting an additional term obtained by switching $\ell_1$ and $\ell_2$ in the first term.

\subsection{Quadratic Estimate of the Lensing Potential}
Eq. (\ref{eq:lensing_correlation}) suggests that, in a given realization, an un-normalized estimate of the lensing potential can be obtained to first order from a weighted sum over pairs of multipole moments \citep{okamoto03}:

\begin{align}
\bar{\phi}_{LM} = &\frac{(-1)^M}{2}\sum_{\ell_1 m_1}\sum_{\ell_2 m_2}
\begin{pmatrix}
\ell_1 & \ell_2 & L\\
m_1 & m_2 & -M
\end{pmatrix}
W_{\ell_1\ell_2 L}\bar{T}_{\ell_1 m_1}\bar{T}_{\ell_2 m_2} \; .
\label{eq:phi_biased}
\end{align}
Following \cite{lewis11}, we use the lensed rather than the unlensed CMB spectrum for the QE weights $W_{\ell_1\ell_2 L}$, which provides a better approximation to the true non-perturbative response. The overbar on $T_{\ell m}$ denotes that the temperature field has been inverse-variance filtered to maximize the CMB signal relative to the noise. Specifically, the filter is given by:

\begin{align}
F_{\ell} = \frac{1}{C_{\ell}^{TT}+C_{\ell}^{FG}+N_\ell},
\label{eq:filter}
\end{align}
where $C_{\ell}^{FG}$ and $N_{\ell}$ correspond to the power spectrum of extragalactic foregrounds, and atmospheric and instrumental noise, respectively. \\

\textit{Mean-Field}: The estimate given by Eq.~(\ref{eq:phi_biased}) is biased due to statistical anisotropy induced by non-lensing sources such as the mask and inhomogeneous sky noise, known as mean-field (MF) bias that has to be subtracted from Eq.~(\ref{eq:phi_biased}). This anisotropic signal can be calculated by averaging lensing potential estimates obtained from masked simulations with independent CMB, lensing potential, and instrumental noise realizations:

\ifdefined\PRformat
    \begin{widetext}
        \begin{align}
        \bar{\phi}^{MF}_{LM} = &\frac{(-1)^M}{2}\sum_{\ell_1 m_1}\sum_{\ell_2 m_2}
        \begin{pmatrix}
        \ell_1 & \ell_2 & L\\
        m_1 & m_2 & -M
        \end{pmatrix}
        W_{\ell_1\ell_2L}\langle \bar{T}_{\ell_1m_1}\bar{T}_{\ell_2m_2} \rangle \; .
        \end{align}
        \label{eq:mf_bias}
    \end{widetext}
\else
    \begin{align}
    \bar{\phi}^{MF}_{LM} = &\frac{(-1)^M}{2}\sum_{\ell_1 m_1}\sum_{\ell_2 m_2}
    \begin{pmatrix}
    \ell_1 & \ell_2 & L\\
    m_1 & m_2 & -M
    \end{pmatrix}
    W_{\ell_1\ell_2L}\langle \bar{T}_{\ell_1m_1}\bar{T}_{\ell_2m_2} \rangle \; .
    \label{eq:mf_bias}
    \end{align}
\fi
\textit{Normalization}: The analytical normalization of the estimator is given as the sum of the weights and filters used in Eq.~(\ref{eq:phi_biased}):
\ifdefined\PRformat
    \begin{widetext}
        \begin{align}
        R_{L}^{\text{analytic}} = \frac{1}{2(2L+1)}\sum_{\ell_1}\sum_{\ell_2} W_{\ell_1\ell_2L}\times W_{\ell_1\ell_2L}F_{\ell_1} F_{\ell_2}\; .
        \end{align}
        \label{eq:analytical_response}
    \end{widetext}
\else
        \begin{align}
        R_{L}^{\text{analytic}} = \frac{1}{2(2L+1)}\sum_{\ell_1}\sum_{\ell_2} W_{\ell_1\ell_2L}\times W_{\ell_1\ell_2L}F_{\ell_1} F_{\ell_2}\; .
        \label{eq:analytical_response}
        \end{align}
\fi

Since the inverse variance filter assumes stationarity in the statistics of both the signal and noise, using only the analytical response will lead to a mis-estimation of the lensing potential. To account for this, we calculate a Monte Carlo (MC) response correction as the average of the cross-spectrum between estimated and input lensing potential divided by the auto-spectrum of the input lensing maps:

\begin{align}
R_{L}^{\text{MC}} = \frac{\langle \hat{\phi}^{I}_{LM} \phi^{I*}_{LM}\rangle}{\langle \phi^{I}_{LM} \phi^{I*}_{LM}\rangle}\; ,
\label{eq:mc_response}
\end{align}
where $\phi^{I}_{LM}$ is the input lensing potential map and $
\hat{\phi}^{I}_{LM} = (\bar{\phi}_{LM}-\bar{\phi}_{LM}^{MF})/R^{\text{analytic}}_{L}
$ the corresponding reconstruction. 

The overall normalization of the lensing reconstruction is obtained by combining the analytic and MC responses:

\begin{align}
R_{L} = R_{L}^{\text{analytic}} \times R_{L}^{\text{MC}}.
\label{eq:response}
\end{align}
\\
The normalized, mean-field subtracted estimate of the lensing potential is then given by:

\begin{align}
\hat{\phi}_{LM} = \frac{1}{R_{L}}(\bar{\phi}_{LM}-\bar{\phi}_{LM}^{MF}) \; . 
\label{eq:phi_estimate}
\end{align}

\subsection{Estimating the Lensing Potential Power Spectrum}
\label{sec:power_spectrum}
Using Eq.~(\ref{eq:phi_estimate}), an estimate of the lensing potential power spectrum is given by:

\begin{align}
C_L^{\hat{\phi}\hat{\phi}} = f_\text{mask}^{-1}\langle\hat{\phi}_{LM}^{1}\hat{\phi}_{LM}^{2*}\rangle  \;, 
\label{eq:biased_phi_power_spectrum}
\end{align}
where $f_\text{mask} = \int\text{d}^{2}\hat{\boldsymbol{n}}{M}^4(\hat{\boldsymbol{n}})/4\pi$ corrects for the loss of power due to the application of the analysis mask. The superscripts indicate that we obtain two estimates of the lensing potential, each uses an independent set of simulations for its respective MF subtraction. By calculating the cross-power spectrum between these two maps, we prevent the auto-power spectrum of the MF residual from biasing the final result.

The estimate given by Eq.~(\ref{eq:biased_phi_power_spectrum}) contains contributions beyond the lensing signal. The dominant bias arises from the zeroth-order Gaussian noise term $N_L^{0}$, together with a smaller connected correction $N_L^{1}$ which appears at first order in the lensing power spectrum \citep{kesden03, hanson11}. Higher-order contributions, such as the $N^{3/2}$ term, are negligible at the noise levels considered here \citep{fabbian19}.

The $N_L^{0}$ bias arises through chance correlations in the CMB, astrophysical foregrounds, as well as atmospheric and detector noise. It can be estimated by computing the ensemble average of the lensing power spectra obtained from independent simulation realizations, denoted by $\bar{T}_i$ and $\bar{T}_j$, respectively:

\begin{align}
N_L^{0} = \Big \langle 
C_L^{\hat{\phi}_{\bar{T}_i\bar{T}_j}\hat{\phi}_{\bar{T}_i\bar{T}_j}} + C_L^{\hat{\phi}_{\bar{T}_i\bar{T}_j}\hat{\phi}_{\bar{T}_j\bar{T}_i}}  \Big \rangle_{i,j} \; .
\label{eq:n0_bias}
\end{align}
These realizations represent independent simulations of the total observed CMB sky (including lensed CMB signal, foregrounds, and noise) that do not share a common lensing potential, ensuring the estimate captures only the disconnected bias term. In our case, the two terms in Eq.~(\ref{eq:n0_bias}) are identical because the reconstruction uses only temperature data and the implementation is symmetric between the two input fields.

An improved estimate of the disconnected bias term can be obtained by replacing one of the input maps in Eq.~(\ref{eq:n0_bias}) with data, as this will reduce the difference between the Gaussian power in the data and simulations \citep{namikawa13}. This realization dependent estimate $N_L^{0, \text{RD}}$ is given by:

\ifdefined\PRformat
    \begin{widetext}
    \begin{align}
        N_L^{0, \text{RD}} = &\Big \langle C_L^{\hat{\phi}_{\bar{T}_i\bar{T}}\hat{\phi}_{\bar{T}_i\bar{T}}}+ C_L^{\hat{\phi}_{\bar{T}_i\bar{T}}\hat{\phi}_{\bar{T}\bar{T}_i}}+ C_L^{\hat{\phi}_{\bar{T}\bar{T}_i}\hat{\phi}_{\bar{T}_i\bar{T}}}+  C_L^{\hat{\phi}_{\bar{T}\bar{T}_i}\hat{\phi}_{\bar{T}\bar{T}_i}}\Big \rangle_{ i} - N_L^{0} \; . 
    \label{eq:n0_rd}
    \end{align}
    \end{widetext}
\else
    \begin{align}
        N_L^{0, \text{RD}} = &\Big \langle C_L^{\hat{\phi}_{\bar{T}_i\bar{T}}\hat{\phi}_{\bar{T}_i\bar{T}}}+ C_L^{\hat{\phi}_{\bar{T}_i\bar{T}}\hat{\phi}_{\bar{T}\bar{T}_i}}+ C_L^{\hat{\phi}_{\bar{T}\bar{T}_i}\hat{\phi}_{\bar{T}_i\bar{T}}}+  C_L^{\hat{\phi}_{\bar{T}\bar{T}_i}\hat{\phi}_{\bar{T}\bar{T}_i}}\Big \rangle_{ i} - N_L^{0} \; . 
    \label{eq:n0_rd}
    \end{align}
\fi

The first-order contribution to the bias $N_L^{1}$ is due to the coupling between the CMB and the lensing potential. An estimate of $N_L^{1}$ is given by averaging the lensing power spectra obtained from simulations $T_i^{\phi_i}$ and $T_j^{\phi_i}$, that have been lensed by the same lensing potential but containing different CMB realizations\footnote{To accelerate the convergence of the calculation, due to its smallness, the $N^{1}$ bias is estimated from simulations containing only the lensed CMB. The corresponding $N^{0}$ term is computed from the same type of simulations.}: 

\begin{align} 
\begin{split}
    N_L^{1} = &\Big \langle  C_L^{\hat{\phi}_{\bar{T}_i^{\phi_i}\bar{T}_j^{\phi_i}}\hat{\phi}_{\bar{T}^{\phi_i}_i\bar{T}_j^{\phi_i}}} + C_L^{\hat{\phi}_{\bar{T}_i^{\phi_i}\bar{T}_j^{\phi_i}}\hat{\phi}_{\bar{T}_j^{\phi_i}\bar{T}_i^{\phi_i}}} \Big \rangle_{i,j} - N_L^{0}\; .
\label{eq:n1_bias}
\end{split}
\end{align}

\begin{figure}
    \centering
    \ifdefined\PRformat
        \includegraphics[width=0.5\textwidth, keepaspectratio]{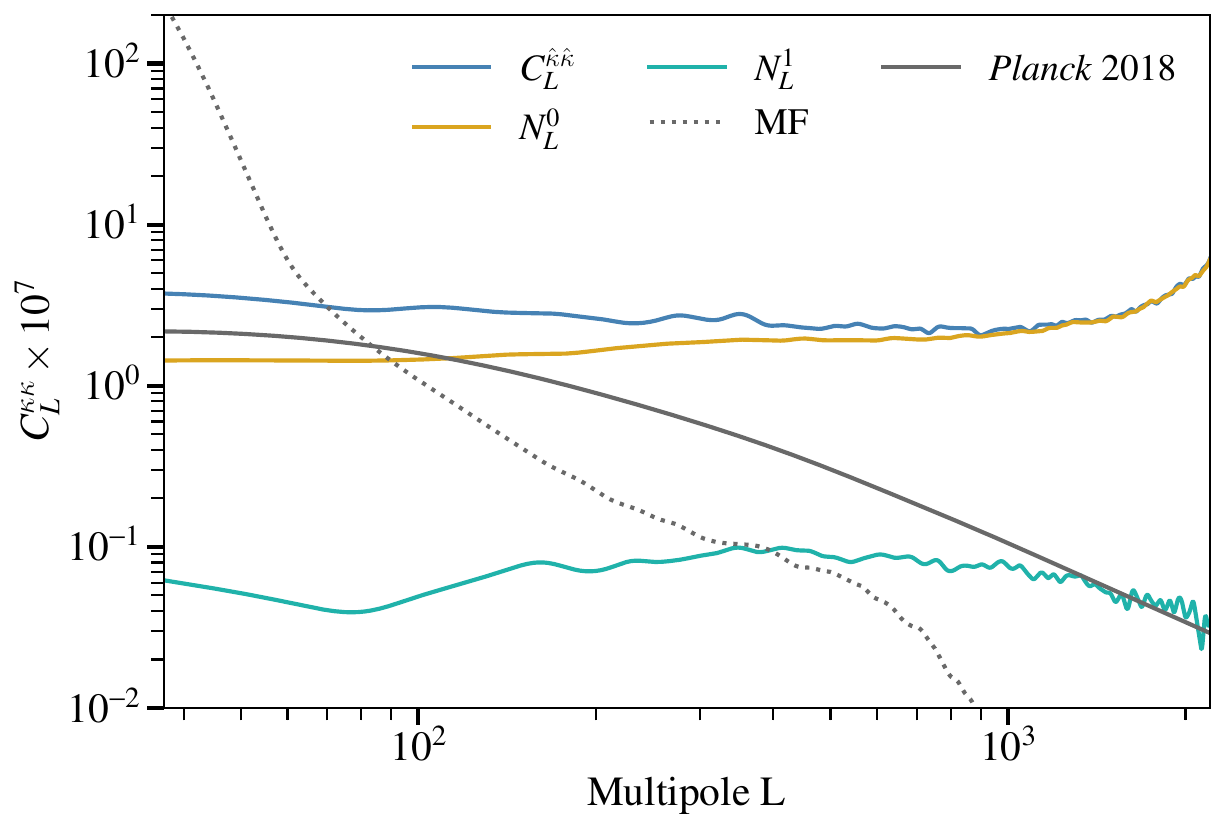}
    \else
        \includegraphics[width=0.7\textwidth, height=0.48\textheight, keepaspectratio]{figures/raw_lensing_spectra.pdf}
    \fi
    \caption{Raw biased lensing power spectrum $C_L^{\hat{\kappa}\hat{\kappa}}$ (blue) as well as the $N_L^{0}$ (yellow) and $N_L^{1}$ (teal) biases for the case of Summer-A. The mean-field contribution MF is shown by the gray dotted line, and the fiducial lensing spectrum $C_{L}^{\kappa\kappa, \rm fid}$ from \textit{Planck} 2018 is shown by the gray solid line. 
    }
\label{fig:lensing_spectra}
\end{figure}

The final estimate of the lensing potential power spectrum is then given by:

\begin{align}
\hat{C}_L^{\phi\phi} = f^{\rm MC}_L(C_L^{\hat{\phi}\hat{\phi}} - N_L^{0,\text{RD}} - N_L^{1})\; ,
\end{align} 
where $f^{\rm MC}_L = C^{\kappa\kappa, \rm fid}_L / \langle\hat{C}^{\kappa \kappa}_L\rangle$ is a multiplicative normalization factor derived from simulations that accounts for residual biases in the reconstruction pipeline, correcting for the difference between the reconstructed and input lensing power spectra. 
For each Summer field, the differences per bin between the simulated reconstructed and input power spectra are found to be $\lesssim 0.4\sigma$, where $\sigma$ denotes the statistical uncertainty of the measured lensing spectrum for that field. 
In this work, we use 125 simulations for each of the two estimates for the MF bias and 250 simulations to compute the MC response, the $N_L^{0}$$, N_L^{0, \rm RD}$, $N_L^{1}$, and $f^{\rm MC}_L$ bias terms, as well as to estimate the statistical uncertainties. Figure~\ref{fig:lensing_spectra} shows the raw, biased spectra, the bias terms, and the mean-field contribution for the case of Summer-A, as well as the fiducial input power spectrum.

\section{Results}
\label{sec:results}
\begin{figure*}
    \centering
    \includegraphics[width=\textwidth, height=\textheight, keepaspectratio]   
    {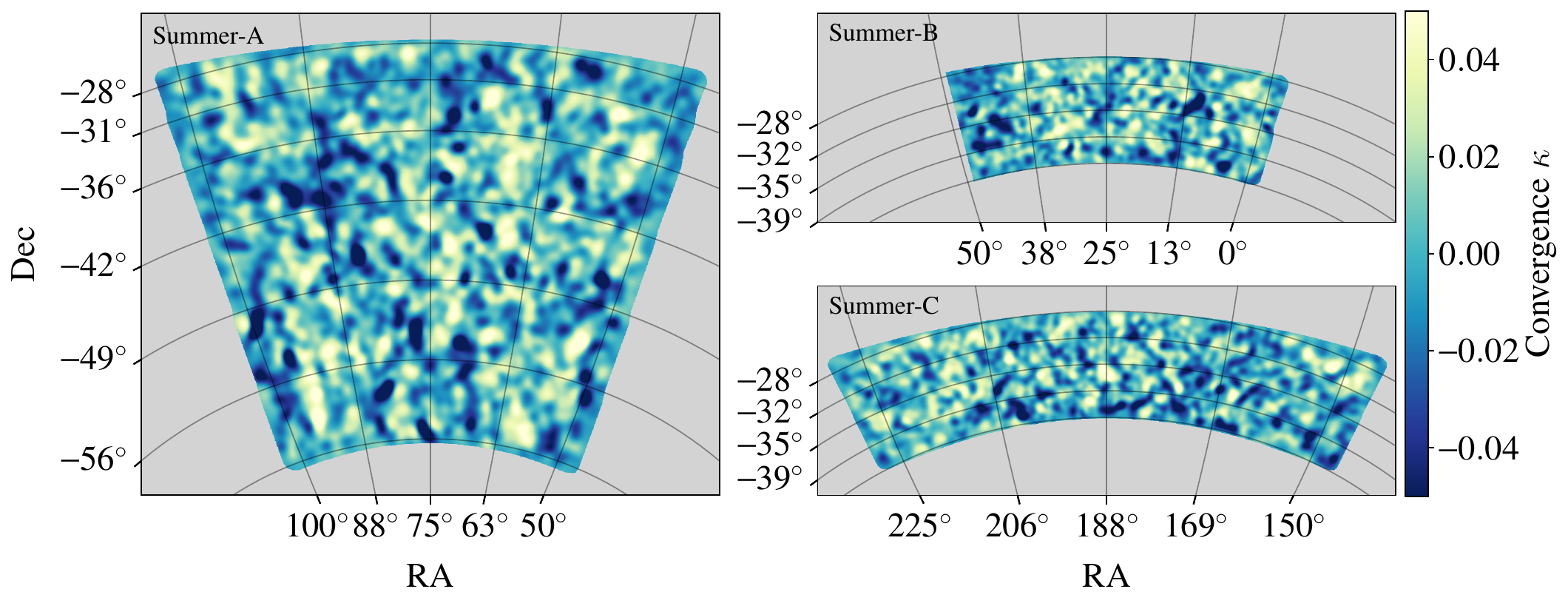}
    \caption{Reconstructed SPT-3G Summer field lensing maps smoothed with a Gaussian kernel with FWHM=1$^\circ$.}
\label{fig:lensing_maps}
\end{figure*}

This section presents high-significance reconstructions of CMB lensing maps and the corresponding power spectra obtained from two years of SPT-3G Summer temperature data. The lensing maps and power spectra are shown in terms of the lensing convergence $\kappa$, which is directly proportional to the projected matter fluctuations along the line of sight.
The reconstructed lensing maps are signal-dominated for multipoles $L \lesssim 120$ and are shown in Figure~{\ref{fig:lensing_maps}}. The maps are displayed on a zenithal equal-area projection and have been smoothed using a Gaussian kernel with FWHM=1$^\circ$ to highlight the large-scale modes. 

{We perform a number of simulation- and data-based validations, including tests on \textsc{Agora} simulations, variations of the analysis choices, and null tests, to ensure that our pipeline returns unbiased results.
We quantify the lensing measurements in terms of the lensing amplitude relative to the fiducial lensing power spectrum of the best-fit spatially flat $\Lambda$CDM \textit{Planck} 2018 cosmology.}
We bin both the measured and fiducial power spectra into 12 logarithmically spaced bins over the range $50 < L < 2000$, designating these vectors as $C_b^{\kappa\kappa, \text{fid}}$ for the fiducial spectrum and $C_b^{\kappa\kappa, i}$ for the measurement on field $i$. 
Averaging over the three Summer fields yields an estimate of the combined lensing amplitude:
\begin{align}
A^{\rm comb} = \frac{\sum_i (\hat{C}_b^{\kappa\kappa, i})^T (\mathbb{C}^{\kappa\kappa, i})^{-1} C_b^{\kappa\kappa, \rm fid}}{\sum_i (C_b^{\kappa\kappa, \rm fid})^T (\mathbb{C}^{\kappa\kappa, i})^{-1} C_b^{\kappa\kappa, \rm fid}} \;,
\label{eq:amp_comb}
\end{align}
where $\mathbb{C}^{\kappa\kappa, i}$ is the covariance matrix for field $i$ obtained from simulations.

\subsection{{Pipeline Validation}}
\label{sec_pipeline_validation}

\begin{figure*}
    \centering
    \includegraphics[width=0.9\textwidth, keepaspectratio]{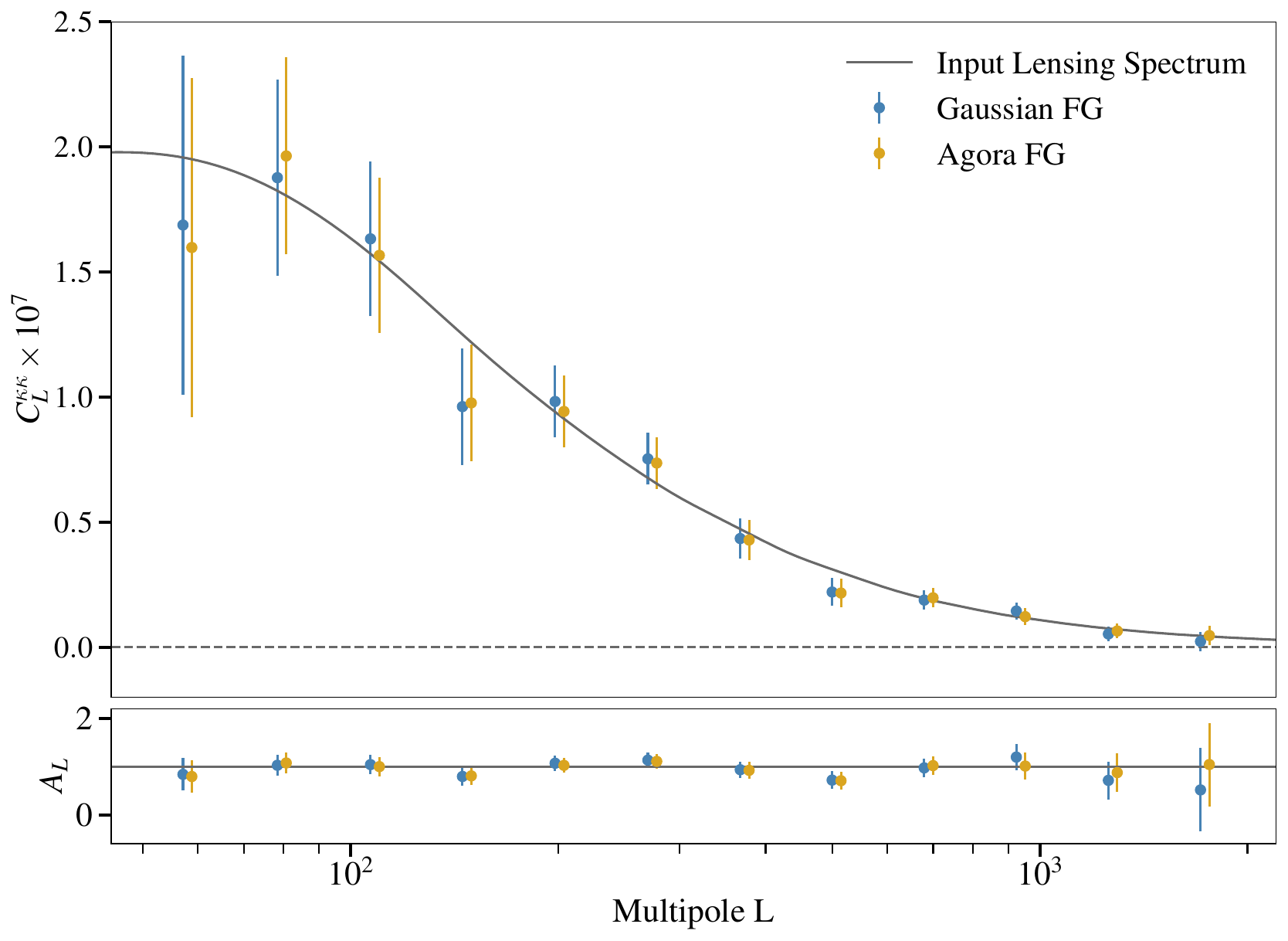}
    \caption{{Lensing power spectra reconstructed from an \textsc{Agora} lensed CMB realization combined with either Gaussian (blue) or \textsc{Agora} (yellow) foregrounds, compared to the input lensing spectrum (gray). The lower panel shows the corresponding lensing amplitudes. Both cases are consistent with the input, showing no significant bias from astrophysical foregrounds.}}
\label{fig_lensing_spectra_sims}
\end{figure*}

{
In Fig.~\ref{fig_lensing_spectra_sims}, we show the lensing bandpowers reconstructed from an \textsc{Agora} lensed CMB realization in the Summer-A field footprint for two tests. In both tests, we include simulated noise realizations corresponding to the Summer-A field. 
In the first test, we add a Gaussian realization of the astrophysical foregrounds (CIB, kSZ, radio, and tSZ) at 90, 150, and 220\,GHz, generated from power spectra consistent with \textsc{Agora} simulations. In the second test, we replace the Gaussian foregrounds with \textsc{Agora} foreground maps in each band.
In both tests, we add noise realizations corresponding to the Summer-A field. Because the Gaussian foregrounds are uncorrelated with the \textsc{Agora} lensed CMB maps, the first test validates that our pipeline returns unbiased results. The second test additionally probes foreground-induced bispectrum and trispectrum biases \citep{engelen12} in the lensing reconstruction, since the \textsc{Agora} foregrounds are correlated with the lensing map used to lens the CMB.
In both tests, the reconstructed bandpowers are consistent with the power spectrum of the input \textsc{Agora} lensing map, with probability-to-exceed (PTE) values of 0.80 and 0.93 for the Gaussian and \textsc{Agora} foreground cases, respectively.
}

\subsection{Analysis Choices}
\label{sec:sys}
\begin{figure}
    \centering
        \ifdefined\PRformat
            \includegraphics[width=0.5\textwidth, keepaspectratio]{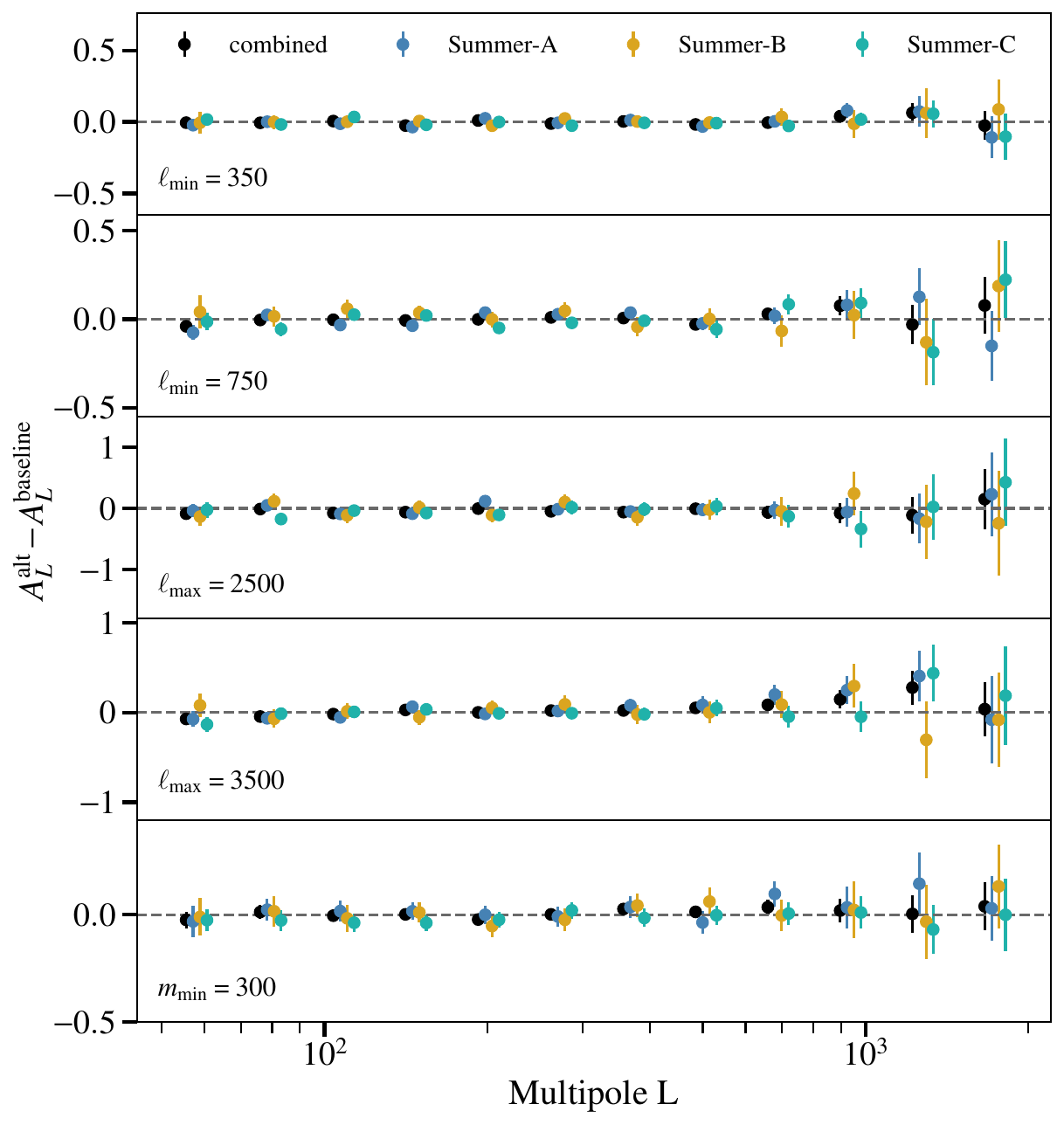}
        \else
            \includegraphics[width=0.7\textwidth,height=\textheight, keepaspectratio]{figures/systematic_tests.pdf}
        \fi
    \caption{Differences in the lensing amplitude between the alternative and baseline $(\ell_{\rm min}=500, \ell_{\rm max}=3000, m_{\rm min}=100)$ analyses {for the Summer-A, B, and C fields (blue, yellow, and teal) and their combination (black). Each panel varies one analysis choice, as indicated in the lower left.} 
    }
\label{fig:systematic_tests}
\end{figure}

To test the robustness of the lensing reconstruction pipeline, we repeat the analysis using different $\ell$ and $m$ cuts. This ensures that removing or adding CMB modes has no significant impact on the results. We quantify the impact of the different settings by calculating the difference $\Delta A \equiv A^{\mathrm{alt}} - A^{\mathrm{baseline}}$ in the lensing amplitude between the alternative and baseline analyses. For the baseline analysis, we set $\ell_{\rm min} = 500$, $\ell_{\rm max} = 3000$, and $m_{\rm min} = 100$. The amplitude differences and the corresponding PTE values to exceed this shift across the set of simulations for each case are summarized in Table~\ref{tab:systematic_tests}. Figure~\ref{fig:systematic_tests} shows the per-bin difference in the lensing amplitude.

\subsubsection{$\ell_{\rm min}$ Cuts}
Low-$\ell$ modes are prone to systematics due to low-frequency atmospheric and instrumental noise leakage, as well as timestream filtering. To analyse the impact of the low-$\ell$ modes, we redo the analysis with $\ell_{\rm min} = 350$ and $\ell_{\rm min} = 750$. We find the resulting lensing spectra to be consistent with the baseline case. Specifically, for the combined case, we find amplitude differences of $\Delta A^{\rm comb} = -0.003 \pm 0.003$ and $\Delta A^{\rm comb} = 0.005 \pm 0.007$ with PTEs of 0.82 and 0.51 for $\ell_{\rm min} = 350$ and $\ell_{\rm min} = 750$, respectively. We do not find a significant change in the lensing SNR compared to the baseline case.

\subsubsection{$\ell_{\rm max}$ Cuts}
High-$\ell$ modes are impacted by the finite angular resolution of the telescope and instrumental noise. Additionally, extragalactic foregrounds dominate over the CMB temperature at $\ell \gtrsim 3000$ which can cause a bias in the lensing potential if not properly accounted for. We rerun the lensing pipeline with $\ell_{\rm max} = 2500$ and $\ell_{\rm max} = 3500$, finding combined amplitude differences of $\Delta A^{\rm comb} = -0.028 \pm 0.029$ and $-0.004 \pm 0.020$ with PTE values of 0.72 and 0.82, respectively. The loss in the lensing SNR is $\sim 13\%$ when comparing the $\ell_{\rm max} = 2500$ analysis to the baseline case. For $\ell_{\rm max} = 3500$, the highest-$\ell$ bins show a small excess of power consistent with foreground contributions. Since the SNR does not increase substantially beyond $\ell \simeq 3000$, we adopt this value as the high-$\ell$ cutoff for our baseline analysis. 
As we do not perform cosmological parameter fits in this work, we do not marginalize over an estimate of the residual foreground contribution as has been done in some previous works \citep{pan23}.

\subsubsection{$m_{\rm min}$ Cut}

The filtering strategy of SPT-3G removes modes corresponding to $\ell_x \leqslant 300$. Since $m \approx \ell_x \cos \delta$ (where $\delta \equiv$ declination), this results in a declination-dependent $m$ cutoff, with lower-declination areas of the map contributing more lower-$m$ modes. 
For the Summer fields, the top edge of the field only retains modes down to $m \approx 300$, whereas the bottom of the field in Summer-A has sensitivity down to $m \approx 100$. 
To quantify the sensitivity of the lensing reconstruction to these lower-$m$ modes, we redo the lensing analysis with $m_{\rm min} = 300$. 
We find $\Delta A^{\rm comb} = 0.010 \pm 0.011$ with a PTE of 0.16 and no significant change in the lensing SNR.

\begin{table*}
\centering
\begin{tabular}{l|cc|cc|cc}
\hline
Field 
& \multicolumn{2}{c|}{$\ell_{\rm min} = 350$} 
& \multicolumn{2}{c|}{$\ell_{\rm min} = 750$} 
\\
\cline{2-5}
 & $\Delta A$ & PTE 
 & $\Delta A$ & PTE \\
\cline{1-5}
Summer-A & $-0.006 \pm 0.005$ & 0.29 & $-0.004 \pm 0.012$ & 0.76  \\
Summer-B & $-0.002 \pm 0.008$ & 0.84 & $0.009 \pm  0.015$ & 0.53 \\
Summer-C & $-0.004 \pm 0.005$ & 0.43 & $-0.016 \pm 0.012$ & 0.20 \\
combined & $-0.003 \pm 0.003$ & 0.82 & $-0.005 \pm 0.007$ & 0.51 \\
\hline
\hline
\hline
& \multicolumn{2}{c|}{$\ell_{\rm max} = 2500$} 
& \multicolumn{2}{c|}{$\ell_{\rm max} = 3500$} 
& \multicolumn{2}{c}{$m_{\rm min} = 300$} \\
\cline{2-7}
 & $\Delta A$ & PTE 
 & $\Delta A$ & PTE 
 & $\Delta A$ & PTE \\
\hline
Summer-A  & $-0.022 \pm 0.044$ & 0.62 & $0.004 \pm 0.037$ & 0.81 & $-0.001 \pm 0.021$ &  0.86 \\
Summer-B  &  $-0.064 \pm 0.058$ & 0.13 & $0.020 \pm 0.039$ & 0.61 & $-0.019 \pm 0.019$ & 0.31 \\
Summer-C &  $-0.047 \pm 0.052$ & 0.36 & $-0.004 \pm 0.030$ & 0.88 & $-0.018 \pm 0.017$ & 0.21 \\
combined  &  $-0.028 \pm 0.029$ & 0.72 & $0.004 \pm 0.020$  & 0.82 & $-0.010 \pm 0.011$ & 0.16 \\
\hline
\end{tabular}
\caption{Amplitude differences $\Delta A$ with corresponding PTEs to quantify the impact of the different analysis choices.}
\label{tab:systematic_tests}
\end{table*}

\subsection{Null Tests}
\label{sec:null}
\begin{figure}
\centering
    \ifdefined\PRformat
        \includegraphics[width=0.5\textwidth, keepaspectratio]{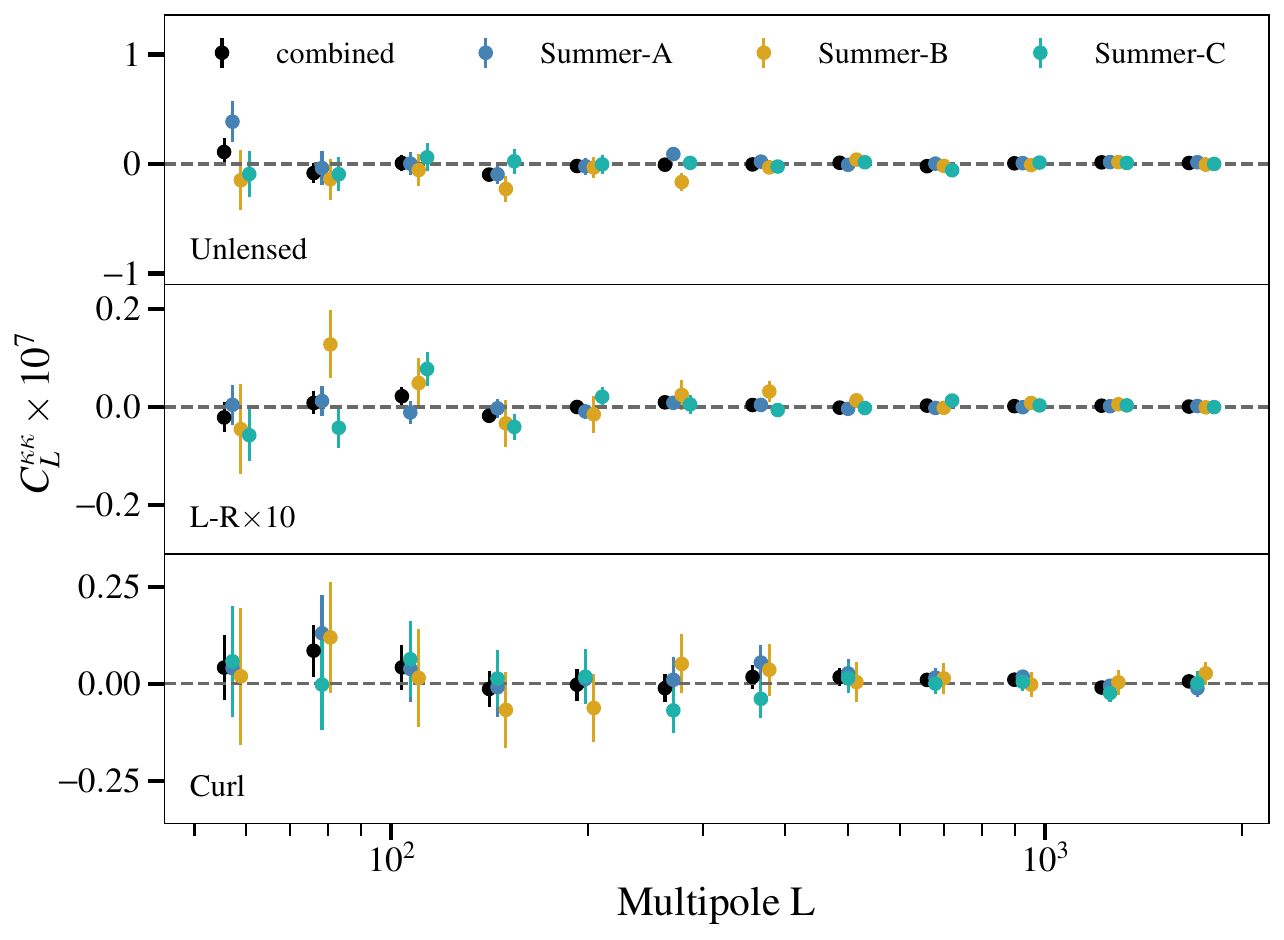}
    \else
        \includegraphics[width=0.7\textwidth,height=\textheight, keepaspectratio]{figures/null_tests.pdf}
    \fi
    \caption{\textbf{Upper panel:} Lensing spectrum for one unlensed simulation to validate the lensing pipeline. \textbf{Middle panel:} Spectrum obtained for a L-R noise realization (multiplied by a factor of 10) to quantify the quality of the noise simulations. \textbf{Lower panel:} Curl spectrum of the data, serving as a null test for residual systematics.}
\label{fig:null_tests}
\end{figure}

The lensing reconstruction obtained from data containing no lensing signal should yield a lensing amplitude consistent with zero. To validate the lensing reconstruction pipeline and to test its robustness to spurious non-Gaussian signals and systematics in the data, we perform several null tests. The corresponding results are illustrated in Figure~\ref{fig:null_tests}. Table~\ref{tab:null_tests} summarizes the best-fit lensing amplitudes, the $\chi^2$ per degree of freedom ($\chi^2$/dof, with $\text{dof}=11$), and the PTE for each null test.

\subsubsection{Unlensed Maps}
As a first null test, we verify that the pipeline does not pick up any signal when applying it to unlensed simulations. We use the same response function as for the lensed case, due to the weighting done in Eq.~(\ref{eq:phi_biased}), and no $N^{1}$ correction is subtracted. We find a best-fit lensing amplitude of $A^{\rm comb} = -0.024 \pm 0.026$ for the combined case with a PTE of 0.41. The measured amplitude is statistically consistent with the null hypothesis, indicating that the pipeline is unbiased in the absence of a true lensing signal.

\subsubsection{L-R Reconstruction}
\label{sec:lr}
To quantify the quality of the noise simulations, we apply the lensing pipeline on a null map obtained by differencing left- from right-going scans. Since the noise simulations that we use are derived from L-R null maps, this null test is particularly stringent since the sky signal has been removed from the MF and $N_0$ components. This test directly probes how well the simulations match the data in terms of noise and systematics, both at the map level (via the MF) and at the power-spectrum level (via $N_0$). From this null test, we obtain a lensing amplitude $A^{\rm comb} = 0.0006 \pm 0.0006$ with a PTE of 0.18, validating the accuracy of the noise simulations.

\subsubsection{Curl}
The lensing field can be decomposed into a gradient $\phi$ and a curl $\Omega$ component. Matter density fluctuations at linear order only produce the gradient mode \citep{namikawa12}. While gravitational waves can introduce a curl component, we do not expect to detect this signal at the given noise levels \citep{cooray05}. Therefore, the presence of curl modes signifies the contamination by non-Gaussian sources such as foregrounds. The curl component can be extracted by replacing the weight function in Eq.~(\ref{eq:phi_biased}) with:

\ifdefined\PRformat
    \begin{widetext}
        \begin{align} \nonumber
        W_{\ell_1L\ell_2}^{\Omega} = -\sqrt{\frac{(2\ell_1+1)(2\ell_2+1)(2L+1)}{4\pi}} \sqrt{L(L+1)\ell_1(\ell_1+1)}\  C^{TT}_{\ell_1} \left( \frac{1-(-1)^{\ell_1+\ell_2+L}}{2} \right) \begin{pmatrix}
        \ell_1 & \ell_2 & L\\
        1 & 0 & -1
        \end{pmatrix}+ (\ell_1 \leftrightarrow \ell_2)\;.
        \end{align}
        \label{eq:curl_weights}
    \end{widetext}    
\else
    \begin{align} \nonumber
    W_{\ell_1L\ell_2}^{\Omega} = &-\sqrt{\frac{(2\ell_1+1)(2\ell_2+1)(2L+1)}{4\pi}} \sqrt{L(L+1)\ell_1(\ell_1+1)} \\ & \times C^{TT}_{\ell_1} \left( \frac{1-(-1)^{\ell_1+\ell_2+L}}{2} \right) \begin{pmatrix}
    \ell_1 & \ell_2 & L\\
    1 & 0 & -1
    \end{pmatrix}+ (\ell_1 \leftrightarrow \ell_2)\;.
    \label{eq:curl_weights}
    \end{align}
\fi
Using this weight function, we rerun the pipeline to extract the curl spectrum. We obtain a combined lensing amplitude of $A_{\rm comb} = {0.019} \pm 0.021$ with PTE = 0.42, which is statistically consistent with the null hypothesis, indicating no evidence for systematic contamination.

\begin{table*}
\begin{adjustbox}{width=1\textwidth}
\centering
\begin{tabular}{l|ccc|ccc|ccc}
\hline
Field
& \multicolumn{3}{c|}{Unlensed} 
& \multicolumn{3}{c|}{L-R} 
& \multicolumn{3}{c}{Curl} \\
\cline{2-10}
 & Ampl. & $\chi^2/\rm{dof}$ & PTE & Ampl. & $\chi^2/\rm{dof}$ & PTE & Ampl. & $\chi^2/\rm{dof}$ & PTE \\
\hline
Summer-A & 0.008 $\pm$ 0.042 & 1.20 & 0.28 & 0.0002 $\pm$ 0.0008 & 0.83 & 0.61 & 0.026 $\pm$ 0.033 & 1.57 & 0.15 \\
Summer-B & -0.094 $\pm$ 0.049 & 0.86 & 0.58 & 0.0014 $\pm$ 0.0016 & 1.19  & 0.29 & 0.002 $\pm$ 0.043 & 0.50 & 0.95 \\
Summer-C & -0.002 $\pm$ 0.046 & 0.96 & 0.48 & 0.0010 $\pm$ 0.0010 & 1.8 & 0.05 & 0.023 $\pm$ 0.035 & 1.04 & 0.40 \\
combined & -0.024 $\pm$ 0.026 & 1.03 & 0.41 & 0.0006 $\pm$ 0.0006 & 1.2 & 0.18 & {0.019} $\pm$ 0.021 & 0.95 & 0.42 \\
\hline
\end{tabular}
\end{adjustbox}
\caption{Summary of null tests performed to check for potential contamination from the lensing reconstruction pipeline, noise simulations, or the data. }
\label{tab:null_tests}
\end{table*}

\subsection{Lensing Power Spectra}
\label{sec:spectra}
\begin{figure*}
    \centering
    \includegraphics[width=0.9\textwidth, keepaspectratio]{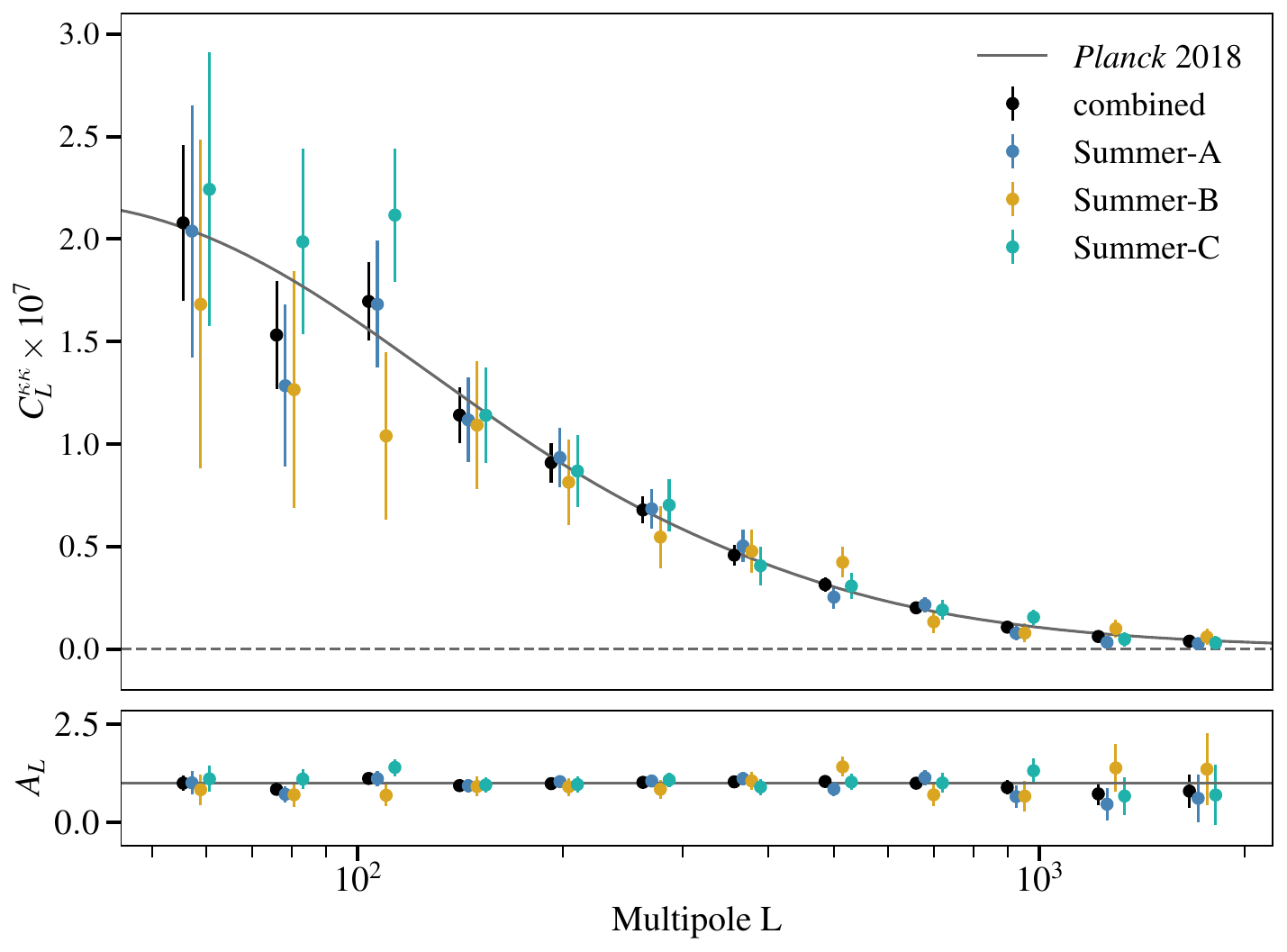}
    \caption{Lensing power spectrum bandpowers (upper panel) and corresponding lensing amplitudes (lower panel) for the individual Summer fields and their combination. The measurements are consistent with the fiducial lensing spectrum predicted by a $\Lambda$CDM cosmology constrained by \textit{Planck} 2018.}
\label{fig:binned_lensing_spectra}
\end{figure*}

\begin{figure*}
    \centering
    \includegraphics[width=0.9\textwidth, keepaspectratio]{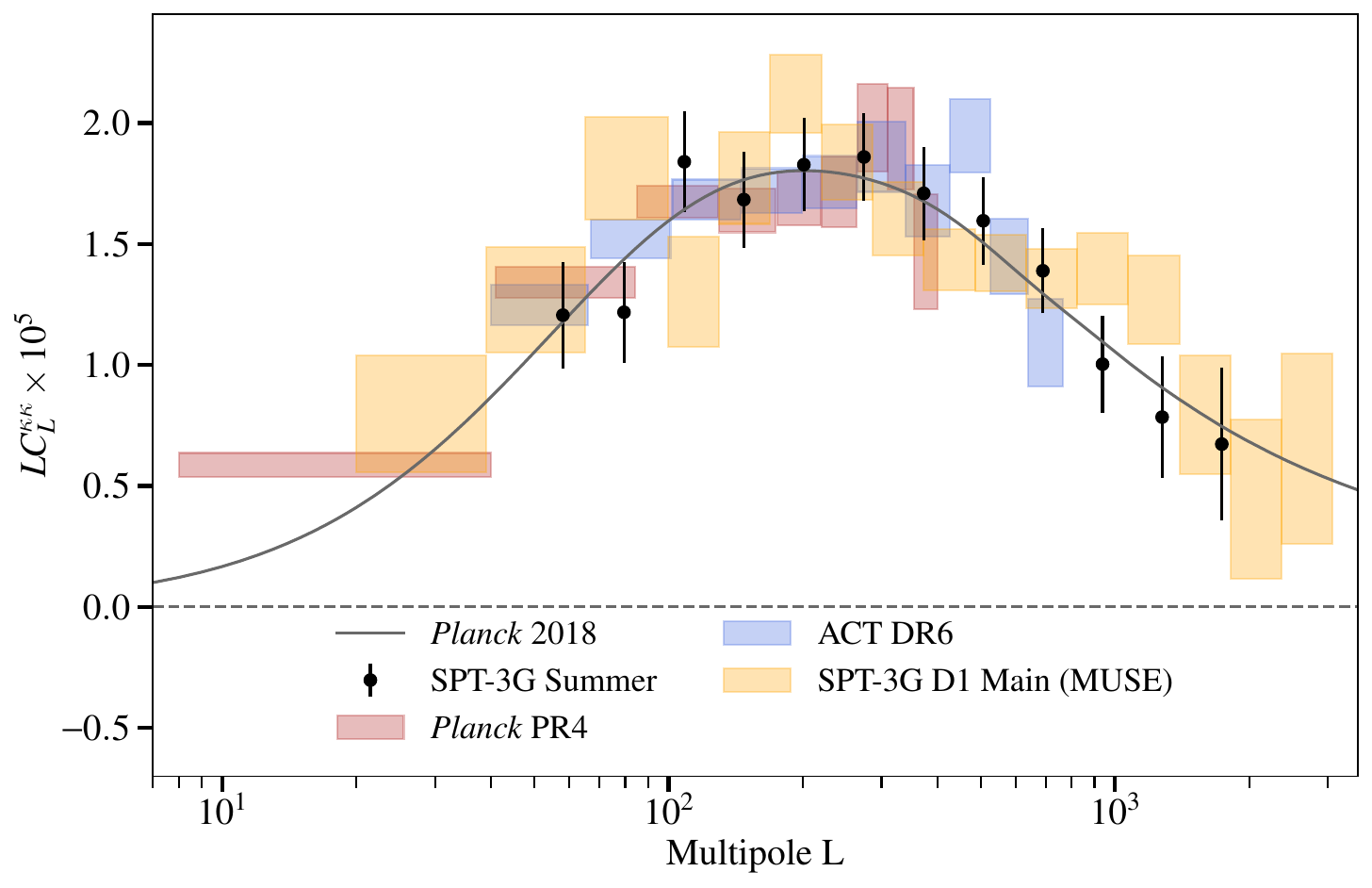}
    \caption{Lensing power spectrum measurements from this work and previous CMB experiments: \textit{\textit{Planck}} PR4 \citep{carron22}, ACT DR6 \citep{qu24}, and SPT-3G D1 polarization-only data using the Marginal Unbiased Score Expansion (MUSE) method \citep{ge25}. The results are consistent with previous CMB lensing power spectrum measurements, demonstrating the robustness of lensing reconstruction in the Summer fields.}
\label{fig:comparison}
\end{figure*}

The binned lensing power spectra for each Summer field, as well as the combined one, are shown in the upper panel of Figure~\ref{fig:binned_lensing_spectra}. The lower panel shows the corresponding binned lensing amplitudes. The band power values in each bin are listed in Table~\ref{tab:bandpowers}, together with the bin edges and bin centers. Combining the lensing spectra of the individual fields results in a best-fit lensing amplitude of

\begin{align}
A^{\rm comb} =  {1.014} \pm 0.053 \;,
\end{align}
with a PTE of {0.82}. This corresponds to a measurement of the lensing amplitude with a SNR of 19 when considering only statistical uncertainty. Table~\ref{tab:amplitudes} summarizes the best-fit lensing amplitudes, the $\chi^2/$dof, and the corresponding PTE values for each individual field as well as for the combined result. Figure~\ref{fig:comparison} compares the combined lensing power spectrum with results from the SPT-3G Main field, as well as results from \textit{\textit{Planck}} and ACT. The agreement between the different measurements provides an important consistency check of the reconstruction. 

\begin{table*}
  \centering
  \begin{adjustbox}{width=\textwidth}      
\begin{tabular}{l|l|l|l|l|l}
    \hline
     $[L_{\rm min}$, $ L_{\rm max}]$ & $L_b$ &
     $\hat{C}_L^{\kappa\kappa,\rm Summer-A} [10^{7}$] & $\hat{C}_L^{\kappa\kappa,\rm Summer-B} [10^{7}$] & $\hat{C}_L^{\kappa\kappa,\rm Summer-C} [10^{7}$] & $\hat{C}_L^{\kappa\kappa,\rm comb} [10^{7}$] \\
    \hline
    $[50, 67]$ & 58 & $2.039 \pm 0.615$  & $1.682 \pm 0.801$ & $2.242 \pm 0.668$  &  {$2.079 \pm 0.381$} \\
    $[68, 92]$  & 79.5  & $1.285 \pm 0.395$ & $1.265 \pm 0.578$  &   $1.987 \pm 0.452$ &  {$1.532 \pm 0.262$}   \\
    $[93, 125]$ & 108.5  & $1.682 \pm 0.309$  & $1.040 \pm 0.410$  & $2.117\pm 0.324$  & {$1.696 \pm 0.192$} \\
    $[126, 170]$ & 147.5  & $1.118 \pm  0.206$  &  $1.092 \pm 0.311$  &  $1.140 \pm 0.232$ &  {$1.141 \pm 0.135$} \\
    $[171, 232]$ & 201  & $0.934 \pm 0.143$  & $0.814 \pm 0.209$ & $0.868 \pm 0.177$ &  {$0.909 \pm 0.096$}   \\
    $[233, 316]$ & 274  & $0.684 \pm 0.095$  & $0.546 \pm 0.152$    &  $0.702 \pm 0.126$ &  {$0.679 \pm 0.066$}      \\
    $[317, 430]$  & 373  & $0.503 \pm 0.079$  & $0.477 \pm 0.107$  &   $0.406 \pm 0.094$  & {$0.458 \pm 0.052$}\\
    $[431, 584]$  & 507  & $0.254 \pm 0.056$  & $0.424 \pm 0.075$  & $0.307 \pm 0.063$ & {$0.315 \pm 0.036$}\\
    $[585, 795]$ & 689.5  & $0.216 \pm 0.038$  & $ 0.133 \pm 0.054$   & $0.191 \pm 0.049$  & {$0.201 \pm 0.025$}\\
    $[796, 1001]$ & 938  & $0.077 \pm 0.034$  & $0.079 \pm 0.046$   & $0.155 \pm 0.037$  & {$0.107 \pm  0.021$}\\
    $[1082, 1470]$ & 1275.5  & $0.033 \pm 0.029$  & $0.100 \pm 0.043$  &   $0.048 \pm 0.036$ &  {$0.062 \pm 0.020$}  \\
    $[1471, 2000]$ & 1735  & $0.027 \pm 0.027$  &  $0.059 \pm 0.040$ &   $0.030 \pm 0.034$ &  {$0.039 \pm  0.018$}  \\
    \hline
\end{tabular}

  \end{adjustbox}
  \caption{Binned bandpowers of the individual and combined lensing spectra. The bins are evenly spaced in log space. The individual bandpowers are reported at the center of each bin.}
  \label{tab:bandpowers}
\end{table*}

\begin{table}
  \centering
  \begin{tabular}{l|l|l|l} 
    \hline
    Field & Ampl. & $\chi^2/\rm{dof}$ & PTE\\
    \hline
    Summer-A & $1.028 \pm 0.079$ & 0.83 & 0.61 \\
    Summer-B & $0.887 \pm 0.115$ & 0.94 & 0.50 \\
    Summer-C & $1.078 \pm 0.093$ & 0.55 & 0.87 \\
    combined & ${1.014} \pm 0.053$ & 0.78 & {0.82} \\
    \hline
  \end{tabular}
  \caption{Best-fit lensing amplitudes, $\chi^2/\rm{dof}$, and PTEs of the individual power spectra as well as the combined spectrum.}
  \label{tab:amplitudes}
\end{table}

\section{Conclusions}
\label{sec:conclusion}
This paper presents the first CMB lensing measurements from the SPT-3G Summer survey. Using two years of temperature data, we reconstruct lensing maps that are signal dominated up to $L \approx 120$.

{We validate our pipeline on \textsc{Agora} simulations and find that the lensing power spectra reconstructed with both Gaussian and \textsc{Agora} foregrounds are consistent with the input, showing no significant bias from the astrophysical foregrounds.
We further} evaluate the robustness of our results by varying the multipole range used in the lensing reconstruction. None of these variations show evidence for significant systematic contamination. Additionally, we perform several null tests to check for potential biases originating from the lensing reconstruction pipeline and from the data. These include lensing reconstruction from unlensed simulations, noise maps, and reconstruction of the curl mode. In all cases, we find results consistent with a null lensing signal. 

We find a combined best-fit lensing amplitude $A^{\rm comb} = {1.014} \pm 0.053$ when considering only statistical uncertainties, which is consistent with the fiducial \textit{Planck} $\Lambda$CDM model. 
Our measurement is in good agreement with previous lensing results from \textit{Planck}, ACT, and SPT-3G across the considered multipole range.

Future lensing reconstructions from the full four-year SPT-3G Summer survey will include polarization data, which is less affected by extragalactic foregrounds and enables the use of smaller angular scales in the input CMB maps. The inclusion of the Summer and Wide fields increases the total SPT-3G survey area to $\approx10~000 \rm{~deg}^2$ (Ext-10K survey), yielding a substantial gain in the lensing SNR. In particular, the Summer fields alone are expected to provide a significant ($\sim$20\%) contribution to the total Ext-10K lensing SNR, highlighting their importance for the final survey sensitivity.

Since the SPT-3G Main and Summer fields have a complete overlap with the Vera C. Rubin Observatory (LSST; \citealp{lsst09}) and the EDF-S field of \textit{Euclid} \citep{euclid25}, the SPT-3G lensing maps will be of great interest for future cross-correlation measurements with galaxy positions and shear measurements. Such measurements enable improved cosmological constraints, calibration of shear and galaxy clustering measurements, and the breaking of redshift–dependent degeneracies.

\section*{Acknowledgments}
\label{sec:acknowledgements}
{We thank the anonymous reviewer for useful feedback that helped in shaping this manuscript better.}
The South Pole Telescope program is supported by the National Science Foundation (NSF) through awards OPP-1852617 and OPP-2332483. Partial support is also provided by the Kavli Institute of Cosmological Physics at the University of Chicago. 
Argonne National Laboratory’s work was supported by the U.S. Department of Energy, Office of High Energy Physics, under contract DE-AC02-06CH11357. 
The UC Davis group acknowledges support from Michael and Ester Vaida. 
This document was prepared by the SPT-3G collaboration using the resources of the Fermi National Accelerator Laboratory (Fermilab), a U.S. Department of Energy, Office of Science, Office of High Energy Physics HEP User Facility. Fermilab is managed by Fermi Forward Discovery Group, LLC, acting under Contract No. 89243024CSC000002.
The Melbourne authors acknowledge support from the Australian Research Council’s Discovery Project scheme (No. DP210102386). 
The Paris group has received funding from the European Research Council (ERC) under the European Union’s Horizon 2020 research and innovation program (grant agreement No 101001897), and funding from the Centre National d’Etudes Spatiales. 
The SLAC group is supported in part by the Department of Energy at SLAC National Accelerator Laboratory, under contract DE-AC02-76SF00515.

\bibliography{spt_summer_2y_lensing}

@ARTICLE{prabhu24,
       author = {{Prabhu}, K. and {Raghunathan}, S. and {Millea}, M. and {Lynch}, G.~P. and {Ade}, P.~A.~R. and {Anderes}, E. and {Anderson}, A.~J. and {Ansarinejad}, B. and {Archipley}, M. and {Balkenhol}, L. and {Benabed}, K. and {Bender}, A.~N. and {Benson}, B.~A. and {Bianchini}, F. and {Bleem}, L.~E. and {Bouchet}, F.~R. and {Bryant}, L. and {Camphuis}, E. and {Carlstrom}, J.~E. and {Cecil}, T.~W. and {Chang}, C.~L. and {Chaubal}, P. and {Chichura}, P.~M. and {Chokshi}, A. and {Chou}, T. -L. and {Coerver}, A. and {Crawford}, T.~M. and {Cukierman}, A. and {Daley}, C. and {de Haan}, T. and {Dibert}, K.~R. and {Dobbs}, M.~A. and {Doussot}, A. and {Dutcher}, D. and {Everett}, W. and {Feng}, C. and {Ferguson}, K.~R. and {Fichman}, K. and {Foster}, A. and {Galli}, S. and {Gambrel}, A.~E. and {Gardner}, R.~W. and {Ge}, F. and {Goeckner-Wald}, N. and {Gualtieri}, R. and {Guidi}, F. and {Guns}, S. and {Halverson}, N.~W. and {Hivon}, E. and {Holder}, G.~P. and {Holzapfel}, W.~L. and {Hood}, J.~C. and {Hryciuk}, A. and {Huang}, N. and {K{\'e}ruzor{\'e}}, F. and {Knox}, L. and {Korman}, M. and {Kornoelje}, K. and {Kuo}, C. -L. and {Lee}, A.~T. and {Levy}, K. and {Lowitz}, A.~E. and {Lu}, C. and {Maniyar}, A. and {Menanteau}, F. and {Montgomery}, J. and {Nakato}, Y. and {Natoli}, T. and {Noble}, G.~I. and {Novosad}, V. and {Omori}, Y. and {Padin}, S. and {Pan}, Z. and {Paschos}, P. and {Phadke}, K.~A. and {Pollak}, A.~W. and {Quan}, W. and {Rahimi}, M. and {Rahlin}, A. and {Reichardt}, C.~L. and {Rouble}, M. and {Ruhl}, J.~E. and {Schiappucci}, E. and {Smecher}, G. and {Sobrin}, J.~A. and {Stark}, A.~A. and {Stephen}, J. and {Suzuki}, A. and {Tandoi}, C. and {Thompson}, K.~L. and {Thorne}, B. and {Trendafilova}, C. and {Tucker}, C. and {Umilta}, C. and {Vitrier}, A. and {Vieira}, J.~D. and {Wan}, Y. and {Wang}, G. and {Whitehorn}, N. and {Wu}, W.~L.~K. and {Yefremenko}, V. and {Young}, M.~R. and {Zebrowski}, J.~A.},
        title = "{Testing the {\ensuremath{\Lambda}}CDM Cosmological Model with Forthcoming Measurements of the Cosmic Microwave Background with SPT-3G}",
      journal = {\apj},
         year = 2024,
        month = sep,
       volume = {973},
       number = {1},
          eid = {4},
        pages = {4},
          doi = {10.3847/1538-4357/ad5ff1},
archivePrefix = {arXiv},
       eprint = {2403.17925},
 primaryClass = {astro-ph.CO},
       adsurl = {https://ui.adsabs.harvard.edu/abs/2024ApJ...973....4P}
}

@ARTICLE{quan26,
       author = {{Quan}, W. and {Camphuis}, E. and {Daley}, C. and {Huang}, N. and {Omori}, Y. and {Guidi}, F. and {Anderes}, E. and {Anderson}, A.~J. and {Ansarinejad}, B. and {Archipley}, M. and {Balkenhol}, L. and {Barron}, D.~R. and {Benabed}, K. and {Bender}, A.~N. and {Benson}, B.~A. and {Bianchini}, F. and {Bleem}, L.~E. and {Bocquet}, S. and {Bouchet}, F.~R. and {Campitiello}, M.~G. and {Carlstrom}, J.~E. and {Carron}, J. and {Chang}, C.~L. and {Chichura}, P.~M. and {Chokshi}, A. and {Chou}, T.-L. and {Coerver}, A. and {Crawford}, T.~M. and {de Haan}, T. and {Dibert}, K.~R. and {Dobbs}, M.~A. and {Doohan}, M. and {Dutcher}, D. and {Feng}, C. and {Ferguson}, K.~R. and {Ferree}, N.~C. and {Fichman}, K. and {Foster}, A. and {Galli}, S. and {Gambrel}, A.~E. and {Gao}, A.~K. and {Ge}, F. and {Guns}, S. and {Halverson}, N.~W. and {Hivon}, E. and {Holder}, G.~P. and {Holzapfel}, W.~L. and {Hood}, J.~C. and {Hryciuk}, A. and {Jhaveri}, T. and {K{\'e}ruzor{\'e}}, F. and {Khalife}, A.~R. and {Knox}, L. and {Kornoelje}, K. and {Kuo}, C.-L. and {Levy}, K. and {Li}, Y. and {Lowitz}, A.~E. and {Lu}, C. and {Lynch}, G.~P. and {Maccarone}, T.~J. and {Maniyar}, A.~S. and {Martsen}, E.~S. and {Menanteau}, F. and {Millea}, M. and {Montgomery}, J. and {Nakato}, Y. and {Natoli}, T. and {Ouellette}, A. and {Pan}, Z. and {Paschos}, P. and {Phadke}, K.~A. and {Pollak}, A.~W. and {Prabhu}, K. and {Raghunathan}, S. and {Rahimi}, M. and {Rahlin}, A. and {Reichardt}, C.~L. and {Rouble}, M. and {Ruhl}, J.~E. and {Silva Oliveira}, A.~C. and {Simpson}, A. and {Sobrin}, J.~A. and {Stark}, A.~A. and {Stephen}, J. and {Tandoi}, C. and {Trendafilova}, C. and {Vieira}, J.~D. and {Vieregg}, A.~G. and {Vitrier}, A. and {Wan}, Y. and {Whitehorn}, N. and {Wu}, W.~L.~K. and {Young}, M.~R. and {Zebrowski}, J.~A.},
        title = "{SPT-3G D1: Maps of the millimeter-wave sky from 2019 and 2020 observations of the SPT-3G Main field}",
      journal = {arXiv e-prints},
         year = 2026,
        month = mar,
          eid = {arXiv:2603.20163},
        pages = {arXiv:2603.20163},
          doi = {10.48550/arXiv.2603.20163},
archivePrefix = {arXiv},
       eprint = {2603.20163},
 primaryClass = {astro-ph.CO},
       adsurl = {https://ui.adsabs.harvard.edu/abs/2026arXiv260320163Q}
}

@ARTICLE{madhavacheril24,
       author = {{Madhavacheril}, Mathew S. and {Qu}, Frank J. and {Sherwin}, Blake D. and {MacCrann}, Niall and {Li}, Yaqiong and {Abril-Cabezas}, Irene and {Ade}, Peter A.~R. and {Aiola}, Simone and {Alford}, Tommy and {Amiri}, Mandana and {Amodeo}, Stefania and {An}, Rui and {Atkins}, Zachary and {Austermann}, Jason E. and {Battaglia}, Nicholas and {Battistelli}, Elia Stefano and {Beall}, James A. and {Bean}, Rachel and {Beringue}, Benjamin and {Bhandarkar}, Tanay and {Biermann}, Emily and {Bolliet}, Boris and {Bond}, J. Richard and {Cai}, Hongbo and {Calabrese}, Erminia and {Calafut}, Victoria and {Capalbo}, Valentina and {Carrero}, Felipe and {Challinor}, Anthony and {Chesmore}, Grace E. and {Cho}, Hsiao-mei and {Choi}, Steve K. and {Clark}, Susan E. and {C{\'o}rdova Rosado}, Rodrigo and {Cothard}, Nicholas F. and {Coughlin}, Kevin and {Coulton}, William and {Crowley}, Kevin T. and {Dalal}, Roohi and {Darwish}, Omar and {Devlin}, Mark J. and {Dicker}, Simon and {Doze}, Peter and {Duell}, Cody J. and {Duff}, Shannon M. and {Duivenvoorden}, Adriaan J. and {Dunkley}, Jo and {D{\"u}nner}, Rolando and {Fanfani}, Valentina and {Fankhanel}, Max and {Farren}, Gerrit and {Ferraro}, Simone and {Freundt}, Rodrigo and {Fuzia}, Brittany and {Gallardo}, Patricio A. and {Garrido}, Xavier and {Givans}, Jahmour and {Gluscevic}, Vera and {Golec}, Joseph E. and {Guan}, Yilun and {Hall}, Kirsten R. and {Halpern}, Mark and {Han}, Dongwon and {Harrison}, Ian and {Hasselfield}, Matthew and {Healy}, Erin and {Henderson}, Shawn and {Hensley}, Brandon and {Herv{\'\i}as-Caimapo}, Carlos and {Hill}, J. Colin and {Hilton}, Gene C. and {Hilton}, Matt and {Hincks}, Adam D. and {Hlo{\v{z}}ek}, Ren{\'e}e and {Ho}, Shuay-Pwu Patty and {Huber}, Zachary B. and {Hubmayr}, Johannes and {Huffenberger}, Kevin M. and {Hughes}, John P. and {Irwin}, Kent and {Isopi}, Giovanni and {Jense}, Hidde T. and {Keller}, Ben and {Kim}, Joshua and {Knowles}, Kenda and {Koopman}, Brian J. and {Kosowsky}, Arthur and {Kramer}, Darby and {Kusiak}, Aleksandra and {La Posta}, Adrien and {Lague}, Alex and {Lakey}, Victoria and {Lee}, Eunseong and {Li}, Zack and {Limon}, Michele and {Lokken}, Martine and {Louis}, Thibaut and {Lungu}, Marius and {MacInnis}, Amanda and {Maldonado}, Diego and {Maldonado}, Felipe and {Mallaby-Kay}, Maya and {Marques}, Gabriela A. and {McMahon}, Jeff and {Mehta}, Yogesh and {Menanteau}, Felipe and {Moodley}, Kavilan and {Morris}, Thomas W. and {Mroczkowski}, Tony and {Naess}, Sigurd and {Namikawa}, Toshiya and {Nati}, Federico and {Newburgh}, Laura and {Nicola}, Andrina and {Niemack}, Michael D. and {Nolta}, Michael R. and {Orlowski-Scherer}, John and {Page}, Lyman A. and {Pandey}, Shivam and {Partridge}, Bruce and {Prince}, Heather and {Puddu}, Roberto and {Radiconi}, Federico and {Robertson}, Naomi and {Rojas}, Felipe and {Sakuma}, Tai and {Salatino}, Maria and {Schaan}, Emmanuel and {Schmitt}, Benjamin L. and {Sehgal}, Neelima and {Shaikh}, Shabbir and {Sierra}, Carlos and {Sievers}, Jon and {Sif{\'o}n}, Crist{\'o}bal and {Simon}, Sara and {Sonka}, Rita and {Spergel}, David N. and {Staggs}, Suzanne T. and {Storer}, Emilie and {Switzer}, Eric R. and {Tampier}, Niklas and {Thornton}, Robert and {Trac}, Hy and {Treu}, Jesse and {Tucker}, Carole and {Ullom}, Joel and {Vale}, Leila R. and {Van Engelen}, Alexander and {Van Lanen}, Jeff and {van Marrewijk}, Joshiwa and {Vargas}, Cristian and {Vavagiakis}, Eve M. and {Wagoner}, Kasey and {Wang}, Yuhan and {Wenzl}, Lukas and {Wollack}, Edward J. and {Xu}, Zhilei and {Zago}, Fernando and {Zheng}, Kaiwen},
        title = "{The Atacama Cosmology Telescope: DR6 Gravitational Lensing Map and Cosmological Parameters}",
      journal = {\apj},
         year = 2024,
        month = feb,
       volume = {962},
       number = {2},
          eid = {113},
        pages = {113},
          doi = {10.3847/1538-4357/acff5f},
archivePrefix = {arXiv},
       eprint = {2304.05203},
 primaryClass = {astro-ph.CO},
       adsurl = {https://ui.adsabs.harvard.edu/abs/2024ApJ...962..113M}
}

@ARTICLE{lewis06,
       author = {{Lewis}, Antony and {Challinor}, Anthony},
        title = "{Weak gravitational lensing of the CMB}",
      journal = {\physrep},
         year = 2006,
        month = jun,
       volume = {429},
       number = {1},
        pages = {1-65},
          doi = {10.1016/j.physrep.2006.03.002},
archivePrefix = {arXiv},
       eprint = {astro-ph/0601594},
 primaryClass = {astro-ph},
       adsurl = {https://ui.adsabs.harvard.edu/abs/2006PhR...429....1L}
}

@ARTICLE{okamoto03,
       author = {{Okamoto}, Takemi and {Hu}, Wayne},
        title = "{Cosmic microwave background lensing reconstruction on the full sky}",
      journal = {\prd},
         year = 2003,
        month = apr,
       volume = {67},
       number = {8},
          eid = {083002},
        pages = {083002},
          doi = {10.1103/PhysRevD.67.083002},
archivePrefix = {arXiv},
       eprint = {astro-ph/0301031},
 primaryClass = {astro-ph},
       adsurl = {https://ui.adsabs.harvard.edu/abs/2003PhRvD..67h3002O}
}

@ARTICLE{omori17,
       author = {{Omori}, Y. and {Chown}, R. and {Simard}, G. and {Story}, K.~T. and {Aylor}, K. and {Baxter}, E.~J. and {Benson}, B.~A. and {Bleem}, L.~E. and {Carlstrom}, J.~E. and {Chang}, C.~L. and {Cho}, H. -M. and {Crawford}, T.~M. and {Crites}, A.~T. and {de Haan}, T. and {Dobbs}, M.~A. and {Everett}, W.~B. and {George}, E.~M. and {Halverson}, N.~W. and {Harrington}, N.~L. and {Holder}, G.~P. and {Hou}, Z. and {Holzapfel}, W.~L. and {Hrubes}, J.~D. and {Knox}, L. and {Lee}, A.~T. and {Leitch}, E.~M. and {Luong-Van}, D. and {Manzotti}, A. and {Marrone}, D.~P. and {McMahon}, J.~J. and {Meyer}, S.~S. and {Mocanu}, L.~M. and {Mohr}, J.~J. and {Natoli}, T. and {Padin}, S. and {Pryke}, C. and {Reichardt}, C.~L. and {Ruhl}, J.~E. and {Sayre}, J.~T. and {Schaffer}, K.~K. and {Shirokoff}, E. and {Staniszewski}, Z. and {Stark}, A.~A. and {Vanderlinde}, K. and {Vieira}, J.~D. and {Williamson}, R. and {Zahn}, O.},
        title = "{A 2500 deg$^{2}$ CMB Lensing Map from Combined South Pole Telescope and Planck Data}",
      journal = {\apj},
         year = 2017,
        month = nov,
       volume = {849},
       number = {2},
          eid = {124},
        pages = {124},
          doi = {10.3847/1538-4357/aa8d1d},
archivePrefix = {arXiv},
       eprint = {1705.00743},
 primaryClass = {astro-ph.CO},
       adsurl = {https://ui.adsabs.harvard.edu/abs/2017ApJ...849..124O}
}

@ARTICLE{carlstrom11,
       author = {{Carlstrom}, J.~E. and {Ade}, P.~A.~R. and {Aird}, K.~A. and {Benson}, B.~A. and {Bleem}, L.~E. and {Busetti}, S. and {Chang}, C.~L. and {Chauvin}, E. and {Cho}, H. -M. and {Crawford}, T.~M. and {Crites}, A.~T. and {Dobbs}, M.~A. and {Halverson}, N.~W. and {Heimsath}, S. and {Holzapfel}, W.~L. and {Hrubes}, J.~D. and {Joy}, M. and {Keisler}, R. and {Lanting}, T.~M. and {Lee}, A.~T. and {Leitch}, E.~M. and {Leong}, J. and {Lu}, W. and {Lueker}, M. and {Luong-Van}, D. and {McMahon}, J.~J. and {Mehl}, J. and {Meyer}, S.~S. and {Mohr}, J.~J. and {Montroy}, T.~E. and {Padin}, S. and {Plagge}, T. and {Pryke}, C. and {Ruhl}, J.~E. and {Schaffer}, K.~K. and {Schwan}, D. and {Shirokoff}, E. and {Spieler}, H.~G. and {Staniszewski}, Z. and {Stark}, A.~A. and {Tucker}, C. and {Vanderlinde}, K. and {Vieira}, J.~D. and {Williamson}, R.},
        title = "{The 10 Meter South Pole Telescope}",
      journal = {\pasp},
         year = 2011,
        month = may,
       volume = {123},
       number = {903},
        pages = {568},
          doi = {10.1086/659879},
archivePrefix = {arXiv},
       eprint = {0907.4445},
 primaryClass = {astro-ph.IM},
       adsurl = {https://ui.adsabs.harvard.edu/abs/2011PASP..123..568C}
}

@INPROCEEDINGS{benson14,
       author = {{Benson}, B.~A. and {Ade}, P.~A.~R. and {Ahmed}, Z. and {Allen}, S.~W. and {Arnold}, K. and {Austermann}, J.~E. and {Bender}, A.~N. and {Bleem}, L.~E. and {Carlstrom}, J.~E. and {Chang}, C.~L. and {Cho}, H.~M. and {Cliche}, J.~F. and {Crawford}, T.~M. and {Cukierman}, A. and {de Haan}, T. and {Dobbs}, M.~A. and {Dutcher}, D. and {Everett}, W. and {Gilbert}, A. and {Halverson}, N.~W. and {Hanson}, D. and {Harrington}, N.~L. and {Hattori}, K. and {Henning}, J.~W. and {Hilton}, G.~C. and {Holder}, G.~P. and {Holzapfel}, W.~L. and {Irwin}, K.~D. and {Keisler}, R. and {Knox}, L. and {Kubik}, D. and {Kuo}, C.~L. and {Lee}, A.~T. and {Leitch}, E.~M. and {Li}, D. and {McDonald}, M. and {Meyer}, S.~S. and {Montgomery}, J. and {Myers}, M. and {Natoli}, T. and {Nguyen}, H. and {Novosad}, V. and {Padin}, S. and {Pan}, Z. and {Pearson}, J. and {Reichardt}, C. and {Ruhl}, J.~E. and {Saliwanchik}, B.~R. and {Simard}, G. and {Smecher}, G. and {Sayre}, J.~T. and {Shirokoff}, E. and {Stark}, A.~A. and {Story}, K. and {Suzuki}, A. and {Thompson}, K.~L. and {Tucker}, C. and {Vanderlinde}, K. and {Vieira}, J.~D. and {Vikhlinin}, A. and {Wang}, G. and {Yefremenko}, V. and {Yoon}, K.~W.},
        title = "{SPT-3G: a next-generation cosmic microwave background polarization experiment on the South Pole telescope}",
    booktitle = {Millimeter, Submillimeter, and Far-Infrared Detectors and Instrumentation for Astronomy VII},
         year = 2014,
       editor = {{Holland}, Wayne S. and {Zmuidzinas}, Jonas},
       series = {Society of Photo-Optical Instrumentation Engineers (SPIE) Conference Series},
       volume = {9153},
        month = jul,
          eid = {91531P},
        pages = {91531P},
          doi = {10.1117/12.2057305},
archivePrefix = {arXiv},
       eprint = {1407.2973},
 primaryClass = {astro-ph.IM},
       adsurl = {https://ui.adsabs.harvard.edu/abs/2014SPIE.9153E..1PB}
}

@INPROCEEDINGS{bender18,
       author = {{Bender}, A.~N. and {Ade}, P.~A.~R. and {Ahmed}, Z. and {Anderson}, A.~J. and {Avva}, J.~S. and {Aylor}, K. and {Barry}, P.~S. and {Basu Thakur}, R. and {Benson}, B.~A. and {Bleem}, L.~S. and {Bocquet}, S. and {Byrum}, K. and {Carlstrom}, J.~E. and {Carter}, F.~W. and {Cecil}, T.~W. and {Chang}, C.~L. and {Cho}, H. -M. and {Cliche}, J.~F. and {Crawford}, T.~M. and {Cukierman}, A. and {de Haan}, T. and {Denison}, E.~V. and {Ding}, J. and {Dobbs}, M.~A. and {Dodelson}, S. and {Dutcher}, D. and {Everett}, W. and {Foster}, A. and {Gallicchio}, J. and {Gilbert}, A. and {Groh}, J.~C. and {Guns}, S.~T. and {Halverson}, N.~W. and {Harke-Hosemann}, A.~H. and {Harrington}, N.~L. and {Henning}, J.~W. and {Hilton}, G.~C. and {Holder}, G.~P. and {Holzapfel}, W.~L. and {Huang}, N. and {Irwin}, K.~D. and {Jeong}, O.~B. and {Jonas}, M. and {Jones}, A. and {Khaire}, T.~S. and {Knox}, L. and {Kofman}, A.~M. and {Korman}, M. and {Kubik}, D.~L. and {Kuhlmann}, S. and {Kuo}, C. -L. and {Lee}, A.~T. and {Leitch}, E.~M. and {Lowitz}, A.~E. and {Meyer}, S.~S. and {Michalik}, D. and {Montgomery}, J. and {Nadolski}, A. and {Natoli}, T. and {Ngyuen}, H. and {Noble}, G.~I. and {Novosad}, V. and {Padin}, S. and {Pan}, Z. and {Pearson}, J. and {Posada}, C.~M. and {Quan}, W. and {Raghunathan}, S. and {Rahlin}, A. and {Reichardt}, C.~L. and {Ruhl}, J.~E. and {Sayre}, J.~T. and {Shirokoff}, E. and {Smecher}, G. and {Sobrin}, J.~A. and {Stark}, A.~A. and {Story}, K.~T. and {Suzuki}, A. and {Thompson}, K.~L. and {Tucker}, C. and {Vale}, L.~R. and {Vanderlinde}, K. and {Vieira}, J.~D. and {Wang}, G. and {Whitehorn}, N. and {Wu}, W.~L.~K. and {Yefremenko}, V. and {Yoon}, K.~W. and {Young}, M.~R.},
        title = "{Year two instrument status of the SPT-3G cosmic microwave background receiver}",
    booktitle = {Millimeter, Submillimeter, and Far-Infrared Detectors and Instrumentation for Astronomy IX},
         year = 2018,
       editor = {{Zmuidzinas}, Jonas and {Gao}, Jian-Rong},
       series = {Society of Photo-Optical Instrumentation Engineers (SPIE) Conference Series},
       volume = {10708},
        month = jul,
          eid = {1070803},
        pages = {1070803},
          doi = {10.1117/12.2312426},
archivePrefix = {arXiv},
       eprint = {1809.00036},
 primaryClass = {astro-ph.IM},
       adsurl = {https://ui.adsabs.harvard.edu/abs/2018SPIE10708E..03B}
}

@ARTICLE{sobrin22,
       author = {{Sobrin}, J.~A. and {Anderson}, A.~J. and {Bender}, A.~N. and {Benson}, B.~A. and {Dutcher}, D. and {Foster}, A. and {Goeckner-Wald}, N. and {Montgomery}, J. and {Nadolski}, A. and {Rahlin}, A. and {Ade}, P.~A.~R. and {Ahmed}, Z. and {Anderes}, E. and {Archipley}, M. and {Austermann}, J.~E. and {Avva}, J.~S. and {Aylor}, K. and {Balkenhol}, L. and {Barry}, P.~S. and {Thakur}, R. Basu and {Benabed}, K. and {Bianchini}, F. and {Bleem}, L.~E. and {Bouchet}, F.~R. and {Bryant}, L. and {Byrum}, K. and {Carlstrom}, J.~E. and {Carter}, F.~W. and {Cecil}, T.~W. and {Chang}, C.~L. and {Chaubal}, P. and {Chen}, G. and {Cho}, H. -M. and {Chou}, T. -L. and {Cliche}, J. -F. and {Crawford}, T.~M. and {Cukierman}, A. and {Daley}, C. and {de Haan}, T. and {Denison}, E.~V. and {Dibert}, K. and {Ding}, J. and {Dobbs}, M.~A. and {Everett}, W. and {Feng}, C. and {Ferguson}, K.~R. and {Fu}, J. and {Galli}, S. and {Gambrel}, A.~E. and {Gardner}, R.~W. and {Gualtieri}, R. and {Guns}, S. and {Gupta}, N. and {Guyser}, R. and {Halverson}, N.~W. and {Harke-Hosemann}, A.~H. and {Harrington}, N.~L. and {Henning}, J.~W. and {Hilton}, G.~C. and {Hivon}, E. and {Holder}, G.~P. and {Holzapfel}, W.~L. and {Hood}, J.~C. and {Howe}, D. and {Huang}, N. and {Irwin}, K.~D. and {Jeong}, O.~B. and {Jonas}, M. and {Jones}, A. and {Khaire}, T.~S. and {Knox}, L. and {Kofman}, A.~M. and {Korman}, M. and {Kubik}, D.~L. and {Kuhlmann}, S. and {Kuo}, C. -L. and {Lee}, A.~T. and {Leitch}, E.~M. and {Lowitz}, A.~E. and {Lu}, C. and {Meyer}, S.~S. and {Michalik}, D. and {Millea}, M. and {Natoli}, T. and {Nguyen}, H. and {Noble}, G.~I. and {Novosad}, V. and {Omori}, Y. and {Padin}, S. and {Pan}, Z. and {Paschos}, P. and {Pearson}, J. and {Posada}, C.~M. and {Prabhu}, K. and {Quan}, W. and {Reichardt}, C.~L. and {Riebel}, D. and {Riedel}, B. and {Rouble}, M. and {Ruhl}, J.~E. and {Saliwanchik}, B. and {Sayre}, J.~T. and {Schiappucci}, E. and {Shirokoff}, E. and {Smecher}, G. and {Stark}, A.~A. and {Stephen}, J. and {Story}, K.~T. and {Suzuki}, A. and {Tandoi}, C. and {Thompson}, K.~L. and {Thorne}, B. and {Tucker}, C. and {Umilta}, C. and {Vale}, L.~R. and {Vanderlinde}, K. and {Vieira}, J.~D. and {Wang}, G. and {Whitehorn}, N. and {Wu}, W.~L.~K. and {Yefremenko}, V. and {Yoon}, K.~W. and {Young}, M.~R.},
        title = "{The Design and Integrated Performance of SPT-3G}",
      journal = {\apjs},
         year = 2022,
        month = feb,
       volume = {258},
       number = {2},
          eid = {42},
        pages = {42},
          doi = {10.3847/1538-4365/ac374f},
archivePrefix = {arXiv},
       eprint = {2106.11202},
 primaryClass = {astro-ph.IM},
       adsurl = {https://ui.adsabs.harvard.edu/abs/2022ApJS..258...42S}
}

@ARTICLE{lewis00,
       author = {{Lewis}, Antony and {Challinor}, Anthony and {Lasenby}, Anthony},
        title = "{Efficient Computation of Cosmic Microwave Background Anisotropies in Closed Friedmann-Robertson-Walker Models}",
      journal = {\apj},
         year = 2000,
        month = aug,
       volume = {538},
       number = {2},
        pages = {473-476},
          doi = {10.1086/309179},
archivePrefix = {arXiv},
       eprint = {astro-ph/9911177},
 primaryClass = {astro-ph},
       adsurl = {https://ui.adsabs.harvard.edu/abs/2000ApJ...538..473L}
}

@ARTICLE{lewis05,
       author = {{Lewis}, Antony},
        title = "{Lensed CMB simulation and parameter estimation}",
      journal = {\prd},
         year = 2005,
        month = apr,
       volume = {71},
       number = {8},
          eid = {083008},
        pages = {083008},
          doi = {10.1103/PhysRevD.71.083008},
archivePrefix = {arXiv},
       eprint = {astro-ph/0502469},
 primaryClass = {astro-ph},
       adsurl = {https://ui.adsabs.harvard.edu/abs/2005PhRvD..71h3008L}
}

@ARTICLE{planck18-6,
       author = {{Planck Collaboration} and {Aghanim}, N. and {Akrami}, Y. and {Ashdown}, M. and {Aumont}, J. and {Baccigalupi}, C. and {Ballardini}, M. and {Banday}, A.~J. and {Barreiro}, R.~B. and {Bartolo}, N. and {Basak}, S. and {Battye}, R. and {Benabed}, K. and {Bernard}, J. -P. and {Bersanelli}, M. and {Bielewicz}, P. and {Bock}, J.~J. and {Bond}, J.~R. and {Borrill}, J. and {Bouchet}, F.~R. and {Boulanger}, F. and {Bucher}, M. and {Burigana}, C. and {Butler}, R.~C. and {Calabrese}, E. and {Cardoso}, J. -F. and {Carron}, J. and {Challinor}, A. and {Chiang}, H.~C. and {Chluba}, J. and {Colombo}, L.~P.~L. and {Combet}, C. and {Contreras}, D. and {Crill}, B.~P. and {Cuttaia}, F. and {de Bernardis}, P. and {de Zotti}, G. and {Delabrouille}, J. and {Delouis}, J. -M. and {Di Valentino}, E. and {Diego}, J.~M. and {Dor{\'e}}, O. and {Douspis}, M. and {Ducout}, A. and {Dupac}, X. and {Dusini}, S. and {Efstathiou}, G. and {Elsner}, F. and {En{\ss}lin}, T.~A. and {Eriksen}, H.~K. and {Fantaye}, Y. and {Farhang}, M. and {Fergusson}, J. and {Fernandez-Cobos}, R. and {Finelli}, F. and {Forastieri}, F. and {Frailis}, M. and {Fraisse}, A.~A. and {Franceschi}, E. and {Frolov}, A. and {Galeotta}, S. and {Galli}, S. and {Ganga}, K. and {G{\'e}nova-Santos}, R.~T. and {Gerbino}, M. and {Ghosh}, T. and {Gonz{\'a}lez-Nuevo}, J. and {G{\'o}rski}, K.~M. and {Gratton}, S. and {Gruppuso}, A. and {Gudmundsson}, J.~E. and {Hamann}, J. and {Handley}, W. and {Hansen}, F.~K. and {Herranz}, D. and {Hildebrandt}, S.~R. and {Hivon}, E. and {Huang}, Z. and {Jaffe}, A.~H. and {Jones}, W.~C. and {Karakci}, A. and {Keih{\"a}nen}, E. and {Keskitalo}, R. and {Kiiveri}, K. and {Kim}, J. and {Kisner}, T.~S. and {Knox}, L. and {Krachmalnicoff}, N. and {Kunz}, M. and {Kurki-Suonio}, H. and {Lagache}, G. and {Lamarre}, J. -M. and {Lasenby}, A. and {Lattanzi}, M. and {Lawrence}, C.~R. and {Le Jeune}, M. and {Lemos}, P. and {Lesgourgues}, J. and {Levrier}, F. and {Lewis}, A. and {Liguori}, M. and {Lilje}, P.~B. and {Lilley}, M. and {Lindholm}, V. and {L{\'o}pez-Caniego}, M. and {Lubin}, P.~M. and {Ma}, Y. -Z. and {Mac{\'\i}as-P{\'e}rez}, J.~F. and {Maggio}, G. and {Maino}, D. and {Mandolesi}, N. and {Mangilli}, A. and {Marcos-Caballero}, A. and {Maris}, M. and {Martin}, P.~G. and {Martinelli}, M. and {Mart{\'\i}nez-Gonz{\'a}lez}, E. and {Matarrese}, S. and {Mauri}, N. and {McEwen}, J.~D. and {Meinhold}, P.~R. and {Melchiorri}, A. and {Mennella}, A. and {Migliaccio}, M. and {Millea}, M. and {Mitra}, S. and {Miville-Desch{\^e}nes}, M. -A. and {Molinari}, D. and {Montier}, L. and {Morgante}, G. and {Moss}, A. and {Natoli}, P. and {N{\o}rgaard-Nielsen}, H.~U. and {Pagano}, L. and {Paoletti}, D. and {Partridge}, B. and {Patanchon}, G. and {Peiris}, H.~V. and {Perrotta}, F. and {Pettorino}, V. and {Piacentini}, F. and {Polastri}, L. and {Polenta}, G. and {Puget}, J. -L. and {Rachen}, J.~P. and {Reinecke}, M. and {Remazeilles}, M. and {Renzi}, A. and {Rocha}, G. and {Rosset}, C. and {Roudier}, G. and {Rubi{\~n}o-Mart{\'\i}n}, J.~A. and {Ruiz-Granados}, B. and {Salvati}, L. and {Sandri}, M. and {Savelainen}, M. and {Scott}, D. and {Shellard}, E.~P.~S. and {Sirignano}, C. and {Sirri}, G. and {Spencer}, L.~D. and {Sunyaev}, R. and {Suur-Uski}, A. -S. and {Tauber}, J.~A. and {Tavagnacco}, D. and {Tenti}, M. and {Toffolatti}, L. and {Tomasi}, M. and {Trombetti}, T. and {Valenziano}, L. and {Valiviita}, J. and {Van Tent}, B. and {Vibert}, L. and {Vielva}, P. and {Villa}, F. and {Vittorio}, N. and {Wandelt}, B.~D. and {Wehus}, I.~K. and {White}, M. and {White}, S.~D.~M. and {Zacchei}, A. and {Zonca}, A.},
        title = "{Planck 2018 results. VI. Cosmological parameters}",
      journal = {\aap},
         year = 2020,
        month = sep,
       volume = {641},
          eid = {A6},
        pages = {A6},
          doi = {10.1051/0004-6361/201833910},
archivePrefix = {arXiv},
       eprint = {1807.06209},
 primaryClass = {astro-ph.CO},
       adsurl = {https://ui.adsabs.harvard.edu/abs/2020A&A...641A...6P}
}

@ARTICLE{planck18-5,
       author = {{Planck Collaboration} and {Aghanim}, N. and {Akrami}, Y. and {Ashdown}, M. and {Aumont}, J. and {Baccigalupi}, C. and {Ballardini}, M. and {Banday}, A.~J. and {Barreiro}, R.~B. and {Bartolo}, N. and {Basak}, S. and {Benabed}, K. and {Bernard}, J. -P. and {Bersanelli}, M. and {Bielewicz}, P. and {Bock}, J.~J. and {Bond}, J.~R. and {Borrill}, J. and {Bouchet}, F.~R. and {Boulanger}, F. and {Bucher}, M. and {Burigana}, C. and {Butler}, R.~C. and {Calabrese}, E. and {Cardoso}, J. -F. and {Carron}, J. and {Casaponsa}, B. and {Challinor}, A. and {Chiang}, H.~C. and {Colombo}, L.~P.~L. and {Combet}, C. and {Crill}, B.~P. and {Cuttaia}, F. and {de Bernardis}, P. and {de Rosa}, A. and {de Zotti}, G. and {Delabrouille}, J. and {Delouis}, J. -M. and {Di Valentino}, E. and {Diego}, J.~M. and {Dor{\'e}}, O. and {Douspis}, M. and {Ducout}, A. and {Dupac}, X. and {Dusini}, S. and {Efstathiou}, G. and {Elsner}, F. and {En{\ss}lin}, T.~A. and {Eriksen}, H.~K. and {Fantaye}, Y. and {Fernandez-Cobos}, R. and {Finelli}, F. and {Frailis}, M. and {Fraisse}, A.~A. and {Franceschi}, E. and {Frolov}, A. and {Galeotta}, S. and {Galli}, S. and {Ganga}, K. and {G{\'e}nova-Santos}, R.~T. and {Gerbino}, M. and {Ghosh}, T. and {Giraud-H{\'e}raud}, Y. and {Gonz{\'a}lez-Nuevo}, J. and {G{\'o}rski}, K.~M. and {Gratton}, S. and {Gruppuso}, A. and {Gudmundsson}, J.~E. and {Hamann}, J. and {Handley}, W. and {Hansen}, F.~K. and {Herranz}, D. and {Hivon}, E. and {Huang}, Z. and {Jaffe}, A.~H. and {Jones}, W.~C. and {Keih{\"a}nen}, E. and {Keskitalo}, R. and {Kiiveri}, K. and {Kim}, J. and {Kisner}, T.~S. and {Krachmalnicoff}, N. and {Kunz}, M. and {Kurki-Suonio}, H. and {Lagache}, G. and {Lamarre}, J. -M. and {Lasenby}, A. and {Lattanzi}, M. and {Lawrence}, C.~R. and {Le Jeune}, M. and {Levrier}, F. and {Lewis}, A. and {Liguori}, M. and {Lilje}, P.~B. and {Lilley}, M. and {Lindholm}, V. and {L{\'o}pez-Caniego}, M. and {Lubin}, P.~M. and {Ma}, Y. -Z. and {Mac{\'\i}as-P{\'e}rez}, J.~F. and {Maggio}, G. and {Maino}, D. and {Mandolesi}, N. and {Mangilli}, A. and {Marcos-Caballero}, A. and {Maris}, M. and {Martin}, P.~G. and {Mart{\'\i}nez-Gonz{\'a}lez}, E. and {Matarrese}, S. and {Mauri}, N. and {McEwen}, J.~D. and {Meinhold}, P.~R. and {Melchiorri}, A. and {Mennella}, A. and {Migliaccio}, M. and {Millea}, M. and {Miville-Desch{\^e}nes}, M. -A. and {Molinari}, D. and {Moneti}, A. and {Montier}, L. and {Morgante}, G. and {Moss}, A. and {Natoli}, P. and {N{\o}rgaard-Nielsen}, H.~U. and {Pagano}, L. and {Paoletti}, D. and {Partridge}, B. and {Patanchon}, G. and {Peiris}, H.~V. and {Perrotta}, F. and {Pettorino}, V. and {Piacentini}, F. and {Polenta}, G. and {Puget}, J. -L. and {Rachen}, J.~P. and {Reinecke}, M. and {Remazeilles}, M. and {Renzi}, A. and {Rocha}, G. and {Rosset}, C. and {Roudier}, G. and {Rubi{\~n}o-Mart{\'\i}n}, J.~A. and {Ruiz-Granados}, B. and {Salvati}, L. and {Sandri}, M. and {Savelainen}, M. and {Scott}, D. and {Shellard}, E.~P.~S. and {Sirignano}, C. and {Sirri}, G. and {Spencer}, L.~D. and {Sunyaev}, R. and {Suur-Uski}, A. -S. and {Tauber}, J.~A. and {Tavagnacco}, D. and {Tenti}, M. and {Toffolatti}, L. and {Tomasi}, M. and {Trombetti}, T. and {Valiviita}, J. and {Van Tent}, B. and {Vielva}, P. and {Villa}, F. and {Vittorio}, N. and {Wandelt}, B.~D. and {Wehus}, I.~K. and {Zacchei}, A. and {Zonca}, A.},
        title = "{Planck 2018 results. V. CMB power spectra and likelihoods}",
      journal = {\aap},
         year = 2020,
        month = sep,
       volume = {641},
          eid = {A5},
        pages = {A5},
          doi = {10.1051/0004-6361/201936386},
archivePrefix = {arXiv},
       eprint = {1907.12875},
 primaryClass = {astro-ph.CO},
       adsurl = {https://ui.adsabs.harvard.edu/abs/2020A&A...641A...5P}
}

@ARTICLE{kaiser92,
       author = {{Kaiser}, Nick},
        title = "{Weak Gravitational Lensing of Distant Galaxies}",
      journal = {\apj},
         year = 1992,
        month = apr,
       volume = {388},
        pages = {272},
          doi = {10.1086/171151},
       adsurl = {https://ui.adsabs.harvard.edu/abs/1992ApJ...388..272K}
}

@ARTICLE{bernardeau97,
       author = {{Bernardeau}, F. and {van Waerbeke}, L. and {Mellier}, Y.},
        title = "{Weak lensing statistics as a probe of \{OMEGA\} and power spectrum.}",
      journal = {\aap},
         year = 1997,
        month = jun,
       volume = {322},
        pages = {1-18},
          doi = {10.48550/arXiv.astro-ph/9609122},
archivePrefix = {arXiv},
       eprint = {astro-ph/9609122},
 primaryClass = {astro-ph},
       adsurl = {https://ui.adsabs.harvard.edu/abs/1997A&A...322....1B}
}

@ARTICLE{kilbinger15,
       author = {{Kilbinger}, Martin},
        title = "{Cosmology with cosmic shear observations: a review}",
      journal = {Reports on Progress in Physics},
         year = 2015,
        month = jul,
       volume = {78},
       number = {8},
          eid = {086901},
        pages = {086901},
          doi = {10.1088/0034-4885/78/8/086901},
archivePrefix = {arXiv},
       eprint = {1411.0115},
 primaryClass = {astro-ph.CO},
       adsurl = {https://ui.adsabs.harvard.edu/abs/2015RPPh...78h6901K}
}

@ARTICLE{das11,
       author = {{Das}, Sudeep and {Sherwin}, Blake D. and {Aguirre}, Paula and {Appel}, John W. and {Bond}, J. Richard and {Carvalho}, C. Sofia and {Devlin}, Mark J. and {Dunkley}, Joanna and {D{\"u}nner}, Rolando and {Essinger-Hileman}, Thomas and {Fowler}, Joseph W. and {Hajian}, Amir and {Halpern}, Mark and {Hasselfield}, Matthew and {Hincks}, Adam D. and {Hlozek}, Ren{\'e}e and {Huffenberger}, Kevin M. and {Hughes}, John P. and {Irwin}, Kent D. and {Klein}, Jeff and {Kosowsky}, Arthur and {Lupton}, Robert H. and {Marriage}, Tobias A. and {Marsden}, Danica and {Menanteau}, Felipe and {Moodley}, Kavilan and {Niemack}, Michael D. and {Nolta}, Michael R. and {Page}, Lyman A. and {Parker}, Lucas and {Reese}, Erik D. and {Schmitt}, Benjamin L. and {Sehgal}, Neelima and {Sievers}, Jon and {Spergel}, David N. and {Staggs}, Suzanne T. and {Swetz}, Daniel S. and {Switzer}, Eric R. and {Thornton}, Robert and {Visnjic}, Katerina and {Wollack}, Ed},
        title = "{Detection of the Power Spectrum of Cosmic Microwave Background Lensing by the Atacama Cosmology Telescope}",
      journal = {\prl},
         year = 2011,
        month = jul,
       volume = {107},
       number = {2},
          eid = {021301},
        pages = {021301},
          doi = {10.1103/PhysRevLett.107.021301},
archivePrefix = {arXiv},
       eprint = {1103.2124},
 primaryClass = {astro-ph.CO},
       adsurl = {https://ui.adsabs.harvard.edu/abs/2011PhRvL.107b1301D}
}

@ARTICLE{das14,
       author = {{Das}, Sudeep and {Louis}, Thibaut and {Nolta}, Michael R. and {Addison}, Graeme E. and {Battistelli}, Elia S. and {Bond}, J. Richard and {Calabrese}, Erminia and {Crichton}, Devin and {Devlin}, Mark J. and {Dicker}, Simon and {Dunkley}, Joanna and {D{\"u}nner}, Rolando and {Fowler}, Joseph W. and {Gralla}, Megan and {Hajian}, Amir and {Halpern}, Mark and {Hasselfield}, Matthew and {Hilton}, Matt and {Hincks}, Adam D. and {Hlozek}, Ren{\'e}e and {Huffenberger}, Kevin M. and {Hughes}, John P. and {Irwin}, Kent D. and {Kosowsky}, Arthur and {Lupton}, Robert H. and {Marriage}, Tobias A. and {Marsden}, Danica and {Menanteau}, Felipe and {Moodley}, Kavilan and {Niemack}, Michael D. and {Page}, Lyman A. and {Partridge}, Bruce and {Reese}, Erik D. and {Schmitt}, Benjamin L. and {Sehgal}, Neelima and {Sherwin}, Blake D. and {Sievers}, Jonathan L. and {Spergel}, David N. and {Staggs}, Suzanne T. and {Swetz}, Daniel S. and {Switzer}, Eric R. and {Thornton}, Robert and {Trac}, Hy and {Wollack}, Ed},
        title = "{The Atacama Cosmology Telescope: temperature and gravitational lensing power spectrum measurements from three seasons of data}",
      journal = {\jcap},
         year = 2014,
        month = apr,
       volume = {2014},
       number = {4},
          eid = {014},
        pages = {014},
          doi = {10.1088/1475-7516/2014/04/014},
archivePrefix = {arXiv},
       eprint = {1301.1037},
 primaryClass = {astro-ph.CO},
       adsurl = {https://ui.adsabs.harvard.edu/abs/2014JCAP...04..014D}
}

@ARTICLE{sherwin17,
       author = {{Sherwin}, Blake D. and {van Engelen}, Alexander and {Sehgal}, Neelima and {Madhavacheril}, Mathew and {Addison}, Graeme E. and {Aiola}, Simone and {Allison}, Rupert and {Battaglia}, Nicholas and {Becker}, Daniel T. and {Beall}, James A. and {Bond}, J. Richard and {Calabrese}, Erminia and {Datta}, Rahul and {Devlin}, Mark J. and {D{\"u}nner}, Rolando and {Dunkley}, Joanna and {Fox}, Anna E. and {Gallardo}, Patricio and {Halpern}, Mark and {Hasselfield}, Matthew and {Henderson}, Shawn and {Hill}, J. Colin and {Hilton}, Gene C. and {Hubmayr}, Johannes and {Hughes}, John P. and {Hincks}, Adam D. and {Hlozek}, Ren{\'e}e and {Huffenberger}, Kevin M. and {Koopman}, Brian and {Kosowsky}, Arthur and {Louis}, Thibaut and {Maurin}, Lo{\"\i}c and {McMahon}, Jeff and {Moodley}, Kavilan and {Naess}, Sigurd and {Nati}, Federico and {Newburgh}, Laura and {Niemack}, Michael D. and {Page}, Lyman A. and {Sievers}, Jonathan and {Spergel}, David N. and {Staggs}, Suzanne T. and {Thornton}, Robert J. and {Van Lanen}, Jeff and {Vavagiakis}, Eve and {Wollack}, Edward J.},
        title = "{Two-season Atacama Cosmology Telescope polarimeter lensing power spectrum}",
      journal = {\prd},
         year = 2017,
        month = jun,
       volume = {95},
       number = {12},
          eid = {123529},
        pages = {123529},
          doi = {10.1103/PhysRevD.95.123529},
archivePrefix = {arXiv},
       eprint = {1611.09753},
 primaryClass = {astro-ph.CO},
       adsurl = {https://ui.adsabs.harvard.edu/abs/2017PhRvD..95l3529S}
}

@ARTICLE{darwish21,
       author = {{Darwish}, Omar and {Madhavacheril}, Mathew S. and {Sherwin}, Blake D. and {Aiola}, Simone and {Battaglia}, Nicholas and {Beall}, James A. and {Becker}, Daniel T. and {Bond}, J. Richard and {Calabrese}, Erminia and {Choi}, Steve K. and {Devlin}, Mark J. and {Dunkley}, Jo and {D{\"u}nner}, Rolando and {Ferraro}, Simone and {Fox}, Anna E. and {Gallardo}, Patricio A. and {Guan}, Yilun and {Halpern}, Mark and {Han}, Dongwon and {Hasselfield}, Matthew and {Hill}, J. Colin and {Hilton}, Gene C. and {Hilton}, Matt and {Hincks}, Adam D. and {Patty Ho}, Shuay-Pwu and {Hubmayr}, J. and {Hughes}, John P. and {Koopman}, Brian J. and {Kosowsky}, Arthur and {Van Lanen}, J. and {Louis}, Thibaut and {Lungu}, Marius and {MacInnis}, Amanda and {Maurin}, Lo{\"\i}c and {McMahon}, Jeffrey and {Moodley}, Kavilan and {Naess}, Sigurd and {Namikawa}, Toshiya and {Nati}, Federico and {Newburgh}, Laura and {Nibarger}, John P. and {Niemack}, Michael D. and {Page}, Lyman A. and {Partridge}, Bruce and {Qu}, Frank J. and {Robertson}, Naomi and {Schillaci}, Alessandro and {Schmitt}, Benjamin and {Sehgal}, Neelima and {Sif{\'o}n}, Crist{\'o}bal and {Spergel}, David N. and {Staggs}, Suzanne and {Storer}, Emilie and {van Engelen}, Alexander and {Wollack}, Edward J.},
        title = "{The Atacama Cosmology Telescope: a CMB lensing mass map over 2100 square degrees of sky and its cross-correlation with BOSS-CMASS galaxies}",
      journal = {\mnras},
         year = 2021,
        month = jan,
       volume = {500},
       number = {2},
        pages = {2250-2263},
          doi = {10.1093/mnras/staa3438},
archivePrefix = {arXiv},
       eprint = {2004.01139},
 primaryClass = {astro-ph.CO},
       adsurl = {https://ui.adsabs.harvard.edu/abs/2021MNRAS.500.2250D}
}

@ARTICLE{qu24,
       author = {{Qu}, Frank J. and {Sherwin}, Blake D. and {Madhavacheril}, Mathew S. and {Han}, Dongwon and {Crowley}, Kevin T. and {Abril-Cabezas}, Irene and {Ade}, Peter A.~R. and {Aiola}, Simone and {Alford}, Tommy and {Amiri}, Mandana and {Amodeo}, Stefania and {An}, Rui and {Atkins}, Zachary and {Austermann}, Jason E. and {Battaglia}, Nicholas and {Battistelli}, Elia Stefano and {Beall}, James A. and {Bean}, Rachel and {Beringue}, Benjamin and {Bhandarkar}, Tanay and {Biermann}, Emily and {Bolliet}, Boris and {Bond}, J. Richard and {Cai}, Hongbo and {Calabrese}, Erminia and {Calafut}, Victoria and {Capalbo}, Valentina and {Carrero}, Felipe and {Carron}, Julien and {Challinor}, Anthony and {Chesmore}, Grace E. and {Cho}, Hsiao-mei and {Choi}, Steve K. and {Clark}, Susan E. and {Rosado}, Rodrigo C{\'o}rdova and {Cothard}, Nicholas F. and {Coughlin}, Kevin and {Coulton}, William and {Dalal}, Roohi and {Darwish}, Omar and {Devlin}, Mark J. and {Dicker}, Simon and {Doze}, Peter and {Duell}, Cody J. and {Duff}, Shannon M. and {Duivenvoorden}, Adriaan J. and {Dunkley}, Jo and {D{\"u}nner}, Rolando and {Fanfani}, Valentina and {Fankhanel}, Max and {Farren}, Gerrit and {Ferraro}, Simone and {Freundt}, Rodrigo and {Fuzia}, Brittany and {Gallardo}, Patricio A. and {Garrido}, Xavier and {Gluscevic}, Vera and {Golec}, Joseph E. and {Guan}, Yilun and {Halpern}, Mark and {Harrison}, Ian and {Hasselfield}, Matthew and {Healy}, Erin and {Henderson}, Shawn and {Hensley}, Brandon and {Herv{\'\i}as-Caimapo}, Carlos and {Hill}, J. Colin and {Hilton}, Gene C. and {Hilton}, Matt and {Hincks}, Adam D. and {Hlo{\v{z}}ek}, Ren{\'e}e and {Ho}, Shuay-Pwu Patty and {Huber}, Zachary B. and {Hubmayr}, Johannes and {Huffenberger}, Kevin M. and {Hughes}, John P. and {Irwin}, Kent and {Isopi}, Giovanni and {Jense}, Hidde T. and {Keller}, Ben and {Kim}, Joshua and {Knowles}, Kenda and {Koopman}, Brian J. and {Kosowsky}, Arthur and {Kramer}, Darby and {Kusiak}, Aleksandra and {La Posta}, Adrien and {Lague}, Alex and {Lakey}, Victoria and {Lee}, Eunseong and {Li}, Zack and {Li}, Yaqiong and {Limon}, Michele and {Lokken}, Martine and {Louis}, Thibaut and {Lungu}, Marius and {MacCrann}, Niall and {MacInnis}, Amanda and {Maldonado}, Diego and {Maldonado}, Felipe and {Mallaby-Kay}, Maya and {Marques}, Gabriela A. and {McMahon}, Jeff and {Mehta}, Yogesh and {Menanteau}, Felipe and {Moodley}, Kavilan and {Morris}, Thomas W. and {Mroczkowski}, Tony and {Naess}, Sigurd and {Namikawa}, Toshiya and {Nati}, Federico and {Newburgh}, Laura and {Nicola}, Andrina and {Niemack}, Michael D. and {Nolta}, Michael R. and {Orlowski-Scherer}, John and {Page}, Lyman A. and {Pandey}, Shivam and {Partridge}, Bruce and {Prince}, Heather and {Puddu}, Roberto and {Radiconi}, Federico and {Robertson}, Naomi and {Rojas}, Felipe and {Sakuma}, Tai and {Salatino}, Maria and {Schaan}, Emmanuel and {Schmitt}, Benjamin L. and {Sehgal}, Neelima and {Shaikh}, Shabbir and {Sierra}, Carlos and {Sievers}, Jon and {Sif{\'o}n}, Crist{\'o}bal and {Simon}, Sara and {Sonka}, Rita and {Spergel}, David N. and {Staggs}, Suzanne T. and {Storer}, Emilie and {Switzer}, Eric R. and {Tampier}, Niklas and {Thornton}, Robert and {Trac}, Hy and {Treu}, Jesse and {Tucker}, Carole and {Ullom}, Joel and {Vale}, Leila R. and {Van Engelen}, Alexander and {Van Lanen}, Jeff and {van Marrewijk}, Joshiwa and {Vargas}, Cristian and {Vavagiakis}, Eve M. and {Wagoner}, Kasey and {Wang}, Yuhan and {Wenzl}, Lukas and {Wollack}, Edward J. and {Xu}, Zhilei and {Zago}, Fernando and {Zheng}, Kaiwen},
        title = "{The Atacama Cosmology Telescope: A Measurement of the DR6 CMB Lensing Power Spectrum and Its Implications for Structure Growth}",
      journal = {\apj},
         year = 2024,
        month = feb,
       volume = {962},
       number = {2},
          eid = {112},
        pages = {112},
          doi = {10.3847/1538-4357/acfe06},
archivePrefix = {arXiv},
       eprint = {2304.05202},
 primaryClass = {astro-ph.CO},
       adsurl = {https://ui.adsabs.harvard.edu/abs/2024ApJ...962..112Q}
}

@ARTICLE{planck14-17,
       author = {{Planck Collaboration} and {Ade}, P.~A.~R. and {Aghanim}, N. and {Armitage-Caplan}, C. and {Arnaud}, M. and {Ashdown}, M. and {Atrio-Barandela}, F. and {Aumont}, J. and {Baccigalupi}, C. and {Banday}, A.~J. and {Barreiro}, R.~B. and {Bartlett}, J.~G. and {Basak}, S. and {Battaner}, E. and {Benabed}, K. and {Beno{\^\i}t}, A. and {Benoit-L{\'e}vy}, A. and {Bernard}, J. -P. and {Bersanelli}, M. and {Bielewicz}, P. and {Bobin}, J. and {Bock}, J.~J. and {Bonaldi}, A. and {Bonavera}, L. and {Bond}, J.~R. and {Borrill}, J. and {Bouchet}, F.~R. and {Bridges}, M. and {Bucher}, M. and {Burigana}, C. and {Butler}, R.~C. and {Cardoso}, J. -F. and {Catalano}, A. and {Challinor}, A. and {Chamballu}, A. and {Chiang}, H.~C. and {Chiang}, L. -Y. and {Christensen}, P.~R. and {Church}, S. and {Clements}, D.~L. and {Colombi}, S. and {Colombo}, L.~P.~L. and {Couchot}, F. and {Coulais}, A. and {Crill}, B.~P. and {Curto}, A. and {Cuttaia}, F. and {Danese}, L. and {Davies}, R.~D. and {Davis}, R.~J. and {de Bernardis}, P. and {de Rosa}, A. and {de Zotti}, G. and {D{\'e}chelette}, T. and {Delabrouille}, J. and {Delouis}, J. -M. and {D{\'e}sert}, F. -X. and {Dickinson}, C. and {Diego}, J.~M. and {Dole}, H. and {Donzelli}, S. and {Dor{\'e}}, O. and {Douspis}, M. and {Dunkley}, J. and {Dupac}, X. and {Efstathiou}, G. and {En{\ss}lin}, T.~A. and {Eriksen}, H.~K. and {Finelli}, F. and {Forni}, O. and {Frailis}, M. and {Franceschi}, E. and {Galeotta}, S. and {Ganga}, K. and {Giard}, M. and {Giardino}, G. and {Giraud-H{\'e}raud}, Y. and {Gonz{\'a}lez-Nuevo}, J. and {G{\'o}rski}, K.~M. and {Gratton}, S. and {Gregorio}, A. and {Gruppuso}, A. and {Gudmundsson}, J.~E. and {Hansen}, F.~K. and {Hanson}, D. and {Harrison}, D. and {Henrot-Versill{\'e}}, S. and {Hern{\'a}ndez-Monteagudo}, C. and {Herranz}, D. and {Hildebrandt}, S.~R. and {Hivon}, E. and {Ho}, S. and {Hobson}, M. and {Holmes}, W.~A. and {Hornstrup}, A. and {Hovest}, W. and {Huffenberger}, K.~M. and {Jaffe}, A.~H. and {Jaffe}, T.~R. and {Jones}, W.~C. and {Juvela}, M. and {Keih{\"a}nen}, E. and {Keskitalo}, R. and {Kisner}, T.~S. and {Kneissl}, R. and {Knoche}, J. and {Knox}, L. and {Kunz}, M. and {Kurki-Suonio}, H. and {Lagache}, G. and {L{\"a}hteenm{\"a}ki}, A. and {Lamarre}, J. -M. and {Lasenby}, A. and {Laureijs}, R.~J. and {Lavabre}, A. and {Lawrence}, C.~R. and {Leahy}, J.~P. and {Leonardi}, R. and {Le{\'o}n-Tavares}, J. and {Lesgourgues}, J. and {Lewis}, A. and {Liguori}, M. and {Lilje}, P.~B. and {Linden-V{\o}rnle}, M. and {L{\'o}pez-Caniego}, M. and {Lubin}, P.~M. and {Mac{\'\i}as-P{\'e}rez}, J.~F. and {Maffei}, B. and {Maino}, D. and {Mandolesi}, N. and {Mangilli}, A. and {Maris}, M. and {Marshall}, D.~J. and {Martin}, P.~G. and {Mart{\'\i}nez-Gonz{\'a}lez}, E. and {Masi}, S. and {Massardi}, M. and {Matarrese}, S. and {Matthai}, F. and {Mazzotta}, P. and {Melchiorri}, A. and {Mendes}, L. and {Mennella}, A. and {Migliaccio}, M. and {Mitra}, S. and {Miville-Desch{\^e}nes}, M. -A. and {Moneti}, A. and {Montier}, L. and {Morgante}, G. and {Mortlock}, D. and {Moss}, A. and {Munshi}, D. and {Murphy}, J.~A. and {Naselsky}, P. and {Nati}, F. and {Natoli}, P. and {Netterfield}, C.~B. and {N{\o}rgaard-Nielsen}, H.~U. and {Noviello}, F. and {Novikov}, D. and {Novikov}, I. and {Osborne}, S. and {Oxborrow}, C.~A. and {Paci}, F. and {Pagano}, L. and {Pajot}, F. and {Paoletti}, D. and {Partridge}, B. and {Pasian}, F. and {Patanchon}, G. and {Perdereau}, O. and {Perotto}, L. and {Perrotta}, F. and {Piacentini}, F. and {Piat}, M. and {Pierpaoli}, E. and {Pietrobon}, D. and {Plaszczynski}, S. and {Pointecouteau}, E. and {Polenta}, G. and {Ponthieu}, N. and {Popa}, L. and {Poutanen}, T. and {Pratt}, G.~W. and {Pr{\'e}zeau}, G. and {Prunet}, S. and {Puget}, J. -L. and {Pullen}, A.~R. and {Rachen}, J.~P. and {Rebolo}, R. and {Reinecke}, M. and {Remazeilles}, M. and {Renault}, C. and {Ricciardi}, S. and {Riller}, T. and {Ristorcelli}, I. and {Rocha}, G. and {Rosset}, C. and {Roudier}, G. and {Rowan-Robinson}, M. and {Rubi{\~n}o-Mart{\'\i}n}, J.~A. and {Rusholme}, B. and {Sandri}, M. and {Santos}, D. and {Savini}, G. and {Scott}, D. and {Seiffert}, M.~D. and {Shellard}, E.~P.~S. and {Smith}, K. and {Spencer}, L.~D. and {Starck}, J. -L. and {Stolyarov}, V. and {Stompor}, R. and {Sudiwala}, R. and {Sunyaev}, R. and {Sureau}, F. and {Sutton}, D. and {Suur-Uski}, A. -S. and {Sygnet}, J. -F. and {Tauber}, J.~A. and {Tavagnacco}, D. and {Terenzi}, L. and {Toffolatti}, L. and {Tomasi}, M. and {Tristram}, M. and {Tucci}, M. and {Tuovinen}, J. and {Umana}, G. and {Valenziano}, L. and {Valiviita}, J. and {Van Tent}, B. and {Vielva}, P. and {Villa}, F. and {Vittorio}, N. and {Wade}, L.~A. and {Wandelt}, B.~D. and {White}, M. and {White}, S.~D.~M. and {Yvon}, D. and {Zacchei}, A. and {Zonca}, A.},
        title = "{Planck 2013 results. XVII. Gravitational lensing by large-scale structure}",
      journal = {\aap},
         year = 2014,
        month = nov,
       volume = {571},
          eid = {A17},
        pages = {A17},
          doi = {10.1051/0004-6361/201321543},
archivePrefix = {arXiv},
       eprint = {1303.5077},
 primaryClass = {astro-ph.CO},
       adsurl = {https://ui.adsabs.harvard.edu/abs/2014A&A...571A..17P}
}

@ARTICLE{planck16-15,
       author = {{Planck Collaboration} and {Ade}, P.~A.~R. and {Aghanim}, N. and {Arnaud}, M. and {Ashdown}, M. and {Aumont}, J. and {Baccigalupi}, C. and {Banday}, A.~J. and {Barreiro}, R.~B. and {Bartlett}, J.~G. and {Bartolo}, N. and {Basak}, S. and {Battaner}, E. and {Benabed}, K. and {Beno{\^\i}t}, A. and {Benoit-L{\'e}vy}, A. and {Bernard}, J. -P. and {Bersanelli}, M. and {Bielewicz}, P. and {Bock}, J.~J. and {Bonaldi}, A. and {Bonavera}, L. and {Bond}, J.~R. and {Borrill}, J. and {Bouchet}, F.~R. and {Boulanger}, F. and {Bucher}, M. and {Burigana}, C. and {Butler}, R.~C. and {Calabrese}, E. and {Cardoso}, J. -F. and {Catalano}, A. and {Challinor}, A. and {Chamballu}, A. and {Chiang}, H.~C. and {Christensen}, P.~R. and {Church}, S. and {Clements}, D.~L. and {Colombi}, S. and {Colombo}, L.~P.~L. and {Combet}, C. and {Couchot}, F. and {Coulais}, A. and {Crill}, B.~P. and {Curto}, A. and {Cuttaia}, F. and {Danese}, L. and {Davies}, R.~D. and {Davis}, R.~J. and {de Bernardis}, P. and {de Rosa}, A. and {de Zotti}, G. and {Delabrouille}, J. and {D{\'e}sert}, F. -X. and {Diego}, J.~M. and {Dole}, H. and {Donzelli}, S. and {Dor{\'e}}, O. and {Douspis}, M. and {Ducout}, A. and {Dunkley}, J. and {Dupac}, X. and {Efstathiou}, G. and {Elsner}, F. and {En{\ss}lin}, T.~A. and {Eriksen}, H.~K. and {Fergusson}, J. and {Finelli}, F. and {Forni}, O. and {Frailis}, M. and {Fraisse}, A.~A. and {Franceschi}, E. and {Frejsel}, A. and {Galeotta}, S. and {Galli}, S. and {Ganga}, K. and {Giard}, M. and {Giraud-H{\'e}raud}, Y. and {Gjerl{\o}w}, E. and {Gonz{\'a}lez-Nuevo}, J. and {G{\'o}rski}, K.~M. and {Gratton}, S. and {Gregorio}, A. and {Gruppuso}, A. and {Gudmundsson}, J.~E. and {Hansen}, F.~K. and {Hanson}, D. and {Harrison}, D.~L. and {Henrot-Versill{\'e}}, S. and {Hern{\'a}ndez-Monteagudo}, C. and {Herranz}, D. and {Hildebrandt}, S.~R. and {Hivon}, E. and {Hobson}, M. and {Holmes}, W.~A. and {Hornstrup}, A. and {Hovest}, W. and {Huffenberger}, K.~M. and {Hurier}, G. and {Jaffe}, A.~H. and {Jaffe}, T.~R. and {Jones}, W.~C. and {Juvela}, M. and {Keih{\"a}nen}, E. and {Keskitalo}, R. and {Kisner}, T.~S. and {Kneissl}, R. and {Knoche}, J. and {Kunz}, M. and {Kurki-Suonio}, H. and {Lagache}, G. and {L{\"a}hteenm{\"a}ki}, A. and {Lamarre}, J. -M. and {Lasenby}, A. and {Lattanzi}, M. and {Lawrence}, C.~R. and {Leonardi}, R. and {Lesgourgues}, J. and {Levrier}, F. and {Lewis}, A. and {Liguori}, M. and {Lilje}, P.~B. and {Linden-V{\o}rnle}, M. and {L{\'o}pez-Caniego}, M. and {Lubin}, P.~M. and {Mac{\'\i}as-P{\'e}rez}, J.~F. and {Maggio}, G. and {Maino}, D. and {Mandolesi}, N. and {Mangilli}, A. and {Maris}, M. and {Martin}, P.~G. and {Mart{\'\i}nez-Gonz{\'a}lez}, E. and {Masi}, S. and {Matarrese}, S. and {McGehee}, P. and {Meinhold}, P.~R. and {Melchiorri}, A. and {Mendes}, L. and {Mennella}, A. and {Migliaccio}, M. and {Mitra}, S. and {Miville-Desch{\^e}nes}, M. -A. and {Moneti}, A. and {Montier}, L. and {Morgante}, G. and {Mortlock}, D. and {Moss}, A. and {Munshi}, D. and {Murphy}, J.~A. and {Naselsky}, P. and {Nati}, F. and {Natoli}, P. and {Netterfield}, C.~B. and {N{\o}rgaard-Nielsen}, H.~U. and {Noviello}, F. and {Novikov}, D. and {Novikov}, I. and {Oxborrow}, C.~A. and {Paci}, F. and {Pagano}, L. and {Pajot}, F. and {Paoletti}, D. and {Pasian}, F. and {Patanchon}, G. and {Perdereau}, O. and {Perotto}, L. and {Perrotta}, F. and {Pettorino}, V. and {Piacentini}, F. and {Piat}, M. and {Pierpaoli}, E. and {Pietrobon}, D. and {Plaszczynski}, S. and {Pointecouteau}, E. and {Polenta}, G. and {Popa}, L. and {Pratt}, G.~W. and {Pr{\'e}zeau}, G. and {Prunet}, S. and {Puget}, J. -L. and {Rachen}, J.~P. and {Reach}, W.~T. and {Rebolo}, R. and {Reinecke}, M. and {Remazeilles}, M. and {Renault}, C. and {Renzi}, A. and {Ristorcelli}, I. and {Rocha}, G. and {Rosset}, C. and {Rossetti}, M. and {Roudier}, G. and {Rowan-Robinson}, M. and {Rubi{\~n}o-Mart{\'\i}n}, J.~A. and {Rusholme}, B. and {Sandri}, M. and {Santos}, D. and {Savelainen}, M. and {Savini}, G. and {Scott}, D. and {Seiffert}, M.~D. and {Shellard}, E.~P.~S. and {Spencer}, L.~D. and {Stolyarov}, V. and {Stompor}, R. and {Sudiwala}, R. and {Sunyaev}, R. and {Sutton}, D. and {Suur-Uski}, A. -S. and {Sygnet}, J. -F. and {Tauber}, J.~A. and {Terenzi}, L. and {Toffolatti}, L. and {Tomasi}, M. and {Tristram}, M. and {Tucci}, M. and {Tuovinen}, J. and {Valenziano}, L. and {Valiviita}, J. and {Van Tent}, B. and {Vielva}, P. and {Villa}, F. and {Wade}, L.~A. and {Wandelt}, B.~D. and {Wehus}, I.~K. and {White}, M. and {Yvon}, D. and {Zacchei}, A. and {Zonca}, A.},
        title = "{Planck 2015 results. XV. Gravitational lensing}",
      journal = {\aap},
         year = 2016,
        month = sep,
       volume = {594},
          eid = {A15},
        pages = {A15},
          doi = {10.1051/0004-6361/201525941},
archivePrefix = {arXiv},
       eprint = {1502.01591},
 primaryClass = {astro-ph.CO},
       adsurl = {https://ui.adsabs.harvard.edu/abs/2016A&A...594A..15P}
}

@ARTICLE{planck20-8,
       author = {{Planck Collaboration} and {Aghanim}, N. and {Akrami}, Y. and {Ashdown}, M. and {Aumont}, J. and {Baccigalupi}, C. and {Ballardini}, M. and {Banday}, A.~J. and {Barreiro}, R.~B. and {Bartolo}, N. and {Basak}, S. and {Benabed}, K. and {Bernard}, J. -P. and {Bersanelli}, M. and {Bielewicz}, P. and {Bock}, J.~J. and {Bond}, J.~R. and {Borrill}, J. and {Bouchet}, F.~R. and {Boulanger}, F. and {Bucher}, M. and {Burigana}, C. and {Calabrese}, E. and {Cardoso}, J. -F. and {Carron}, J. and {Challinor}, A. and {Chiang}, H.~C. and {Colombo}, L.~P.~L. and {Combet}, C. and {Crill}, B.~P. and {Cuttaia}, F. and {de Bernardis}, P. and {de Zotti}, G. and {Delabrouille}, J. and {Di Valentino}, E. and {Diego}, J.~M. and {Dor{\'e}}, O. and {Douspis}, M. and {Ducout}, A. and {Dupac}, X. and {Efstathiou}, G. and {Elsner}, F. and {En{\ss}lin}, T.~A. and {Eriksen}, H.~K. and {Fantaye}, Y. and {Fernandez-Cobos}, R. and {Finelli}, F. and {Forastieri}, F. and {Frailis}, M. and {Fraisse}, A.~A. and {Franceschi}, E. and {Frolov}, A. and {Galeotta}, S. and {Galli}, S. and {Ganga}, K. and {G{\'e}nova-Santos}, R.~T. and {Gerbino}, M. and {Ghosh}, T. and {Gonz{\'a}lez-Nuevo}, J. and {G{\'o}rski}, K.~M. and {Gratton}, S. and {Gruppuso}, A. and {Gudmundsson}, J.~E. and {Hamann}, J. and {Handley}, W. and {Hansen}, F.~K. and {Herranz}, D. and {Hivon}, E. and {Huang}, Z. and {Jaffe}, A.~H. and {Jones}, W.~C. and {Karakci}, A. and {Keih{\"a}nen}, E. and {Keskitalo}, R. and {Kiiveri}, K. and {Kim}, J. and {Knox}, L. and {Krachmalnicoff}, N. and {Kunz}, M. and {Kurki-Suonio}, H. and {Lagache}, G. and {Lamarre}, J. -M. and {Lasenby}, A. and {Lattanzi}, M. and {Lawrence}, C.~R. and {Le Jeune}, M. and {Levrier}, F. and {Lewis}, A. and {Liguori}, M. and {Lilje}, P.~B. and {Lindholm}, V. and {L{\'o}pez-Caniego}, M. and {Lubin}, P.~M. and {Ma}, Y. -Z. and {Mac{\'\i}as-P{\'e}rez}, J.~F. and {Maggio}, G. and {Maino}, D. and {Mandolesi}, N. and {Mangilli}, A. and {Marcos-Caballero}, A. and {Maris}, M. and {Martin}, P.~G. and {Mart{\'\i}nez-Gonz{\'a}lez}, E. and {Matarrese}, S. and {Mauri}, N. and {McEwen}, J.~D. and {Melchiorri}, A. and {Mennella}, A. and {Migliaccio}, M. and {Miville-Desch{\^e}nes}, M. -A. and {Molinari}, D. and {Moneti}, A. and {Montier}, L. and {Morgante}, G. and {Moss}, A. and {Natoli}, P. and {Pagano}, L. and {Paoletti}, D. and {Partridge}, B. and {Patanchon}, G. and {Perrotta}, F. and {Pettorino}, V. and {Piacentini}, F. and {Polastri}, L. and {Polenta}, G. and {Puget}, J. -L. and {Rachen}, J.~P. and {Reinecke}, M. and {Remazeilles}, M. and {Renzi}, A. and {Rocha}, G. and {Rosset}, C. and {Roudier}, G. and {Rubi{\~n}o-Mart{\'\i}n}, J.~A. and {Ruiz-Granados}, B. and {Salvati}, L. and {Sandri}, M. and {Savelainen}, M. and {Scott}, D. and {Sirignano}, C. and {Sunyaev}, R. and {Suur-Uski}, A. -S. and {Tauber}, J.~A. and {Tavagnacco}, D. and {Tenti}, M. and {Toffolatti}, L. and {Tomasi}, M. and {Trombetti}, T. and {Valiviita}, J. and {Van Tent}, B. and {Vielva}, P. and {Villa}, F. and {Vittorio}, N. and {Wandelt}, B.~D. and {Wehus}, I.~K. and {White}, M. and {White}, S.~D.~M. and {Zacchei}, A. and {Zonca}, A.},
        title = "{Planck 2018 results. VIII. Gravitational lensing}",
      journal = {\aap},
         year = 2020,
        month = sep,
       volume = {641},
          eid = {A8},
        pages = {A8},
          doi = {10.1051/0004-6361/201833886},
archivePrefix = {arXiv},
       eprint = {1807.06210},
 primaryClass = {astro-ph.CO},
       adsurl = {https://ui.adsabs.harvard.edu/abs/2020A&A...641A...8P}
}

@ARTICLE{carron22,
       author = {{Carron}, Julien and {Mirmelstein}, Mark and {Lewis}, Antony},
        title = "{CMB lensing from Planck PR4 maps}",
      journal = {\jcap},
         year = 2022,
        month = sep,
       volume = {2022},
       number = {9},
          eid = {039},
        pages = {039},
          doi = {10.1088/1475-7516/2022/09/039},
archivePrefix = {arXiv},
       eprint = {2206.07773},
 primaryClass = {astro-ph.CO},
       adsurl = {https://ui.adsabs.harvard.edu/abs/2022JCAP...09..039C}
}

@ARTICLE{engelen12,
       author = {{van Engelen}, A. and {Keisler}, R. and {Zahn}, O. and {Aird}, K.~A. and {Benson}, B.~A. and {Bleem}, L.~E. and {Carlstrom}, J.~E. and {Chang}, C.~L. and {Cho}, H.~M. and {Crawford}, T.~M. and {Crites}, A.~T. and {de Haan}, T. and {Dobbs}, M.~A. and {Dudley}, J. and {George}, E.~M. and {Halverson}, N.~W. and {Holder}, G.~P. and {Holzapfel}, W.~L. and {Hoover}, S. and {Hou}, Z. and {Hrubes}, J.~D. and {Joy}, M. and {Knox}, L. and {Lee}, A.~T. and {Leitch}, E.~M. and {Lueker}, M. and {Luong-Van}, D. and {McMahon}, J.~J. and {Mehl}, J. and {Meyer}, S.~S. and {Millea}, M. and {Mohr}, J.~J. and {Montroy}, T.~E. and {Natoli}, T. and {Padin}, S. and {Plagge}, T. and {Pryke}, C. and {Reichardt}, C.~L. and {Ruhl}, J.~E. and {Sayre}, J.~T. and {Schaffer}, K.~K. and {Shaw}, L. and {Shirokoff}, E. and {Spieler}, H.~G. and {Staniszewski}, Z. and {Stark}, A.~A. and {Story}, K. and {Vanderlinde}, K. and {Vieira}, J.~D. and {Williamson}, R.},
        title = "{A Measurement of Gravitational Lensing of the Microwave Background Using South Pole Telescope Data}",
      journal = {\apj},
         year = 2012,
        month = sep,
       volume = {756},
       number = {2},
          eid = {142},
        pages = {142},
          doi = {10.1088/0004-637X/756/2/142},
archivePrefix = {arXiv},
       eprint = {1202.0546},
 primaryClass = {astro-ph.CO},
       adsurl = {https://ui.adsabs.harvard.edu/abs/2012ApJ...756..142V}
}

@ARTICLE{story15,
       author = {{Story}, K.~T. and {Hanson}, D. and {Ade}, P.~A.~R. and {Aird}, K.~A. and {Austermann}, J.~E. and {Beall}, J.~A. and {Bender}, A.~N. and {Benson}, B.~A. and {Bleem}, L.~E. and {Carlstrom}, J.~E. and {Chang}, C.~L. and {Chiang}, H.~C. and {Cho}, H. -M. and {Citron}, R. and {Crawford}, T.~M. and {Crites}, A.~T. and {de Haan}, T. and {Dobbs}, M.~A. and {Everett}, W. and {Gallicchio}, J. and {Gao}, J. and {George}, E.~M. and {Gilbert}, A. and {Halverson}, N.~W. and {Harrington}, N. and {Henning}, J.~W. and {Hilton}, G.~C. and {Holder}, G.~P. and {Holzapfel}, W.~L. and {Hoover}, S. and {Hou}, Z. and {Hrubes}, J.~D. and {Huang}, N. and {Hubmayr}, J. and {Irwin}, K.~D. and {Keisler}, R. and {Knox}, L. and {Lee}, A.~T. and {Leitch}, E.~M. and {Li}, D. and {Liang}, C. and {Luong-Van}, D. and {McMahon}, J.~J. and {Mehl}, J. and {Meyer}, S.~S. and {Mocanu}, L. and {Montroy}, T.~E. and {Natoli}, T. and {Nibarger}, J.~P. and {Novosad}, V. and {Padin}, S. and {Pryke}, C. and {Reichardt}, C.~L. and {Ruhl}, J.~E. and {Saliwanchik}, B.~R. and {Sayre}, J.~T. and {Schaffer}, K.~K. and {Smecher}, G. and {Stark}, A.~A. and {Tucker}, C. and {Vanderlinde}, K. and {Vieira}, J.~D. and {Wang}, G. and {Whitehorn}, N. and {Yefremenko}, V. and {Zahn}, O.},
        title = "{A Measurement of the Cosmic Microwave Background Gravitational Lensing Potential from 100 Square Degrees of SPTpol Data}",
      journal = {\apj},
         year = 2015,
        month = sep,
       volume = {810},
       number = {1},
          eid = {50},
        pages = {50},
          doi = {10.1088/0004-637X/810/1/50},
archivePrefix = {arXiv},
       eprint = {1412.4760},
 primaryClass = {astro-ph.CO},
       adsurl = {https://ui.adsabs.harvard.edu/abs/2015ApJ...810...50S}
}

@ARTICLE{wu19,
       author = {{Wu}, W.~L.~K. and {Mocanu}, L.~M. and {Ade}, P.~A.~R. and {Anderson}, A.~J. and {Austermann}, J.~E. and {Avva}, J.~S. and {Beall}, J.~A. and {Bender}, A.~N. and {Benson}, B.~A. and {Bianchini}, F. and {Bleem}, L.~E. and {Carlstrom}, J.~E. and {Chang}, C.~L. and {Chiang}, H.~C. and {Citron}, R. and {Corbett Moran}, C. and {Crawford}, T.~M. and {Crites}, A.~T. and {de Haan}, T. and {Dobbs}, M.~A. and {Everett}, W. and {Gallicchio}, J. and {George}, E.~M. and {Gilbert}, A. and {Gupta}, N. and {Halverson}, N.~W. and {Harrington}, N. and {Henning}, J.~W. and {Hilton}, G.~C. and {Holder}, G.~P. and {Holzapfel}, W.~L. and {Hou}, Z. and {Hrubes}, J.~D. and {Huang}, N. and {Hubmayr}, J. and {Irwin}, K.~D. and {Knox}, L. and {Lee}, A.~T. and {Li}, D. and {Lowitz}, A. and {Manzotti}, A. and {McMahon}, J.~J. and {Meyer}, S.~S. and {Millea}, M. and {Montgomery}, J. and {Nadolski}, A. and {Natoli}, T. and {Nibarger}, J.~P. and {Noble}, G.~I. and {Novosad}, V. and {Omori}, Y. and {Padin}, S. and {Patil}, S. and {Pryke}, C. and {Reichardt}, C.~L. and {Ruhl}, J.~E. and {Saliwanchik}, B.~R. and {Sayre}, J.~T. and {Schaffer}, K.~K. and {Sievers}, C. and {Simard}, G. and {Smecher}, G. and {Stark}, A.~A. and {Story}, K.~T. and {Tucker}, C. and {Vanderlinde}, K. and {Veach}, T. and {Vieira}, J.~D. and {Wang}, G. and {Whitehorn}, N. and {Yefremenko}, V.},
        title = "{A Measurement of the Cosmic Microwave Background Lensing Potential and Power Spectrum from 500 deg$^{2}$ of SPTpol Temperature and Polarization Data}",
      journal = {\apj},
         year = 2019,
        month = oct,
       volume = {884},
       number = {1},
          eid = {70},
        pages = {70},
          doi = {10.3847/1538-4357/ab4186},
archivePrefix = {arXiv},
       eprint = {1905.05777},
 primaryClass = {astro-ph.CO},
       adsurl = {https://ui.adsabs.harvard.edu/abs/2019ApJ...884...70W}
}

@ARTICLE{millea21,
       author = {{Millea}, M. and {Daley}, C.~M. and {Chou}, T. -L. and {Anderes}, E. and {Ade}, P.~A.~R. and {Anderson}, A.~J. and {Austermann}, J.~E. and {Avva}, J.~S. and {Beall}, J.~A. and {Bender}, A.~N. and {Benson}, B.~A. and {Bianchini}, F. and {Bleem}, L.~E. and {Carlstrom}, J.~E. and {Chang}, C.~L. and {Chaubal}, P. and {Chiang}, H.~C. and {Citron}, R. and {Moran}, C. Corbett and {Crawford}, T.~M. and {Crites}, A.~T. and {de Haan}, T. and {Dobbs}, M.~A. and {Everett}, W. and {Gallicchio}, J. and {George}, E.~M. and {Goeckner-Wald}, N. and {Guns}, S. and {Gupta}, N. and {Halverson}, N.~W. and {Henning}, J.~W. and {Hilton}, G.~C. and {Holder}, G.~P. and {Holzapfel}, W.~L. and {Hrubes}, J.~D. and {Huang}, N. and {Hubmayr}, J. and {Irwin}, K.~D. and {Knox}, L. and {Lee}, A.~T. and {Li}, D. and {Lowitz}, A. and {McMahon}, J.~J. and {Meyer}, S.~S. and {Mocanu}, L.~M. and {Montgomery}, J. and {Natoli}, T. and {Nibarger}, J.~P. and {Noble}, G. and {Novosad}, V. and {Omori}, Y. and {Padin}, S. and {Patil}, S. and {Pryke}, C. and {Reichardt}, C.~L. and {Ruhl}, J.~E. and {Saliwanchik}, B.~R. and {Schaffer}, K.~K. and {Sievers}, C. and {Smecher}, G. and {Stark}, A.~A. and {Thorne}, B. and {Tucker}, C. and {Veach}, T. and {Vieira}, J.~D. and {Wang}, G. and {Whitehorn}, N. and {Wu}, W.~L.~K. and {Yefremenko}, V.},
        title = "{Optimal Cosmic Microwave Background Lensing Reconstruction and Parameter Estimation with SPTpol Data}",
      journal = {\apj},
         year = 2021,
        month = dec,
       volume = {922},
       number = {2},
          eid = {259},
        pages = {259},
          doi = {10.3847/1538-4357/ac02bb},
archivePrefix = {arXiv},
       eprint = {2012.01709},
 primaryClass = {astro-ph.CO},
       adsurl = {https://ui.adsabs.harvard.edu/abs/2021ApJ...922..259M}
}

@ARTICLE{pan23,
       author = {{Pan}, Z. and {Bianchini}, F. and {Wu}, W.~L.~K. and {Ade}, P.~A.~R. and {Ahmed}, Z. and {Anderes}, E. and {Anderson}, A.~J. and {Ansarinejad}, B. and {Archipley}, M. and {Aylor}, K. and {Balkenhol}, L. and {Barry}, P.~S. and {Basu Thakur}, R. and {Benabed}, K. and {Bender}, A.~N. and {Benson}, B.~A. and {Bleem}, L.~E. and {Bouchet}, F.~R. and {Bryant}, L. and {Byrum}, K. and {Camphuis}, E. and {Carlstrom}, J.~E. and {Carter}, F.~W. and {Cecil}, T.~W. and {Chang}, C.~L. and {Chaubal}, P. and {Chen}, G. and {Chichura}, P.~M. and {Cho}, H. -M. and {Chou}, T. -L. and {Cliche}, J. -F. and {Coerver}, A. and {Crawford}, T.~M. and {Cukierman}, A. and {Daley}, C. and {de Haan}, T. and {Denison}, E.~V. and {Dibert}, K.~R. and {Ding}, J. and {Dobbs}, M.~A. and {Doussot}, A. and {Dutcher}, D. and {Everett}, W. and {Feng}, C. and {Ferguson}, K.~R. and {Fichman}, K. and {Foster}, A. and {Fu}, J. and {Galli}, S. and {Gambrel}, A.~E. and {Gardner}, R.~W. and {Ge}, F. and {Goeckner-Wald}, N. and {Gualtieri}, R. and {Guidi}, F. and {Guns}, S. and {Gupta}, N. and {Halverson}, N.~W. and {Harke-Hosemann}, A.~H. and {Harrington}, N.~L. and {Henning}, J.~W. and {Hilton}, G.~C. and {Hivon}, E. and {Holder}, G.~P. and {Holzapfel}, W.~L. and {Hood}, J.~C. and {Howe}, D. and {Huang}, N. and {Irwin}, K.~D. and {Jeong}, O. and {Jonas}, M. and {Jones}, A. and {K{\'e}ruzor{\'e}}, F. and {Khaire}, T.~S. and {Knox}, L. and {Kofman}, A.~M. and {Korman}, M. and {Kubik}, D.~L. and {Kuhlmann}, S. and {Kuo}, C. -L. and {Lee}, A.~T. and {Leitch}, E.~M. and {Levy}, K. and {Lowitz}, A.~E. and {Lu}, C. and {Maniyar}, A. and {Menanteau}, F. and {Meyer}, S.~S. and {Michalik}, D. and {Millea}, M. and {Montgomery}, J. and {Nadolski}, A. and {Nakato}, Y. and {Natoli}, T. and {Nguyen}, H. and {Noble}, G.~I. and {Novosad}, V. and {Omori}, Y. and {Padin}, S. and {Paschos}, P. and {Pearson}, J. and {Posada}, C.~M. and {Prabhu}, K. and {Quan}, W. and {Raghunathan}, S. and {Rahimi}, M. and {Rahlin}, A. and {Reichardt}, C.~L. and {Riebel}, D. and {Riedel}, B. and {Ruhl}, J.~E. and {Sayre}, J.~T. and {Schiappucci}, E. and {Shirokoff}, E. and {Smecher}, G. and {Sobrin}, J.~A. and {Stark}, A.~A. and {Stephen}, J. and {Story}, K.~T. and {Suzuki}, A. and {Takakura}, S. and {Tandoi}, C. and {Thompson}, K.~L. and {Thorne}, B. and {Trendafilova}, C. and {Tucker}, C. and {Umilta}, C. and {Vale}, L.~R. and {Vanderlinde}, K. and {Vieira}, J.~D. and {Wang}, G. and {Whitehorn}, N. and {Yefremenko}, V. and {Yoon}, K.~W. and {Young}, M.~R. and {Zebrowski}, J.~A.},
        title = "{Measurement of gravitational lensing of the cosmic microwave background using SPT-3G 2018 data}",
      journal = {\prd},
         year = 2023,
        month = dec,
       volume = {108},
       number = {12},
          eid = {122005},
        pages = {122005},
          doi = {10.1103/PhysRevD.108.122005},
archivePrefix = {arXiv},
       eprint = {2308.11608},
 primaryClass = {astro-ph.CO},
       adsurl = {https://ui.adsabs.harvard.edu/abs/2023PhRvD.108l2005P}
}

@ARTICLE{kamionkowski16,
       author = {{Kamionkowski}, Marc and {Kovetz}, Ely D.},
        title = "{The Quest for B Modes from Inflationary Gravitational Waves}",
      journal = {\araa},
         year = 2016,
        month = sep,
       volume = {54},
        pages = {227-269},
          doi = {10.1146/annurev-astro-081915-023433},
archivePrefix = {arXiv},
       eprint = {1510.06042},
 primaryClass = {astro-ph.CO},
       adsurl = {https://ui.adsabs.harvard.edu/abs/2016ARA&A..54..227K}
}

@ARTICLE{ade14,
       author = {{Ade}, P.~A.~R. and {Akiba}, Y. and {Anthony}, A.~E. and {Arnold}, K. and {Atlas}, M. and {Barron}, D. and {Boettger}, D. and {Borrill}, J. and {Chapman}, S. and {Chinone}, Y. and {Dobbs}, M. and {Elleflot}, T. and {Errard}, J. and {Fabbian}, G. and {Feng}, C. and {Flanigan}, D. and {Gilbert}, A. and {Grainger}, W. and {Halverson}, N.~W. and {Hasegawa}, M. and {Hattori}, K. and {Hazumi}, M. and {Holzapfel}, W.~L. and {Hori}, Y. and {Howard}, J. and {Hyland}, P. and {Inoue}, Y. and {Jaehnig}, G.~C. and {Jaffe}, A. and {Keating}, B. and {Kermish}, Z. and {Keskitalo}, R. and {Kisner}, T. and {Le Jeune}, M. and {Lee}, A.~T. and {Linder}, E. and {Leitch}, E.~M. and {Lungu}, M. and {Matsuda}, F. and {Matsumura}, T. and {Meng}, X. and {Miller}, N.~J. and {Morii}, H. and {Moyerman}, S. and {Myers}, M.~J. and {Navaroli}, M. and {Nishino}, H. and {Paar}, H. and {Peloton}, J. and {Quealy}, E. and {Rebeiz}, G. and {Reichardt}, C.~L. and {Richards}, P.~L. and {Ross}, C. and {Schanning}, I. and {Schenck}, D.~E. and {Sherwin}, B. and {Shimizu}, A. and {Shimmin}, C. and {Shimon}, M. and {Siritanasak}, P. and {Smecher}, G. and {Spieler}, H. and {Stebor}, N. and {Steinbach}, B. and {Stompor}, R. and {Suzuki}, A. and {Takakura}, S. and {Tomaru}, T. and {Wilson}, B. and {Yadav}, A. and {Zahn}, O. and {Polarbear Collaboration}},
        title = "{Measurement of the Cosmic Microwave Background Polarization Lensing Power Spectrum with the POLARBEAR Experiment}",
      journal = {\prl},
         year = 2014,
        month = jul,
       volume = {113},
       number = {2},
          eid = {021301},
        pages = {021301},
          doi = {10.1103/PhysRevLett.113.021301},
archivePrefix = {arXiv},
       eprint = {1312.6646},
 primaryClass = {astro-ph.CO},
       adsurl = {https://ui.adsabs.harvard.edu/abs/2014PhRvL.113b1301A}
}

@ARTICLE{faundez20,
       author = {{Fa{\'u}ndez}, M. Aguilar and {Arnold}, K. and {Baccigalupi}, C. and {Barron}, D. and {Beck}, D. and {Beckman}, S. and {Bianchini}, F. and {Carron}, J. and {Cheung}, K. and {Chinone}, Y. and {Bouhargani}, H. El and {Elleflot}, T. and {Errard}, J. and {Fabbian}, G. and {Feng}, C. and {Fujino}, T. and {Goeckner-Wald}, N. and {Hamada}, T. and {Hasegawa}, M. and {Hazumi}, M. and {Hill}, C.~A. and {Hirose}, H. and {Jeong}, O. and {Katayama}, N. and {Keating}, B. and {Kikuchi}, S. and {Kusaka}, A. and {Lee}, A.~T. and {Leon}, D. and {Linder}, E. and {Lowry}, L.~N. and {Matsuda}, F. and {Matsumura}, T. and {Minami}, Y. and {Navaroli}, M. and {Nishino}, H. and {Pham}, A.~T.~P. and {Poletti}, D. and {Puglisi}, G. and {Reichardt}, C.~L. and {Segawa}, Y. and {Sherwin}, B.~D. and {Silva-Feaver}, M. and {Siritanasak}, P. and {Stompor}, R. and {Suzuki}, A. and {Tajima}, O. and {Takatori}, S. and {Tanabe}, D. and {Teply}, G.~P. and {Tsai}, C. and {Polarbear Collaboration}},
        title = "{Measurement of the Cosmic Microwave Background Polarization Lensing Power Spectrum from Two Years of POLARBEAR Data}",
      journal = {\apj},
         year = 2020,
        month = apr,
       volume = {893},
       number = {1},
          eid = {85},
        pages = {85},
          doi = {10.3847/1538-4357/ab7e29},
archivePrefix = {arXiv},
       eprint = {1911.10980},
 primaryClass = {astro-ph.CO},
       adsurl = {https://ui.adsabs.harvard.edu/abs/2020ApJ...893...85F}
}

@ARTICLE{bicep2keck16,
       author = {{BICEP2 Collaboration} and {Keck Array Collaboration} and {Ade}, P.~A.~R. and {Ahmed}, Z. and {Aikin}, R.~W. and {Alexander}, K.~D. and {Barkats}, D. and {Benton}, S.~J. and {Bischoff}, C.~A. and {Bock}, J.~J. and {Bowens-Rubin}, R. and {Brevik}, J.~A. and {Buder}, I. and {Bullock}, E. and {Buza}, V. and {Connors}, J. and {Crill}, B.~P. and {Duband}, L. and {Dvorkin}, C. and {Filippini}, J.~P. and {Fliescher}, S. and {Grayson}, J. and {Halpern}, M. and {Harrison}, S. and {Hildebrandt}, S.~R. and {Hilton}, G.~C. and {Hui}, H. and {Irwin}, K.~D. and {Kang}, J. and {Karkare}, K.~S. and {Karpel}, E. and {Kaufman}, J.~P. and {Keating}, B.~G. and {Kefeli}, S. and {Kernasovskiy}, S.~A. and {Kovac}, J.~M. and {Kuo}, C.~L. and {Leitch}, E.~M. and {Lueker}, M. and {Megerian}, K.~G. and {Namikawa}, T. and {Netterfield}, C.~B. and {Nguyen}, H.~T. and {O'Brient}, R. and {Ogburn}, R.~W., IV and {Orlando}, A. and {Pryke}, C. and {Richter}, S. and {Schwarz}, R. and {Sheehy}, C.~D. and {Staniszewski}, Z.~K. and {Steinbach}, B. and {Sudiwala}, R.~V. and {Teply}, G.~P. and {Thompson}, K.~L. and {Tolan}, J.~E. and {Tucker}, C. and {Turner}, A.~D. and {Vieregg}, A.~G. and {Weber}, A.~C. and {Wiebe}, D.~V. and {Willmert}, J. and {Wong}, C.~L. and {Wu}, W.~L.~K. and {Yoon}, K.~W.},
        title = "{BICEP2/Keck Array  VIII: Measurement of Gravitational Lensing from Large-scale B-mode Polarization}",
      journal = {\apj},
         year = 2016,
        month = dec,
       volume = {833},
       number = {2},
          eid = {228},
        pages = {228},
          doi = {10.3847/1538-4357/833/2/228},
archivePrefix = {arXiv},
       eprint = {1606.01968},
 primaryClass = {astro-ph.CO},
       adsurl = {https://ui.adsabs.harvard.edu/abs/2016ApJ...833..228B}
}

@ARTICLE{ade23,
       author = {{Ade}, P.~A.~R. and {Ahmed}, Z. and {Amiri}, M. and {Barkats}, D. and {Thakur}, R. Basu and {Bischoff}, C.~A. and {Beck}, D. and {Bock}, J.~J. and {Boenish}, H. and {Bullock}, E. and {Buza}, V. and {Cheshire}, J.~R., IV and {Connors}, J. and {Cornelison}, J. and {Crumrine}, M. and {Cukierman}, A. and {Denison}, E.~V. and {Dierickx}, M. and {Duband}, L. and {Eiben}, M. and {Fatigoni}, S. and {Filippini}, J.~P. and {Fliescher}, S. and {Giannakopoulos}, C. and {Goeckner-Wald}, N. and {Goldfinger}, D.~C. and {Grayson}, J. and {Grimes}, P. and {Hall}, G. and {Halal}, G. and {Halpern}, M. and {Hand}, E. and {Harrison}, S. and {Henderson}, S. and {Hildebrandt}, S.~R. and {Hubmayr}, J. and {Hui}, H. and {Irwin}, K.~D. and {Kang}, J. and {Karkare}, K.~S. and {Karpel}, E. and {Kefeli}, S. and {Kernasovskiy}, S.~A. and {Kovac}, J.~M. and {Kuo}, C.~L. and {Lau}, K. and {Leitch}, E.~M. and {Lennox}, A. and {Megerian}, K.~G. and {Minutolo}, L. and {Moncelsi}, L. and {Nakato}, Y. and {Namikawa}, T. and {Nguyen}, H.~T. and {O'Brient}, R. and {Ogburn}, R.~W., IV and {Palladino}, S. and {Petroff}, M. and {Prouve}, T. and {Pryke}, C. and {Racine}, B. and {Reintsema}, C.~D. and {Richter}, S. and {Schillaci}, A. and {Schwarz}, R. and {Schmitt}, B.~L. and {Sheehy}, C.~D. and {Singari}, B. and {Soliman}, A. and {St. Germaine}, T. and {Steinbach}, B. and {Sudiwala}, R.~V. and {Teply}, G.~P. and {Thompson}, K.~L. and {Tolan}, J.~E. and {Tucker}, C. and {Turner}, A.~D. and {Umilt{\`a}}, C. and {Verg{\`e}s}, C. and {Vieregg}, A.~G. and {Wandui}, A. and {Weber}, A.~C. and {Wiebe}, D.~V. and {Willmert}, J. and {Wong}, C.~L. and {Wu}, W.~L.~K. and {Yang}, H. and {Yoon}, K.~W. and {Young}, E. and {Yu}, C. and {Zeng}, L. and {Zhang}, C. and {Zhang}, S. and {Bicep/Keck Collaboration}},
        title = "{BICEP/Keck. XVII. Line-of-sight Distortion Analysis: Estimates of Gravitational Lensing, Anisotropic Cosmic Birefringence, Patchy Reionization, and Systematic Errors}",
      journal = {\apj},
         year = 2023,
        month = jun,
       volume = {949},
       number = {2},
          eid = {43},
        pages = {43},
          doi = {10.3847/1538-4357/acc85c},
archivePrefix = {arXiv},
       eprint = {2210.08038},
 primaryClass = {astro-ph.CO},
       adsurl = {https://ui.adsabs.harvard.edu/abs/2023ApJ...949...43A}
}

@ARTICLE{blanchard87,
       author = {{Blanchard}, A. and {Schneider}, J.},
        title = "{Gravitational lensing effect on the fluctuations of the cosmic background radiation}",
      journal = {\aap},
         year = 1987,
        month = oct,
       volume = {184},
       number = {1-2},
        pages = {1-6},
       adsurl = {https://ui.adsabs.harvard.edu/abs/1987A&A...184....1B}
}

@ARTICLE{kesden03,
       author = {{Kesden}, Michael and {Cooray}, Asantha and {Kamionkowski}, Marc},
        title = "{Lensing reconstruction with CMB temperature and polarization}",
      journal = {\prd},
         year = 2003,
        month = jun,
       volume = {67},
       number = {12},
          eid = {123507},
        pages = {123507},
          doi = {10.1103/PhysRevD.67.123507},
archivePrefix = {arXiv},
       eprint = {astro-ph/0302536},
 primaryClass = {astro-ph},
       adsurl = {https://ui.adsabs.harvard.edu/abs/2003PhRvD..67l3507K}
}

@ARTICLE{hanson11,
       author = {{Hanson}, Duncan and {Challinor}, Anthony and {Efstathiou}, George and {Bielewicz}, Pawel},
        title = "{CMB temperature lensing power reconstruction}",
      journal = {\prd},
         year = 2011,
        month = feb,
       volume = {83},
       number = {4},
          eid = {043005},
        pages = {043005},
          doi = {10.1103/PhysRevD.83.043005},
archivePrefix = {arXiv},
       eprint = {1008.4403},
 primaryClass = {astro-ph.CO},
       adsurl = {https://ui.adsabs.harvard.edu/abs/2011PhRvD..83d3005H}
}

@ARTICLE{namikawa13,
       author = {{Namikawa}, Toshiya and {Hanson}, Duncan and {Takahashi}, Ryuichi},
        title = "{Bias-hardened CMB lensing}",
      journal = {\mnras},
         year = 2013,
        month = may,
       volume = {431},
       number = {1},
        pages = {609-620},
          doi = {10.1093/mnras/stt195},
archivePrefix = {arXiv},
       eprint = {1209.0091},
 primaryClass = {astro-ph.CO},
       adsurl = {https://ui.adsabs.harvard.edu/abs/2013MNRAS.431..609N}
}

@ARTICLE{dutcher21,
       author = {{Dutcher}, D. and {Balkenhol}, L. and {Ade}, P.~A.~R. and {Ahmed}, Z. and {Anderes}, E. and {Anderson}, A.~J. and {Archipley}, M. and {Avva}, J.~S. and {Aylor}, K. and {Barry}, P.~S. and {Basu Thakur}, R. and {Benabed}, K. and {Bender}, A.~N. and {Benson}, B.~A. and {Bianchini}, F. and {Bleem}, L.~E. and {Bouchet}, F.~R. and {Bryant}, L. and {Byrum}, K. and {Carlstrom}, J.~E. and {Carter}, F.~W. and {Cecil}, T.~W. and {Chang}, C.~L. and {Chaubal}, P. and {Chen}, G. and {Cho}, H. -M. and {Chou}, T. -L. and {Cliche}, J. -F. and {Crawford}, T.~M. and {Cukierman}, A. and {Daley}, C. and {de Haan}, T. and {Denison}, E.~V. and {Dibert}, K. and {Ding}, J. and {Dobbs}, M.~A. and {Everett}, W. and {Feng}, C. and {Ferguson}, K.~R. and {Foster}, A. and {Fu}, J. and {Galli}, S. and {Gambrel}, A.~E. and {Gardner}, R.~W. and {Goeckner-Wald}, N. and {Gualtieri}, R. and {Guns}, S. and {Gupta}, N. and {Guyser}, R. and {Halverson}, N.~W. and {Harke-Hosemann}, A.~H. and {Harrington}, N.~L. and {Henning}, J.~W. and {Hilton}, G.~C. and {Hivon}, E. and {Holder}, G.~P. and {Holzapfel}, W.~L. and {Hood}, J.~C. and {Howe}, D. and {Huang}, N. and {Irwin}, K.~D. and {Jeong}, O.~B. and {Jonas}, M. and {Jones}, A. and {Khaire}, T.~S. and {Knox}, L. and {Kofman}, A.~M. and {Korman}, M. and {Kubik}, D.~L. and {Kuhlmann}, S. and {Kuo}, C. -L. and {Lee}, A.~T. and {Leitch}, E.~M. and {Lowitz}, A.~E. and {Lu}, C. and {Meyer}, S.~S. and {Michalik}, D. and {Millea}, M. and {Montgomery}, J. and {Nadolski}, A. and {Natoli}, T. and {Nguyen}, H. and {Noble}, G.~I. and {Novosad}, V. and {Omori}, Y. and {Padin}, S. and {Pan}, Z. and {Paschos}, P. and {Pearson}, J. and {Posada}, C.~M. and {Prabhu}, K. and {Quan}, W. and {Raghunathan}, S. and {Rahlin}, A. and {Reichardt}, C.~L. and {Riebel}, D. and {Riedel}, B. and {Rouble}, M. and {Ruhl}, J.~E. and {Sayre}, J.~T. and {Schiappucci}, E. and {Shirokoff}, E. and {Smecher}, G. and {Sobrin}, J.~A. and {Stark}, A.~A. and {Stephen}, J. and {Story}, K.~T. and {Suzuki}, A. and {Thompson}, K.~L. and {Thorne}, B. and {Tucker}, C. and {Umilta}, C. and {Vale}, L.~R. and {Vanderlinde}, K. and {Vieira}, J.~D. and {Wang}, G. and {Whitehorn}, N. and {Wu}, W.~L.~K. and {Yefremenko}, V. and {Yoon}, K.~W. and {Young}, M.~R. and {SPT-3G Collaboration}},
        title = "{Measurements of the E -mode polarization and temperature-E -mode correlation of the CMB from SPT-3G 2018 data}",
      journal = {\prd},
         year = 2021,
        month = jul,
       volume = {104},
       number = {2},
          eid = {022003},
        pages = {022003},
          doi = {10.1103/PhysRevD.104.022003},
archivePrefix = {arXiv},
       eprint = {2101.01684},
 primaryClass = {astro-ph.CO},
       adsurl = {https://ui.adsabs.harvard.edu/abs/2021PhRvD.104b2003D}
}

@ARTICLE{benoit13,
       author = {{Benoit-L{\'e}vy}, A. and {D{\'e}chelette}, T. and {Benabed}, K. and {Cardoso}, J. -F. and {Hanson}, D. and {Prunet}, S.},
        title = "{Full-sky CMB lensing reconstruction in presence of sky-cuts}",
      journal = {\aap},
         year = 2013,
        month = jul,
       volume = {555},
          eid = {A37},
        pages = {A37},
          doi = {10.1051/0004-6361/201321048},
archivePrefix = {arXiv},
       eprint = {1301.4145},
 primaryClass = {astro-ph.CO},
       adsurl = {https://ui.adsabs.harvard.edu/abs/2013A&A...555A..37B}
}

@ARTICLE{raghunathan19,
       author = {{Raghunathan}, Srinivasan and {Holder}, Gilbert P. and {Bartlett}, James G. and {Patil}, SanjayKumar and {Reichardt}, Christian L. and {Whitehorn}, Nathan},
        title = "{An inpainting approach to tackle the kinematic and thermal SZ induced biases in CMB-cluster lensing estimators}",
      journal = {\jcap},
         year = 2019,
        month = nov,
       volume = {2019},
       number = {11},
          eid = {037},
        pages = {037},
          doi = {10.1088/1475-7516/2019/11/037},
archivePrefix = {arXiv},
       eprint = {1904.13392},
 primaryClass = {astro-ph.CO},
       adsurl = {https://ui.adsabs.harvard.edu/abs/2019JCAP...11..037R}
}

@ARTICLE{gorski05,
       author = {{G{\'o}rski}, K.~M. and {Hivon}, E. and {Banday}, A.~J. and {Wandelt}, B.~D. and {Hansen}, F.~K. and {Reinecke}, M. and {Bartelmann}, M.},
        title = "{HEALPix: A Framework for High-Resolution Discretization and Fast Analysis of Data Distributed on the Sphere}",
      journal = {\apj},
         year = 2005,
        month = apr,
       volume = {622},
       number = {2},
        pages = {759-771},
          doi = {10.1086/427976},
archivePrefix = {arXiv},
       eprint = {astro-ph/0409513},
 primaryClass = {astro-ph},
       adsurl = {https://ui.adsabs.harvard.edu/abs/2005ApJ...622..759G}
}

@ARTICLE{namikawa12,
       author = {{Namikawa}, Toshiya and {Yamauchi}, Daisuke and {Taruya}, Atsushi},
        title = "{Full-sky lensing reconstruction of gradient and curl modes from CMB maps}",
      journal = {\jcap},
         year = 2012,
        month = jan,
       volume = {2012},
       number = {1},
          eid = {007},
        pages = {007},
          doi = {10.1088/1475-7516/2012/01/007},
archivePrefix = {arXiv},
       eprint = {1110.1718},
 primaryClass = {astro-ph.CO},
       adsurl = {https://ui.adsabs.harvard.edu/abs/2012JCAP...01..007N}
}

@ARTICLE{cooray05,
       author = {{Cooray}, Asantha and {Kamionkowski}, Marc and {Caldwell}, Robert R.},
        title = "{Cosmic shear of the microwave background: The curl diagnostic}",
      journal = {\prd},
         year = 2005,
        month = jun,
       volume = {71},
       number = {12},
          eid = {123527},
        pages = {123527},
          doi = {10.1103/PhysRevD.71.123527},
archivePrefix = {arXiv},
       eprint = {astro-ph/0503002},
 primaryClass = {astro-ph},
       adsurl = {https://ui.adsabs.harvard.edu/abs/2005PhRvD..71l3527C}
}

@ARTICLE{fabbian19,
       author = {{Fabbian}, Giulio and {Lewis}, Antony and {Beck}, Dominic},
        title = "{CMB lensing reconstruction biases in cross-correlation with large-scale structure probes}",
      journal = {\jcap},
         year = 2019,
        month = oct,
       volume = {2019},
       number = {10},
          eid = {057},
        pages = {057},
          doi = {10.1088/1475-7516/2019/10/057},
archivePrefix = {arXiv},
       eprint = {1906.08760},
 primaryClass = {astro-ph.CO},
       adsurl = {https://ui.adsabs.harvard.edu/abs/2019JCAP...10..057F}
}

@ARTICLE{omori24,
       author = {{Omori}, Yuuki},
        title = "{AGORA: Multicomponent simulation for cross-survey science}",
      journal = {\mnras},
         year = 2024,
        month = jun,
       volume = {530},
       number = {4},
        pages = {5030-5068},
          doi = {10.1093/mnras/stae1031},
archivePrefix = {arXiv},
       eprint = {2212.07420},
 primaryClass = {astro-ph.CO},
       adsurl = {https://ui.adsabs.harvard.edu/abs/2024MNRAS.530.5030O}
}

@ARTICLE{gupta19,
       author = {{Gupta}, N. and {Reichardt}, C.~L. and {Ade}, P.~A.~R. and {Anderson}, A.~J. and {Archipley}, M. and {Austermann}, J.~E. and {Avva}, J.~S. and {Beall}, J.~A. and {Bender}, A.~N. and {Benson}, B.~A. and {Bianchini}, F. and {Bleem}, L.~E. and {Carlstrom}, J.~E. and {Chang}, C.~L. and {Chiang}, H.~C. and {Citron}, R. and {Moran}, C. Corbett and {Crawford}, T.~M. and {Crites}, A.~T. and {de Haan}, T. and {Dobbs}, M.~A. and {Everett}, W. and {Feng}, C. and {Gallicchio}, J. and {George}, E.~M. and {Gilbert}, A. and {Halverson}, N.~W. and {Harrington}, N. and {Henning}, J.~W. and {Hilton}, G.~C. and {Holder}, G.~P. and {Holzapfel}, W.~L. and {Hou}, Z. and {Hrubes}, J.~D. and {Huang}, N. and {Hubmayr}, J. and {Irwin}, K.~D. and {Knox}, L. and {Lee}, A.~T. and {Li}, D. and {Lowitz}, A. and {Luong-Van}, D. and {Marrone}, D.~P. and {McMahon}, J.~J. and {Meyer}, S.~S. and {Mocanu}, L.~M. and {Mohr}, J.~J. and {Montgomery}, J. and {Nadolski}, A. and {Natoli}, T. and {Nibarger}, J.~P. and {Noble}, G.~I. and {Novosad}, V. and {Padin}, S. and {Patil}, S. and {Pryke}, C. and {Ruhl}, J.~E. and {Saliwanchik}, B.~R. and {Sayre}, J.~T. and {Schaffer}, K.~K. and {Shirokoff}, E. and {Sievers}, C. and {Smecher}, G. and {Staniszewski}, Z. and {Stark}, A.~A. and {Story}, K.~T. and {Switzer}, E.~R. and {Tucker}, C. and {Vanderlinde}, K. and {Veach}, T. and {Vieira}, J.~D. and {Wang}, G. and {Whitehorn}, N. and {Williamson}, R. and {Wu}, W.~L.~K. and {Yefremenko}, V. and {Zhang}, L.},
        title = "{Fractional polarization of extragalactic sources in the 500 deg$^{2}$ SPTpol survey}",
      journal = {\mnras},
         year = 2019,
        month = dec,
       volume = {490},
       number = {4},
        pages = {5712-5721},
          doi = {10.1093/mnras/stz2905},
archivePrefix = {arXiv},
       eprint = {1907.02156},
 primaryClass = {astro-ph.CO},
       adsurl = {https://ui.adsabs.harvard.edu/abs/2019MNRAS.490.5712G}
}

@ARTICLE{datta19,
       author = {{Datta}, Rahul and {Aiola}, Simone and {Choi}, Steve K. and {Devlin}, Mark and {Dunkley}, Joanna and {D{\"u}nner}, Rolando and {Gallardo}, Patricio A. and {Gralla}, Megan and {Halpern}, Mark and {Hasselfield}, Matthew and {Hilton}, Matt and {Hincks}, Adam D. and {Ho}, Shuay-Pwu P. and {Hubmayr}, Johannes and {Huffenberger}, Kevin M. and {Hughes}, John P. and {Kosowsky}, Arthur and {L{\'o}pez-Caraballo}, Carlos H. and {Louis}, Thibaut and {Lungu}, Marius and {Marriage}, Tobias and {Maurin}, Lo{\"\i}c and {McMahon}, Jeff and {Moodley}, Kavilan and {Naess}, Sigurd K. and {Nati}, Federico and {Niemack}, Michael D. and {Page}, Lyman A. and {Partridge}, Bruce and {Prince}, Heather and {Staggs}, Suzanne T. and {Switzer}, Eric R. and {Wollack}, Edward J. and {Farren}, Gerrit},
        title = "{The Atacama Cosmology Telescope: two-season ACTPol extragalactic point sources and their polarization properties}",
      journal = {\mnras},
         year = 2019,
        month = jul,
       volume = {486},
       number = {4},
        pages = {5239-5262},
          doi = {10.1093/mnras/sty2934},
archivePrefix = {arXiv},
       eprint = {1811.01854},
 primaryClass = {astro-ph.CO},
       adsurl = {https://ui.adsabs.harvard.edu/abs/2019MNRAS.486.5239D}
}

@ARTICLE{smith07,
       author = {{Smith}, Kendrick M. and {Zahn}, Oliver and {Dor{\'e}}, Olivier},
        title = "{Detection of gravitational lensing in the cosmic microwave background}",
      journal = {\prd},
         year = 2007,
        month = aug,
       volume = {76},
       number = {4},
          eid = {043510},
        pages = {043510},
          doi = {10.1103/PhysRevD.76.043510},
archivePrefix = {arXiv},
       eprint = {0705.3980},
 primaryClass = {astro-ph},
       adsurl = {https://ui.adsabs.harvard.edu/abs/2007PhRvD..76d3510S}
}

@ARTICLE{atkins23,
       author = {{Atkins}, Zachary and {Duivenvoorden}, Adriaan J. and {Coulton}, William R. and {Qu}, Frank J. and {Aiola}, Simone and {Calabrese}, Erminia and {Chesmore}, Grace E. and {Choi}, Steve K. and {Devlin}, Mark J. and {Dunkley}, Jo and {Herv{\'\i}as-Caimapo}, Carlos and {Guan}, Yilun and {Posta}, Adrien La and {Li}, Zack and {Louis}, Thibaut and {Madhavacheril}, Mathew S. and {Moodley}, Kavilan and {Naess}, Sigurd and {Nati}, Federico and {Niemack}, Michael D. and {Page}, Lyman and {Puddu}, Roberto and {Salatino}, Maria and {Sif{\'o}n}, Crist{\'o}bal and {Staggs}, Suzanne T. and {Vargas}, Cristian and {Vavagiakis}, Eve M. and {Wollack}, Edward J.},
        title = "{The Atacama Cosmology Telescope: map-based noise simulations for DR6}",
      journal = {\jcap},
         year = 2023,
        month = nov,
       volume = {2023},
       number = {11},
          eid = {073},
        pages = {073},
          doi = {10.1088/1475-7516/2023/11/073},
archivePrefix = {arXiv},
       eprint = {2303.04180},
 primaryClass = {astro-ph.CO},
       adsurl = {https://ui.adsabs.harvard.edu/abs/2023JCAP...11..073A}
}

@ARTICLE{camphuis25,
       author = {{Camphuis}, E. and {Quan}, W. and {Balkenhol}, L. and {Khalife}, A.~R. and {Ge}, F. and {Guidi}, F. and {Huang}, N. and {Lynch}, G.~P. and {Omori}, Y. and {Trendafilova}, C. and {Anderson}, A.~J. and {Ansarinejad}, B. and {Archipley}, M. and {Barry}, P.~S. and {Benabed}, K. and {Bender}, A.~N. and {Benson}, B.~A. and {Bianchini}, F. and {Bleem}, L.~E. and {Bouchet}, F.~R. and {Bryant}, L. and {Campitiello}, M.~G. and {Carlstrom}, J.~E. and {Chang}, C.~L. and {Chaubal}, P. and {Chichura}, P.~M. and {Chokshi}, A. and {Chou}, T. -L. and {Coerver}, A. and {Crawford}, T.~M. and {Daley}, C. and {de Haan}, T. and {Dibert}, K.~R. and {Dobbs}, M.~A. and {Doohan}, M. and {Doussot}, A. and {Dutcher}, D. and {Everett}, W. and {Feng}, C. and {Ferguson}, K.~R. and {Fichman}, K. and {Foster}, A. and {Galli}, S. and {Gambrel}, A.~E. and {Gardner}, R.~W. and {Goeckner-Wald}, N. and {Gualtieri}, R. and {Guns}, S. and {Halverson}, N.~W. and {Hivon}, E. and {Holder}, G.~P. and {Holzapfel}, W.~L. and {Hood}, J.~C. and {Hryciuk}, A. and {K{\'e}ruzor{\'e}}, F. and {Knox}, L. and {Korman}, M. and {Kornoelje}, K. and {Kuo}, C. -L. and {Levy}, K. and {Lowitz}, A.~E. and {Lu}, C. and {Maniyar}, A. and {Martsen}, E.~S. and {Menanteau}, F. and {Millea}, M. and {Montgomery}, J. and {Nakato}, Y. and {Natoli}, T. and {Noble}, G.~I. and {Ouellette}, A. and {Pan}, Z. and {Paschos}, P. and {Phadke}, K.~A. and {Pollak}, A.~W. and {Prabhu}, K. and {Raghunathan}, S. and {Rahimi}, M. and {Rahlin}, A. and {Reichardt}, C.~L. and {Rouble}, M. and {Ruhl}, J.~E. and {Schiappucci}, E. and {Simpson}, A. and {Sobrin}, J.~A. and {Stark}, A.~A. and {Stephen}, J. and {Tandoi}, C. and {Thorne}, B. and {Umilta}, C. and {Vieira}, J.~D. and {Vitrier}, A. and {Wan}, Y. and {Whitehorn}, N. and {Wu}, W.~L.~K. and {Young}, M.~R. and {Zebrowski}, J.~A.},
        title = "{SPT-3G D1: CMB temperature and polarization power spectra and cosmology from 2019 and 2020 observations of the SPT-3G Main field}",
      journal = {arXiv e-prints},
         year = 2025,
        month = jun,
          eid = {arXiv:2506.20707},
        pages = {arXiv:2506.20707},
          doi = {10.48550/arXiv.2506.20707},
archivePrefix = {arXiv},
       eprint = {2506.20707},
 primaryClass = {astro-ph.CO},
       adsurl = {https://ui.adsabs.harvard.edu/abs/2025arXiv250620707C}
}

@ARTICLE{louis25,
       author = {{Louis}, Thibaut and {La Posta}, Adrien and {Atkins}, Zachary and {Jense}, Hidde T. and {Abril-Cabezas}, Irene and {Addison}, Graeme E. and {Ade}, Peter A.~R. and {Aiola}, Simone and {Alford}, Tommy and {Alonso}, David and {Amiri}, Mandana and {An}, Rui and {Austermann}, Jason E. and {Barbavara}, Eleonora and {Battaglia}, Nicholas and {Battistelli}, Elia Stefano and {Beall}, James A. and {Bean}, Rachel and {Beheshti}, Ali and {Beringue}, Benjamin and {Bhandarkar}, Tanay and {Biermann}, Emily and {Bolliet}, Boris and {Bond}, J Richard and {Calabrese}, Erminia and {Capalbo}, Valentina and {Carrero}, Felipe and {Chen}, Shi-Fan and {Chesmore}, Grace and {Cho}, Hsiao-mei and {Choi}, Steve K. and {Clark}, Susan E. and {Cothard}, Nicholas F. and {Coughlin}, Kevin and {Coulton}, William and {Crichton}, Devin and {Crowley}, Kevin T. and {Darwish}, Omar and {Devlin}, Mark J. and {Dicker}, Simon and {Duell}, Cody J. and {Duff}, Shannon M. and {Duivenvoorden}, Adriaan J. and {Dunkley}, Jo and {Dunner}, Rolando and {Embil Villagra}, Carmen and {Fankhanel}, Max and {Farren}, Gerrit S. and {Ferraro}, Simone and {Foster}, Allen and {Freundt}, Rodrigo and {Fuzia}, Brittany and {Gallardo}, Patricio A. and {Garrido}, Xavier and {Gerbino}, Martina and {Giardiello}, Serena and {Gill}, Ajay and {Givans}, Jahmour and {Gluscevic}, Vera and {Goldstein}, Samuel and {Golec}, Joseph E. and {Gong}, Yulin and {Guan}, Yilun and {Halpern}, Mark and {Harrison}, Ian and {Hasselfield}, Matthew and {Healy}, Erin and {Henderson}, Shawn and {Hensley}, Brandon and {Herv{\'\i}as-Caimapo}, Carlos and {Hill}, J. Colin and {Hilton}, Gene C. and {Hilton}, Matt and {Hincks}, Adam D. and {Hlo{\v{z}}ek}, Ren{\'e}e and {Ho}, Shuay-Pwu Patty and {Hood}, John and {Hornecker}, Erika and {Huber}, Zachary B. and {Hubmayr}, Johannes and {Huffenberger}, Kevin M. and {Hughes}, John P. and {Ikape}, Margaret and {Irwin}, Kent and {Isopi}, Giovanni and {Joshi}, Neha and {Keller}, Ben and {Kim}, Joshua and {Knowles}, Kenda and {Koopman}, Brian J. and {Kosowsky}, Arthur and {Kramer}, Darby and {Kusiak}, Aleksandra and {Lague}, Alex and {Lakey}, Victoria and {Lee}, Eunseong and {Li}, Yaqiong and {Li}, Zack and {Limon}, Michele and {Lokken}, Martine and {Lungu}, Marius and {MacCrann}, Niall and {MacInnis}, Amanda and {Madhavacheril}, Mathew S. and {Maldonado}, Diego and {Maldonado}, Felipe and {Mallaby-Kay}, Maya and {Marques}, Gabriela A. and {van Marrewijk}, Joshiwa and {McCarthy}, Fiona and {McMahon}, Jeff and {Mehta}, Yogesh and {Menanteau}, Felipe and {Moodley}, Kavilan and {Morris}, Thomas W. and {Mroczkowski}, Tony and {Naess}, Sigurd and {Namikawa}, Toshiya and {Nati}, Federico and {Nerval}, Simran K. and {Newburgh}, Laura and {Nicola}, Andrina and {Niemack}, Michael D. and {Nolta}, Michael R. and {Orlowski-Scherer}, John and {Pagano}, Luca and {Page}, Lyman A. and {Pandey}, Shivam and {Partridge}, Bruce and {Perez Sarmiento}, Karen and {Prince}, Heather and {Puddu}, Roberto and {Qu}, Frank J. and {Ragavan}, Damien C. and {Ried Guachalla}, Bernardita and {Rogers}, Keir K. and {Rojas}, Felipe and {Sakuma}, Tai and {Schaan}, Emmanuel and {Schmitt}, Benjamin L. and {Sehgal}, Neelima and {Shaikh}, Shabbir and {Sherwin}, Blake D. and {Sierra}, Carlos and {Sievers}, Jon and {Sif{\'o}n}, Crist{\'o}bal and {Simon}, Sara and {Sonka}, Rita and {Spergel}, David N. and {Staggs}, Suzanne T. and {Storer}, Emilie and {Surrao}, Kristen and {Switzer}, Eric R. and {Tampier}, Niklas and {Thornton}, Robert and {Trac}, Hy and {Tucker}, Carole and {Ullom}, Joel and {Vale}, Leila R. and {Van Engelen}, Alexander and {Van Lanen}, Jeff and {Vargas}, Cristian and {Vavagiakis}, Eve M. and {Wagoner}, Kasey and {Wang}, Yuhan and {Wenzl}, Lukas and {Wollack}, Edward J. and {Zheng}, Kaiwen},
        title = "{The Atacama Cosmology Telescope: DR6 Power Spectra, Likelihoods and $Λ$CDM Parameters}",
      journal = {arXiv e-prints},
         year = 2025,
        month = mar,
          eid = {arXiv:2503.14452},
        pages = {arXiv:2503.14452},
          doi = {10.48550/arXiv.2503.14452},
archivePrefix = {arXiv},
       eprint = {2503.14452},
 primaryClass = {astro-ph.CO},
       adsurl = {https://ui.adsabs.harvard.edu/abs/2025arXiv250314452L}
}

@ARTICLE{madhavacheril20,
       author = {{Madhavacheril}, Mathew S. and {Smith}, Kendrick M. and {Sherwin}, Blake D. and {Naess}, Sigurd},
        title = "{CMB lensing power spectrum estimation without instrument noise bias}",
      journal = {arXiv e-prints},
         year = 2020,
        month = nov,
          eid = {arXiv:2011.02475},
        pages = {arXiv:2011.02475},
          doi = {10.48550/arXiv.2011.02475},
archivePrefix = {arXiv},
       eprint = {2011.02475},
 primaryClass = {astro-ph.CO},
       adsurl = {https://ui.adsabs.harvard.edu/abs/2020arXiv201102475M}
}

@ARTICLE{ge25,
       author = {{Ge}, F. and {Millea}, M. and {Camphuis}, E. and {Daley}, C. and {Huang}, N. and {Omori}, Y. and {Quan}, W. and {Anderes}, E. and {Anderson}, A.~J. and {Ansarinejad}, B. and {Archipley}, M. and {Balkenhol}, L. and {Benabed}, K. and {Bender}, A.~N. and {Benson}, B.~A. and {Bianchini}, F. and {Bleem}, L.~E. and {Bouchet}, F.~R. and {Bryant}, L. and {Carlstrom}, J.~E. and {Chang}, C.~L. and {Chaubal}, P. and {Chen}, G. and {Chichura}, P.~M. and {Chokshi}, A. and {Chou}, T.-L. and {Coerver}, A. and {Crawford}, T.~M. and {de Haan}, T. and {Dibert}, K.~R. and {Dobbs}, M.~A. and {Doohan}, M. and {Doussot}, A. and {Dutcher}, D. and {Everett}, W. and {Feng}, C. and {Ferguson}, K.~R. and {Fichman}, K. and {Foster}, A. and {Galli}, S. and {Gambrel}, A.~E. and {Gardner}, R.~W. and {Goeckner-Wald}, N. and {Gualtieri}, R. and {Guidi}, F. and {Guns}, S. and {Halverson}, N.~W. and {Hivon}, E. and {Holder}, G.~P. and {Holzapfel}, W.~L. and {Hood}, J.~C. and {Howe}, D. and {Hryciuk}, A. and {K{\'e}ruzor{\'e}}, F. and {Khalife}, A.~R. and {Knox}, L. and {Korman}, M. and {Kornoelje}, K. and {Kuo}, C.-L. and {Lee}, A.~T. and {Levy}, K. and {Lowitz}, A.~E. and {Lu}, C. and {Maniyar}, A. and {Martsen}, E.~S. and {Menanteau}, F. and {Montgomery}, J. and {Nakato}, Y. and {Natoli}, T. and {Noble}, G.~I. and {Pan}, Z. and {Paschos}, P. and {Phadke}, K.~A. and {Pollak}, A.~W. and {Prabhu}, K. and {Rahimi}, M. and {Rahlin}, A. and {Reichardt}, C.~L. and {Riebel}, D. and {Rouble}, M. and {Ruhl}, J.~E. and {Schiappucci}, E. and {Sobrin}, J.~A. and {Stark}, A.~A. and {Stephen}, J. and {Tandoi}, C. and {Thorne}, B. and {Trendafilova}, C. and {Umilta}, C. and {Vieira}, J.~D. and {Vitrier}, A. and {Wan}, Y. and {Whitehorn}, N. and {Wu}, W.~L.~K. and {Young}, M.~R. and {Zebrowski}, J.~A. and {SPT-3G Collaboration}},
        title = "{Cosmology from CMB lensing and delensed EE power spectra using 2019{\textendash}2020 SPT-3G polarization data}",
      journal = {\prd},
         year = 2025,
        month = apr,
       volume = {111},
       number = {8},
          eid = {083534},
        pages = {083534},
          doi = {10.1103/PhysRevD.111.083534},
archivePrefix = {arXiv},
       eprint = {2411.06000},
 primaryClass = {astro-ph.CO},
       adsurl = {https://ui.adsabs.harvard.edu/abs/2025PhRvD.111h3534G}
}

@ARTICLE{lewis11,
       author = {{Lewis}, Antony and {Challinor}, Anthony and {Hanson}, Duncan},
        title = "{The shape of the CMB lensing bispectrum}",
      journal = {\jcap},
         year = 2011,
        month = mar,
       volume = {2011},
       number = {3},
          eid = {018},
        pages = {018},
          doi = {10.1088/1475-7516/2011/03/018},
archivePrefix = {arXiv},
       eprint = {1101.2234},
 primaryClass = {astro-ph.CO},
       adsurl = {https://ui.adsabs.harvard.edu/abs/2011JCAP...03..018L}
}
\bibliographystyle{apsrev4-1}
\end{document}